\documentclass[a4paper,12pt]{article}

\usepackage[bbgreekl]{mathbbol}
\usepackage{mathrsfs}
\usepackage{graphicx}
\usepackage{amsmath}
\usepackage{amsfonts}
\usepackage{amssymb}
\usepackage{amsthm}
\usepackage{color}
\usepackage{cancel}
\usepackage{float}
\usepackage{appendix}
\usepackage{bm}
\usepackage{upgreek}
\usepackage{comment}
\usepackage{epstopdf}
\usepackage{multirow}
\usepackage{colortbl}
\usepackage{url}
\usepackage{tikz}
\usepackage{cancel}
\usepackage{mathtools}
\usepackage{multirow}
\usepackage[numbers,sort&compress]{natbib}
\usepackage{subcaption}
\usetikzlibrary{decorations.pathreplacing}
\usepackage[
  colorlinks=true,
  linkcolor=blue,
  citecolor=blue,
  urlcolor=black,
  filecolor=black
]{hyperref}
\usepackage[normalem]{ulem}
\usepackage[dvipsnames]{xcolor}
\usepackage{microtype}

\makeatletter
\AtBeginDocument{%
  \immediate\write\@auxout{\string\citation{apsrev42Control}}%
}
\makeatother

\AtBeginDocument{
  \numberwithin{equation}{section}
  \numberwithin{figure}{section}
  \numberwithin{table}{section}
}

\newcommand{\bvec}[1]{\mathbf{#1}}

\newcommand{\vh}{\bvec{h}}
\newcommand{\im}{\mathrm{i}}

\newcommand{\vk}{\bvec{k}}

\newcommand{\vm}{\bvec{m}}
\newcommand{\vn}{\bvec{n}}

\newcommand{\vq}{\bvec{q}}
\newcommand{\vr}{\bvec{r}}
\newcommand{\vs}{\bvec{s}}

\newcommand{\vu}{\bvec{u}}
\newcommand{\vv}{\bvec{v}}

\newcommand{\vx}{\bvec{x}}
\newcommand{\vy}{\bvec{y}}
\newcommand{\vz}{\bvec{z}}

\newcommand{\vG}{\bvec{G}}

\newcommand{\Z}{\mathbb{Z}}
\newcommand{\R}{\mathbb{R}}

\newcommand{\Nb}{N_{\rm b}}
\newcommand{\C}{\mathbb{C}}
\newcommand{\vzero}{\mathbf{0}}

\newcommand{\AvgTr}{\underline{\operatorname{Tr}}}
\newcommand{\ham}{H}
\newcommand{\dos}{\mathcal{D}}
\newcommand{\ccc}{\mathcal{C}}
\newcommand{\ldos}{\widehat{\mathcal{D}}_{\rm loc}}
\newcommand{\lccc}{\widehat{\mathcal{C}}_{\rm loc}}
\newcommand{\pbz}{{\rm PBZ}_{\epsilon}}
\newcommand{\indicator}{\mathbf{1}}
\newcommand{\diff}{\mathop{}\!\mathrm{d}}

\newcommand{\ee}{\mathrm{e}}
\newcommand{\kg}{g}
\newcommand{\kG}{K}
\newcommand{\gGaus}{g_{E_0,\eta}}
\newcommand{\GFermi}{K_{\rm{F}}}
\newcommand{\chiF}{\chi_{\rm{F}}}

\newcommand{\HeffL}{\hat{\ham}_{\mathrm{eff},L}}
\newcommand{\Htilde}{\widetilde{\ham}_{\rm{eff}}}

\newcommand{\JeffL}{\hat{J}_{\mathrm{eff},L}}
\newcommand{\Jtilde}{\widetilde{J}_{\rm{eff}}}
\newcommand{\Pspec}[1]{\mathsf{P}_{#1}}
\newcommand{\Hbar}{\bar{\ham}}
\newcommand{\Tn}[1]{T_{n}(#1)}
\newcommand{\sds}{\mathcal{G}_{\epsilon}}
\newcommand{\ppara}{P_{\parallel}}
\newcommand{\pperp}{P_{\perp}}
\newcommand{\overlap}{\hat{\mathcal{S}}}
\newcommand{\rhigh}{R_{\rm high}}
\newcommand{\eps}{\epsilon}
\newcommand{\tauorth}{\tau}
\newcommand{\rpara}{R_{\parallel}}
\newcommand{\rperp}{R_{\perp}}

\usepackage{authblk}

\graphicspath{{./figs/}}

\title{Momentum-Resolved Electronic Structure for Quasicrystals: Full-Band Spectra and Chern Number}

\author[1]{Donglin Yang}
\author[1]{Huajie Chen}
\author[1]{Dexuan Zhou}
\author[2]{Xiaoxu Li\thanks{\textit{Corresponding. xiaoxuli@bnu.edu.cn}}}

\affil[1]{School of Mathematical Sciences, Beijing Normal University, Beijing 100875, China}
\affil[2]{Faculty of Arts and Sciences, Beijing Normal University, Zhuhai 519087, China}

\date{}

\begin{document}
\maketitle

\begin{abstract}
Quasicrystals lack the translational symmetry that underlies Bloch decomposition and Brillouin-zone integration, making full-band momentum-resolved electronic structure difficult to formulate and compute. 
We develop a systematically convergent reciprocal-space tight-binding framework for a broad class of quasicrystals. 
The method combines two systematically refinable components: a Fourier-module scattering-channel Hamiltonian that yields local spectral and current-current correlation quantities at each physical momentum, and an expanding hierarchy of pseudo-Brillouin zones that converts the resulting local quantities into bulk thermodynamic observables through an exact local-to-global relation.
Applied to the Penrose and Ammann-Beenker models, the framework uncovers full-band momentum-resolved quasibands and a multichannel mechanism for pseudogap formation, both beyond the scope of low-energy effective models. 
It further resolves Zeeman-driven gap closings and reopenings, quantized Chern plateaus, and the phason invariance of bulk spectral and topological observables.
This framework provides a unified reciprocal-space route to full-band spectral and topological properties of quasicrystals.
\end{abstract}

\section{Introduction}
\label{sec:introduction}

Quasicrystals are a class of ordered structures that exhibit long-range order without translational periodicity. 
Unlike conventional crystals described by primitive unit cells, quasicrystals possess quasiperiodic atomic arrangements and can host unconventional electronic phenomena, including singular spectra, anomalous transport, and nontrivial topological phases \cite{Shechtman1984,LevineSteinhardt1984,KohmotoSutherlandTang1987, RocheFujiwara1998, Kraus2012, HuangLiu2018,Ahn2018,Yao2018,MoonKoshinoSon2019,WangLiuHuang2022}.
Such structures can be generated through various approaches, among which higher-dimensional projection methods provide a unified description of quasiperiodic order. 
Specifically, a broad class of quasicrystalline structures can be constructed by projecting selected points from a higher-dimensional periodic lattice onto the physical space \cite{DuneauKatz1985,BaakeGrimm2013}.
Let $D>d$, $\ppara:\R^D\to\R^d$ and $\pperp:\R^D\to\R^{D-d}$ denote the physical and internal space projections, respectively. 
The corresponding quasicrystalline point set is defined as
\begin{equation}
\label{Lambda:quasiCrystal}
\Lambda = \left\{ \ppara \vn :~ \vn\in\Z^{D},~ \pperp\vn\in W \right\} \subset\R^{d},
\end{equation}
where $W$ is an acceptance window in the internal space. 
Further details of this cut-and-project construction and the quasicrystalline geometries are provided in
Appendix \ref{app:real_reciprocal_geometry}.
For such quasicrystalline structures, theoretical and computational tools capable of resolving spectral, transport, and topological properties remain under development, largely due to the absence of translational symmetry, which prevents a direct application of Bloch theory and conventional momentum-space descriptions.

Tight-binding methods provide an efficient framework for studying electronic structures of large and structurally complex systems by representing electronic states as linear combinations of localized atomic orbitals \cite{SlaterKoster1954,Harrison1980,Goringe1997}.
Recent advances in high-fidelity tight-binding models, including machine-learning based Hamiltonian constructions trained on first-principles data, have further improved their predictive accuracy while preserving computational efficiency \cite{LiWangZou2022,GuZhouyinPandey2024}. 
These approaches are capable of describing important electronic properties, such as density of states and transport properties, with accuracy approaching that of first-principles calculations.
These advantages make tight-binding methods a natural choice for studying quasicrystalline systems, where large structural complexity and the absence of translational symmetry present unique challenges for electronic structure calculations.

Several reciprocal-space approaches have been developed to describe electronic structures of quasicrystalline systems, which can be broadly classified according to how quasiperiodicity is represented. 
The most widely used strategy is the supercell method, where quasiperiodic structures are replaced by large periodic supercells, allowing conventional Bloch theory and first-principles techniques to be applied \cite{Mele2010,Trambly2014}. 
Although effective for capturing local structural and electronic properties, such approximations inevitably introduce artificial periodicity and may become computationally prohibitive for approaching the quasiperiodic limit.
Another strategy is to exploit the higher-dimensional periodic structures underlying cut-and-project quasicrystals, providing a natural framework for describing quasiperiodic order in reciprocal space \cite{JiangZhang2014,JiangLiZhang2024}. 
However, their application to electronic structure calculations often requires additional formulations to connect the higher-dimensional description with physical observables.
More recently, direct reciprocal-space formulations have been developed for incommensurate systems by exploiting the intrinsic periodicity of individual constituents without imposing periodicity on the entire structure \cite{MassattCarrLuskinOrtner2018,ZhouChenZhou2019,MassattCarrLuskin2020,WangChenZhouZhouMassatt2025}. 
Such approaches provide efficient descriptions of electronic states, density of states, and transport properties in various incommensurate systems, including multilayers, but remain closely tied to multilayer geometries.
Complementary to these full-scale approaches, low-energy $k\cdot p$ theories have revealed important quasiperiodic electronic and topological phenomena near selected momenta \cite{WangLiuHuang2022,LopesDosSantos2007,BistritzerMacDonald2011}. 
Despite these advances, a general reciprocal-space framework capable of treating arbitrary quasicrystals and systematic electronic structure calculations remains unavailable.

To address this gap, we develop a systematically convergent local-to-global reciprocal-space tight-binding framework for a broad class of cut-and-project quasicrystals. 
The framework generalizes the two key ingredients of Bloch theory to the quasiperiodic setting: the fixed-momentum Hamiltonian and Brillouin-zone integration.
At each physical momentum, Fourier-module shifts generate a coupled scattering-channel Hamiltonian from which momentum-resolved spectral and current-current correlation quantities are constructed. 
An exact local-to-global identity then expresses the corresponding thermodynamic-limit observables as full reciprocal-space means of these local quantities, which are approximated by an expanding hierarchy of diffraction-guided pseudo-Brillouin zones.
Together, systematic refinement is achieved by enlarging both the scattering-channel space and the pseudo-Brillouin zone, which provides a controlled route for calculating bulk observables without introducing periodic approximants or a low-energy projection. 
We apply the framework to the Penrose and Ammann-Beenker models, revealing full-band momentum-resolved quasiband structures and the multichannel redistribution of spectral weight underlying pseudogaps, including spectral features beyond the validity range of low-energy effective models. 
We further resolve Zeeman-driven gap closings and reopenings together with the associated quantized Chern plateaus, and numerically demonstrate the invariance of bulk spectral and topological observables under uniform phason shifts.

The rest of this paper is organized as follows. Section \ref{sec:tight_binding_real_space} establishes the real-space tight-binding formulation for quasicrystals and introduces the thermodynamic-limit expressions of bulk observables, including the density of states and current-current correlation function.
Section \ref{sec:tight_binding_reciprocal_space} introduces a reciprocal-space framework by constructing local spectral quantities at fixed momentum and averaging them over a hierarchy of diffraction-guided pseudo-Brillouin zones to obtain bulk observables.
Section \ref{sec:applications_simulations} applies the framework to two representative quasicrystals, Penrose and Ammann-Beenker tilings, demonstrating its convergence and resolving their momentum-resolved full-band spectra, Zeeman-driven Chern transitions, and phason-invariant bulk observables.
Section \ref{sec:conclusion} concludes with a summary and possible extensions. 
Explicit quasicrystalline geometries, theoretical derivations, and numerical implementation details are provided in the Appendices.

\section{Real-space formulation of tight-binding observables}
\label{sec:tight_binding_real_space}

Consider a quasicrystal with atomic structure $\Lambda\subset\R^d$ as defined in \eqref{Lambda:quasiCrystal}.
We first establish the real-space tight-binding formulation, which provides the reference definitions of bulk observables that will be reformulated in reciprocal-space in the following section.

Within the tight-binding framework, electronic states are expanded in a set of localized atomic orbitals $\big\{|\phi_{\vr \alpha}\rangle\big\}$, where $\vr\in\Lambda$ denotes the atomic site and $1\leq \alpha\leq \Nb$ labels the orbitals associated with each site.
We assume an orthonormal localized basis for simplicity. 
A nonorthogonal atomic basis can be transformed into this representation through the symmetric L\"owdin orthonormalization \cite{Lowdin1950} with
\begin{align}
\label{L-transform}
|\phi_{\vr\alpha}\rangle^{\rm ortho}  = 
\sum_{\vr'\alpha'} S^{-1/2}_{\vr\alpha,\vr'\alpha'}|\phi_{\vr'\alpha'}\rangle
\qquad{\rm with}\qquad
S_{\vr\alpha,\vr'\alpha'} = \langle\phi_{\vr \alpha} | \phi_{\vr' \alpha'}\rangle .
\end{align}
In the following, the orthonormalized basis is denoted again by $|\phi_{\vr\alpha}\rangle$.
The tight-binding Hamiltonian is represented as
\begin{equation}
\label{ham:real-space}
\ham = \sum_{\vr,\vr'\in\Lambda}
\sum_{1\leq \alpha,\alpha'\leq \Nb}
|\phi_{\vr \alpha}\rangle t_{\vr\alpha,\vr'\alpha'} \langle \phi_{\vr' \alpha'}| ,
\end{equation}
where $t_{\vr\alpha,\vr'\alpha'}=\langle \phi_{\vr \alpha} |\ham| \phi_{\vr' \alpha'}\rangle$ denotes the  onsite or hopping matrix element, with
$t_{\vr\alpha,\vr'\alpha'}=t^*_{\vr'\alpha',\vr\alpha}$.
In tight-binding models, these matrix elements are described by parameterized forms determined from empirical models or fitted to first-principles calculations.
For example, Slater--Koster parametrizations express the hopping amplitudes in terms of orbital types and relative atomic positions \cite{SlaterKoster1954,PapaconstantopoulosMehl2003}, while recent machine-learning Hamiltonians provide high-fidelity representations trained on first-principles data \cite{LiWangZou2022,ZhongYuSuGongXiang2023,GuZhouyinPandey2024}.
For the quasicrystal $\Lambda$, $\ham$ defines an infinite-dimensional operator indexed by $\Lambda\times\{1, \cdots, \Nb\}$, with finite orbital blocks $H_{\vr\vr'}\in\C^{\Nb\times\Nb}$.

Many bulk observables can be characterized through ``averaged" trace of spectral functions of the Hamiltonian.
Let $\Pspec{\ham}$ denote the projection-valued spectral measure of $\ham$ \cite{ReedSimon1980}: for every Borel set $\Delta\subset\R$, $\Pspec{\ham}(\Delta)$ is the projector onto the spectral subspace associated with energies in $\Delta$.
For a spectral test function $\kg:\R\rightarrow\R$, the corresponding bulk quantity is characterized by
\begin{equation}
\label{eq:dos}
\dos(\kg) := \AvgTr\!\left( \int_{\R}\kg(E)\,\Pspec{\ham}(\diff E) \right) 
:= \lim_{R\to\infty} \frac{1}{|\Lambda_R|} \operatorname{Tr}\!\left( \kg\!\left(\ham_{\Lambda_R}\right) \right) ,
\end{equation}
where $\ham_{\Lambda_R}$ is the restriction of Hamiltonian $\ham$ acting on a finite restriction on the quasicrystalline subset $\Lambda_R=\Lambda\cap B_R$. 
We can assume throughout that this limit exists and is independent of the chosen thermodynamic exhaustion, see \cite{LenzStollmann2003,LenzStollmann2005,LenzPeyerimhoffVeselic2007} for theoretical justifications.
The choice of the test function $\kg(E)$ for some physical observable determines both the energy region being probed and the relative weight assigned to its spectral contributions: spectral weight near energy $E$ enters $\dos(\kg)$ with the factor $\kg(E)$. 
For example, choosing $\kg(E)=\delta(E-E_0)$ gives the density of states at energy $E_0$.
The Fermi-Dirac distribution $\kg(E)=f_{\beta,\mu}(E):=(\ee^{\beta(E-\mu)}+1)^{-1}$ describes the electronic occupation, while the kernel involving $\kg(E)=Ef_{\beta,\mu}(E)$ yields the corresponding energy contribution.

While spectral functions characterize equilibrium electronic properties, transport coefficients and topological quantities can be formulated through current-current correlation functions. 
For a two-energy correlation kernel $\kG:\R\times\R\to\R$, the corresponding function is given by the following $d\times d$ tensor
\begin{equation}
\label{eq:current_correlation}
\ccc(\kG) = \bigg(\AvgTr~\Big(\iint_{\R^2} \kG(E,E')\,\Pspec{\ham}(\diff E)\,J_i\, \Pspec{\ham}(\diff E')\,J_j
\Big)\bigg)_{1\leq i,j\leq d} ,
\end{equation}
where $J_i=-ev_i$ is the current operator along the $i$-th direction with $v_i=(\im /\hbar)[\ham,X_i]$ the velocity operator and $X_i = \sum_{\vr\in\Lambda}\sum_{1\leq \alpha\leq\Nb} \vr_i |\phi_{\vr\alpha}\rangle\langle \phi_{\vr\alpha}|$ the position operator associated with the atomic coordinates in the orthonormal localized basis.
Here trace-per-site $\AvgTr$ is the same thermodynamic limit introduced in \eqref{eq:dos}.
In this representation, the spectral projections $\Pspec{\ham}(\diff E)$ and $\Pspec{\ham}(\diff E')$ select the spectral components of the Hamiltonian at energies $E$ and
$E'$, respectively. 
The two-energy correlation kernel $\kG(E,E')$ specifies how correlations between these spectral components contribute to the observable. 
Different choices of $\kG$ therefore lead to different transport or topological quantities.
For example, the Kubo-Greenwood kernel $\kG(E,E')=\delta(E'-E-\hbar\omega) \big(f_{\beta,\mu}(E)-f_{\beta,\mu}(E')\big) / (E'-E)$ produces the correlation term entering the dissipative optical conductivity at frequency $\omega$ \cite{Kubo1957,Greenwood1958,EtterMassattLuskinOrtner2020}.
By contrast, the antisymmetric combination $\ccc_{ij}(\kG)-\ccc_{ji}(\kG)$ of the Fermi kernel 
\begin{align*}
\GFermi(E,E')=\chiF(E)(1-\chiF(E'))/(E-E')^2
\qquad{\rm with}\quad
\chiF(E)=\indicator_{(-\infty,E_{\rm{F}})}(E) 
\end{align*} 
provides the contribution entering the real-space formulation of the
Chern number \cite{Bellissard1994}.

To evaluate the bulk observables for a quasicrystal by the preceding formulas \eqref{eq:dos} and \eqref{eq:current_correlation}, one usually resorts to a supercell approach.
In particular, a finite subset
$\Lambda_R$ is selected and the restricted Hamiltonian $\ham_{\Lambda_R}$ is constructed with appropriate boundary conditions.
However, this type of approach could introduce periodicity or boundary effects and often exhibits slow convergence toward the thermodynamic limit \cite{JohnstoneColbrookNielsenOhbergDuncan2022,ThickeWatsonLu2021,ColbrookHorningThickeWatson2023}.
To overcome these limitations, we reformulate the Hamiltonian and bulk observables in reciprocal space in the following section.
Unlike periodic crystals, where Bloch theory directly provides a momentum representation, quasicrystals lack translational symmetry and therefore do not possess a conventional Brillouin zone.

\section{Momentum-resolved reciprocal-space formulation}
\label{sec:tight_binding_reciprocal_space}

The real-space formulation in Section \ref{sec:tight_binding_real_space} defines bulk spectral and correlation observables through thermodynamic limits. 
In this section, we develop a reciprocal-space formulation of the same quantities for quasicrystals. 
The framework consists of two complementary components: a momentum-resolved local formulation and a global reciprocal-space averaging procedure.

At the local level, for each physical momentum $\vk\in\R^d$, the Fourier module generates coupled scattering channels that define a reciprocal-space tight-binding Hamiltonian and the associated local density of states and local current-current correlation function. 
At the global level, these local quantities are averaged over a hierarchy of diffraction-guided pseudo-Brillouin zones (PBZs), which provides systematically refinable approximations to the reciprocal-space average and recovers the bulk observables defined in Section \ref{sec:tight_binding_real_space}. 
This local-to-global construction is first developed for the density of states and subsequently extended to current-current correlation functions. 
Finally, the relationship with existing reciprocal-space approaches is discussed.

\subsection{Momentum-resolved local density of states}
\label{subsec:kspace_hamiltonian_overlap}

We first introduce reciprocal-space states obtained from the Fourier transformation of localized orbitals. 
Unlike periodic crystals, this construction does not rely on Bloch’s theorem and does not generate eigenstates of the Hamiltonian. 
Instead, it provides a reciprocal-space representation in terms of coupled scattering channels.
For a wave vector $\vk\in\R^d$ in the reciprocal space and local orbital index $\alpha$, we define the normalized plane-wave state on $\Lambda_R$
\begin{equation}
\label{basis:bloch}
|\psi^R_{\vk\alpha}\rangle = \frac{1}{\sqrt{|\Lambda_R|}}\sum_{\vr\in\Lambda_R} \ee^{-\im\vk\cdot\vr} |\phi_{\vr\alpha}\rangle.
\end{equation}
The thermodynamic reciprocal-space quantities below are defined by taking $R\to\infty$ at the level of matrix elements of these states. Moreover, $|\psi^R_{\vk\alpha}\rangle$ represents a reciprocal-space plane-wave state rather than a conventional Bloch state. 
In periodic systems, Bloch states at momenta differing by reciprocal lattice vectors are equivalent because of translational symmetry. 
In contrast, quasicrystals lack a translational lattice, and the above reciprocal-space states generally remain distinct even when their momenta differ by vectors in the Fourier module.

Although the real-space orbitals $|\phi_{\vr\alpha}\rangle$ are assumed to form an orthonormal basis, the reciprocal-space states $|\psi^R_{\vk\alpha}\rangle$ are generally nonorthogonal. 
This nonorthogonality originates from the absence of the discrete reciprocal lattice structure that guarantees the orthogonality relation of Bloch states in periodic crystals.

We next transform the real-space tight-binding Hamiltonian into this reciprocal representation. For a fixed reference momentum $\vk\in\R^d$, the coupling between two reciprocal-space states satisfies a selection rule determined by the Fourier module of the quasicrystal:
\begin{equation*}
\vk'-\vk\in\Lambda^* := \left\{ \ppara^*\vn :~ \vn\in\Z^D \right\} \subset\R^d .
\end{equation*}
Here $\Lambda^*$ is the Fourier module generated by the higher-dimensional construction, with $\ppara^*$ denoting the reciprocal-space projection dual to the real-space projection. A detailed derivation of this coupling rule is provided in Appendix \ref{app:coupling_rule}.

Unlike a reciprocal lattice of a periodic crystal, $\Lambda^*$ is generally not a discrete lattice and is dense in the physical reciprocal space. 
However, only a subset of Fourier components carries significant diffraction amplitudes. 
These vectors determine the dominant scattering channels of the quasiperiodic structure, as illustrated in Figure \ref{fig:quasi-structure}.

We therefore label the coupled reciprocal-space states by
\begin{align*}
   \vk+\vG , \qquad \vG\in\Lambda^* ,
\end{align*}
and construct a reciprocal-space Hamiltonian at each reference momentum $\vk$. 
The matrix elements in this scattering-channel basis are given by
\begin{align*}
\hat{\ham}(\vk)_{\vG \alpha,\vG' \alpha'}
:= & \lim_{R\to\infty} \langle \psi^R_{\vk+\vG \alpha} |\ham| \psi^R_{\vk+\vG' \alpha'} \rangle \notag
\\[1ex]
= & \lim_{R\to\infty} \frac{1}{|\Lambda_R|} \sum_{\vr,\vr'\in\Lambda_R} \ee^{\im(\vk+\vG)\cdot\vr} ~t_{\vr \alpha,\vr' \alpha'}~ \ee^{-\im(\vk+\vG')\cdot\vr'} .
\end{align*}
The above thermodynamic limit average over the quasicrystalline structure corresponds to the trace-per-site convention introduced in the real-space formulation.
The resulting Hamiltonian should be interpreted as a momentum-dependent representation acting on coupled scattering channels, rather than as a finite-dimensional Bloch Hamiltonian.

Because the reciprocal-space states $|\psi^R_{\vk\alpha}\rangle$ are nonorthogonal, the overlap matrix must be included in the formulation. 
The overlap between two scattering channels is given by
\begin{align*}
\hat{S}(\vk)_{\vG\alpha,\vG'\alpha'}
:= \lim_{R\to\infty}\big\langle \psi^R_{\vk+\vG\alpha} \big| \psi^R_{\vk+\vG'\alpha'} \big\rangle
= \delta_{\alpha\alpha'} F(\vG-\vG') ,
\end{align*}
where
\begin{align}
\label{structure_factor}
F(\vG) := & \lim_{R\to\infty} \frac{1}{|\Lambda_R|} \sum_{\vr\in\Lambda_R} \ee^{\im \vG\cdot\vr} 
\end{align}
is the normalized structure factor of the quasicrystalline configuration.
The structure factor satisfies $|F(\vG)|\leq 1$, and becomes a discrete selection rule in periodic systems. 
Importantly, $\hat{S}(\vk)$ depends only on the difference between scattering channels $\vG-\vG'$, but not on the wave vector $\vk$.
We henceforth denote it simply by $\hat{S}\equiv\hat{S}(\vk)$.

We can then follow the L\"owdin orthonormalization procedure to construct an orthonormal reciprocal-space representation.
For definiteness, we define the truncated scattering-channel set $\Lambda_L^* := \{\ppara^*\vn :\vn\in\Z^D\cap B_L \}\subset\Lambda^*$, where $B_L\subset\R^D$ is the higher-dimensional ball of radius $L$. We then denote by $\hat{\ham}_L(\vk)$ and $\hat S_L$ the corresponding finite-dimensional restrictions of $\hat{\ham}(\vk)$ and $\hat S$ to $\Lambda_L^*$.
The ``effective" reciprocal-space Hamiltonian is then given by
\begin{equation}
\label{eq:Heff}
\HeffL(\vk) :=\hat S_L^{-1/2}\hat{\ham}_L(\vk)\hat S_L^{-1/2}.
\end{equation}
Here, $\hat S_L^{-1/2}$ is defined on the nonsingular subspace of the finite overlap matrix $\hat S_L$.
Although periodic systems exhibit exact linear dependencies due to equivalent Bloch states, quasicrystals do not possess such exact degeneracies. 
Nevertheless, the overlap matrix may become highly ill-conditioned as the number of scattering channels increases, requiring appropriate channel truncation or regularization.

With this reciprocal-space representation, we can define local spectral quantities at any momentum $\vk\in\R^d$. For a spectral test function $\kg(E)$, the momentum-resolved local density of states is defined from the central scattering channel $\vG=\vzero$ as
\begin{equation}
\label{eq:ldos_def}
\ldos(\vk;\kg)
:= \lim_{R\to\infty}\sum_{\alpha=1}^{\Nb}\big\langle \psi^R_{\vk\alpha}\big|\kg(\ham)\big|\psi^R_{\vk\alpha} \big\rangle 
= \lim_{L\to\infty}\sum_{\alpha=1}^{\Nb}\left[\hat S_L^{1/2}\kg\!\left(\HeffL(\vk)\right)\hat S_L^{1/2} \right]_{\vzero\alpha,\vzero\alpha}.
\end{equation}
For each finite $L$, this spectral matrix element for the original nonorthogonal $\vG=\vzero$ channel is evaluated in the orthonormalized scattering-channel representation. In this representation, the L\"owdin transformation introduces the factor $\hat S_L^{1/2}$ on each side, while the spectral operator is given by $\kg(\HeffL(\vk))$. The local spectral quantity $\ldos(\vk;\kg)$ is then recovered in the limit $L\to\infty$, with the retained scattering-channel set progressively enlarged.

For numerical implementation, finite scattering-channel sets can be constructed using different systematic truncation schemes.
In our calculations, we consider two typical choices: a higher-dimensional cutoff and a mixed physical/internal-space cutoff, as described in Appendix~\ref{app:reciprocal_truncation}.
The enlargement of the retained scattering-channel family provides a systematic refinement of the momentum-resolved local Hamiltonian and its associated spectral quantities.
The required matrix functions are evaluated using the kernel polynomial method (KPM) rather than direct diagonalization of the reciprocal Hamiltonian at every momentum.
The numerical methods are discussed in detail in Appendix \ref{app:implementation}.

We point out that the above construction reduces to the conventional Bloch formulation in periodic systems. 
In that case, reciprocal vectors differing by reciprocal lattice vectors represent equivalent Bloch states, and the Hamiltonian at each momentum reduces to a finite-dimensional matrix of size $\Nb\times\Nb$.
Diagonalization therefore produces discrete bands and the corresponding momentum-resolved spectral weights.
For quasicrystals, however, the absence of translational symmetry prevents such a finite-dimensional reduction. 
The reciprocal-space Hamiltonian acts on an infinite set of inequivalent scattering channels, resulting in a local spectral distribution characterized by the test function $\kg$. 
We observe that this local distribution exhibits quasiperiodic patterns inherited from the structure of the Fourier module, as illustrated in Figure \ref{fig:local_responses}.
Moreover, reciprocal space itself no longer contains a conventional Brillouin zone, but instead inherits the quasiperiodic structure of the Fourier module.

The remaining challenge is therefore to recover ``global" bulk observables from these momentum-resolved local quantities. 
In periodic crystals, this is achieved by integrating local spectral functions over the first Brillouin zone. 
For quasicrystals, the absence of translational symmetry makes the definition of such a reciprocal-space integration domain nontrivial. 
This motivates the construction of diffraction-guided pseudo-Brillouin zones in the following section.
This is highly nontrivial for quasicrystals, again due to the lack of translational symmetry.

\subsection{Hierarchy of diffraction-guided pseudo-Brillouin zones}
\label{subsec:pbz_quadrature}

With the momentum-resolved local spectral quantities defined above, the corresponding bulk observables can be recovered by averaging over the full reciprocal space. 
For a test function $\kg$, the local spectral quantity in reciprocal-space introduced in Section \ref{subsec:kspace_hamiltonian_overlap} gives
\begin{equation}
\label{eq:dos_limit_def}
\dos(\kg) = \lim_{R\to\infty} \frac{1}{|B_R|} \int_{B_R} \ldos(\vk;\kg)\,\diff \vk .
\end{equation}
This is equivalent to the real-space thermodynamic-limit expression introduced in Section \ref{sec:tight_binding_real_space}.
The corresponding mathematical analysis has been established for incommensurate multilayer systems \cite{MassattCarrLuskinOrtner2018,WangChenZhouZhouMassatt2025}, and the same averaging argument can be extended to the present quasicrystalline setting.
The central question is therefore whether one can identify a finite reciprocal-space region whose quadrature provides a controlled approximation to the full average.

This question is motivated by the observation that the momentum-resolved local spectral quantities of quasicrystals inherit the quasiperiodic structure of the Fourier module, as illustrated in Figure \ref{fig:local_responses}. 
Instead of uniformly sampling the entire reciprocal space, it is therefore desirable to construct a representative region containing inequivalent momentum points while avoiding redundant sampling of approximately equivalent scattering channels.
For periodic crystals, such a role is played by the first Brillouin zone, which is the Wigner-Seitz cell of the reciprocal lattice. 
In quasicrystals, however, the Fourier module does not form a discrete reciprocal lattice, and no exact fundamental reciprocal-space domain exists. 
We therefore introduce diffraction-guided pseudo-Brillouin zones (PBZs) based on the dominant scattering vectors of the quasiperiodic structure.

\begin{figure}[htbp!]
    \centering
    \begin{subfigure}[t]{0.48\textwidth}
        \centering
        \includegraphics[width=\textwidth]{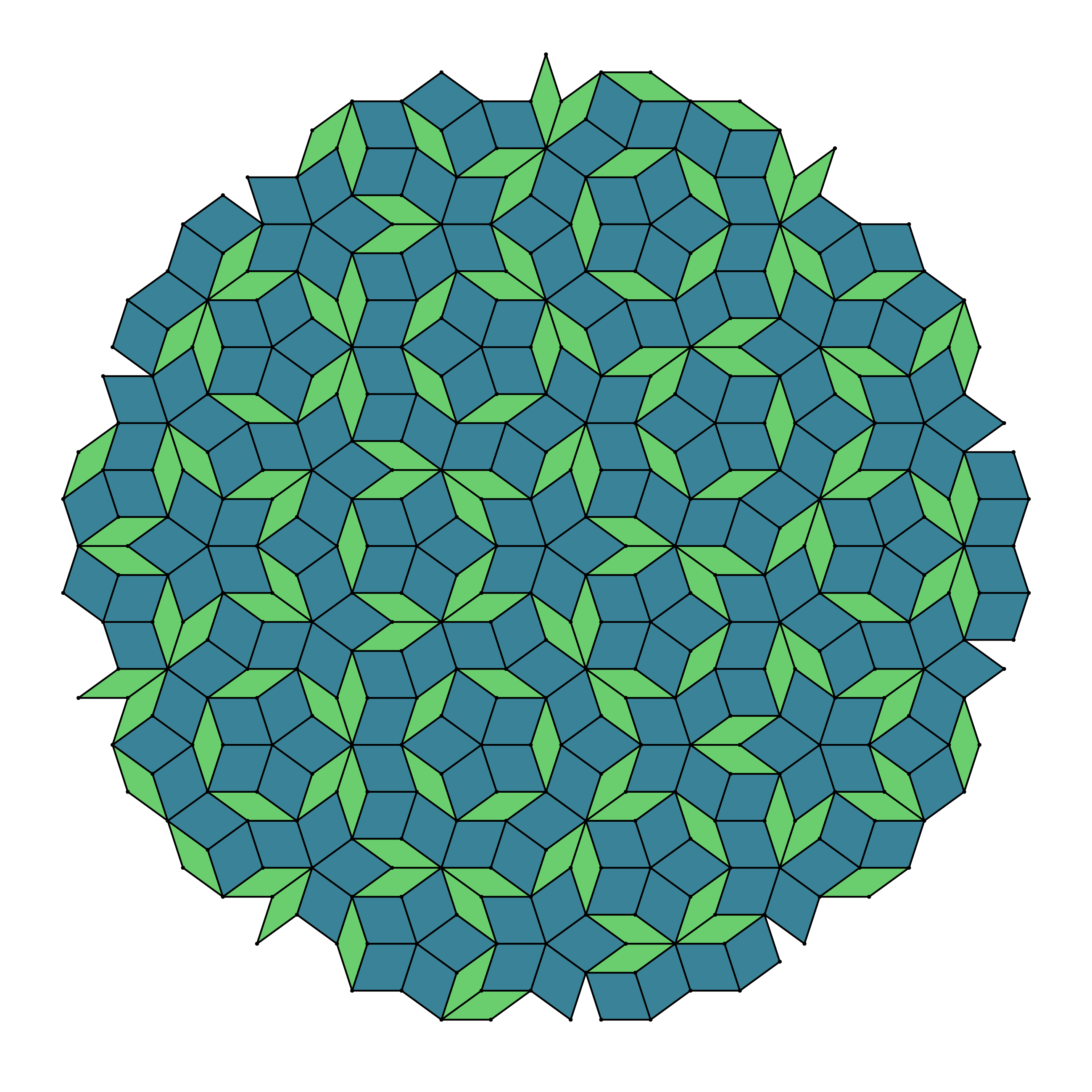}
        \caption{Penrose tiling}
        \label{fig:penrose_real_space_tiling}
    \end{subfigure}
    \hfill
    \begin{subfigure}[t]{0.48\textwidth}
        \centering
        \includegraphics[width=\textwidth]{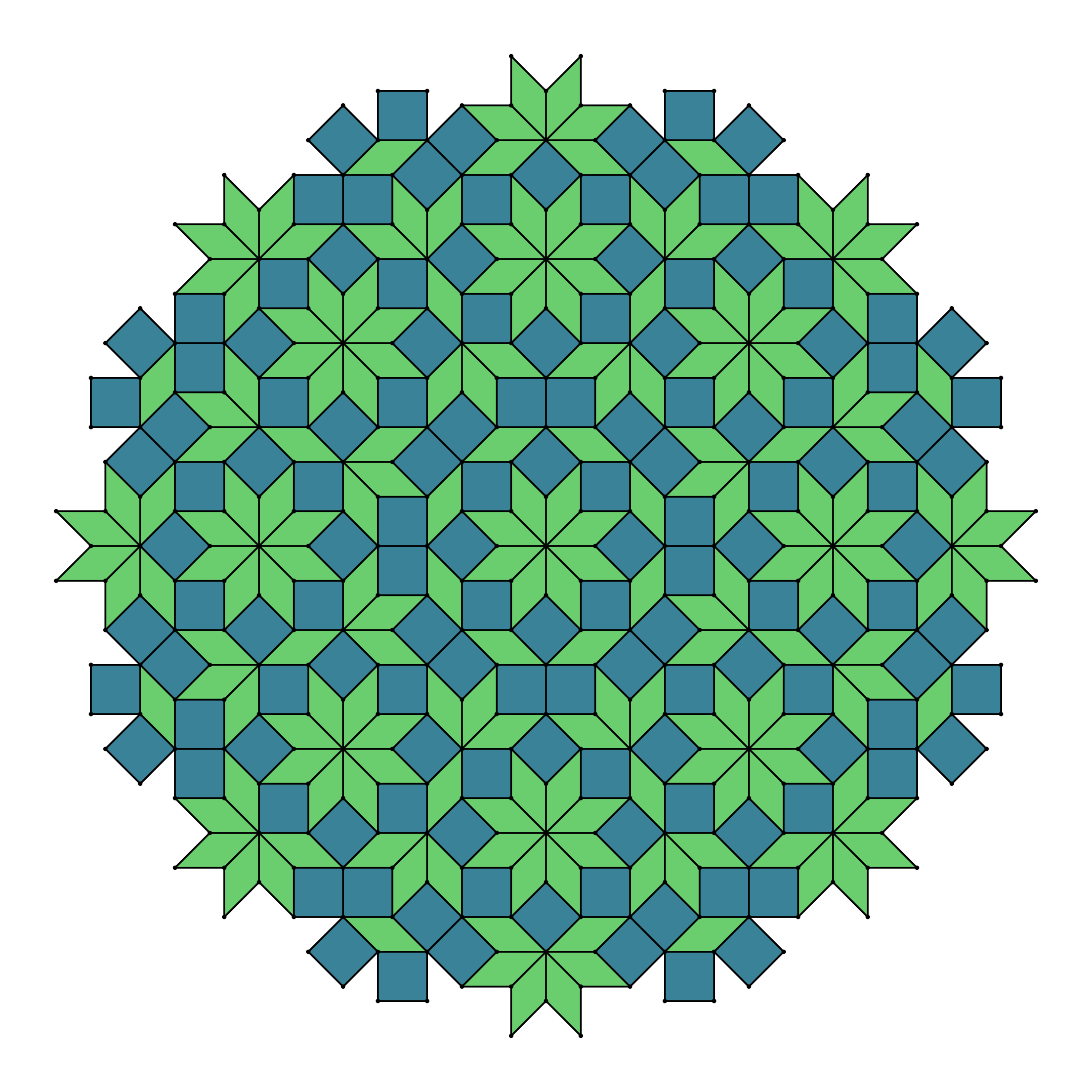}
        \caption{Ammann-Beenker tiling}
        \label{fig:ab_real_space_tiling}
    \end{subfigure}
    \medskip
    \begin{subfigure}[t]{0.48\textwidth}
        \centering
        \includegraphics[width=\textwidth]{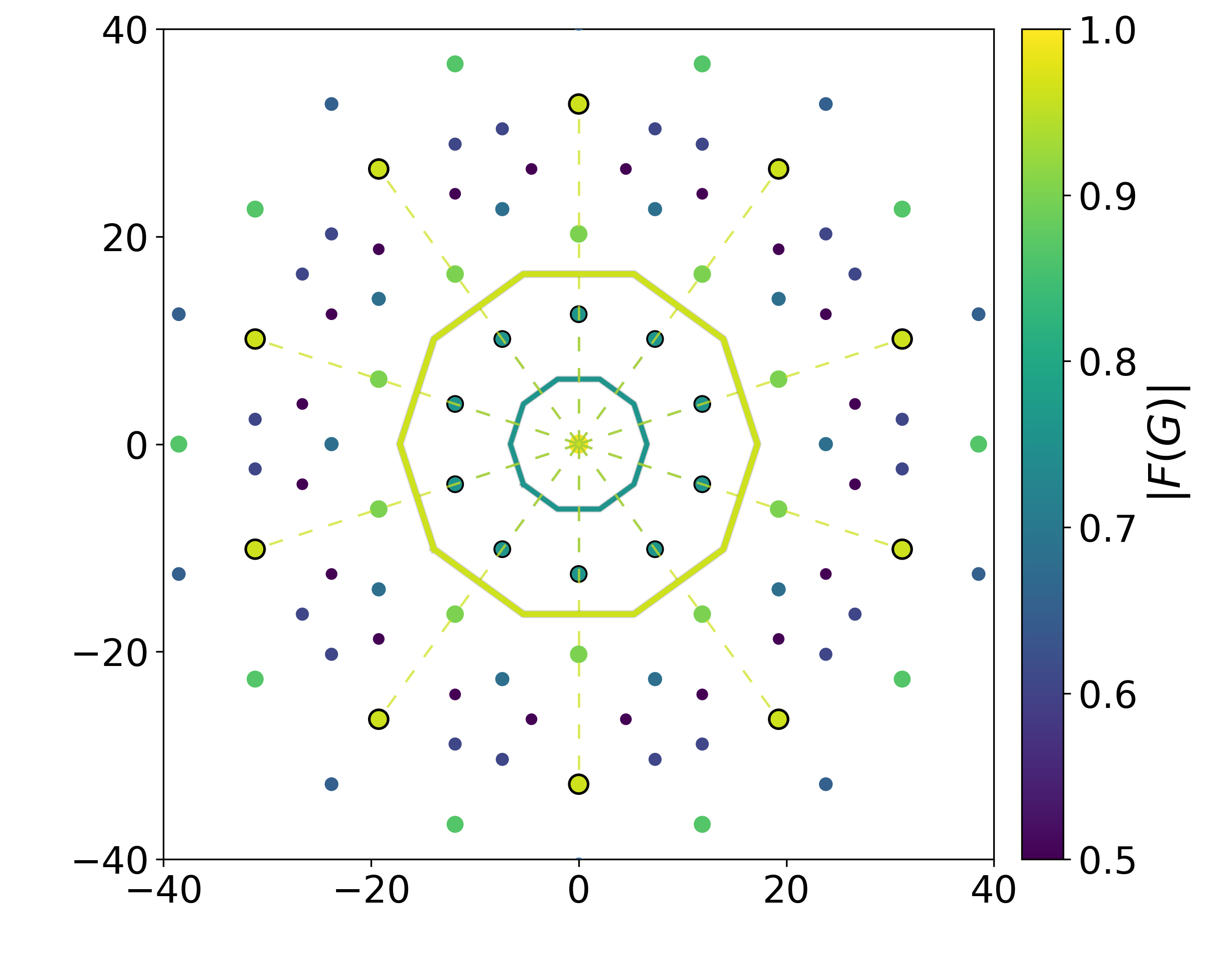}
        \caption{Diffraction of Penrose tiling}
    \label{fig:penrose_diffraction_pattern}
    \end{subfigure}
    \hfill
    \begin{subfigure}[t]{0.48\textwidth}
        \centering
        \includegraphics[width=\textwidth]{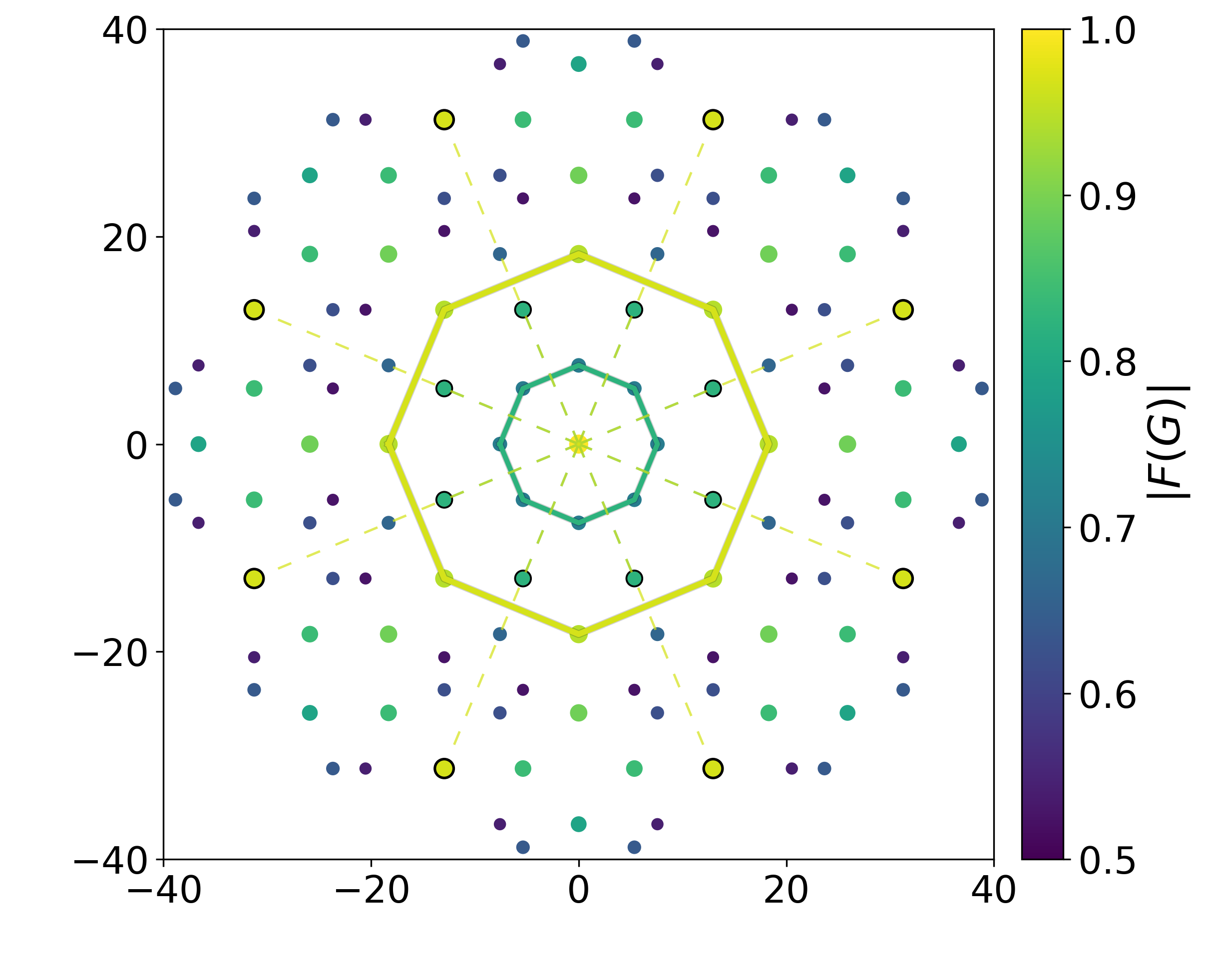}
        \caption{Diffraction of Ammann-Beenker tiling}
        \label{fig:ab_diffraction_pattern}
    \end{subfigure}
    \caption{Real-space configurations and diffraction-guided PBZs of representative quasicrystals. 
    Panels (a)-(b) show the real-space configurations of the Penrose and Ammann-Beenker tilings, respectively. 
    Panels (c)-(d) display the corresponding reciprocal-space diffraction patterns, exhibiting decagonal and octagonal rotational symmetries. The color of each diffraction point represents the normalized diffraction amplitude $|F(\vG)|$.
    Strong diffraction vectors selected according to the threshold parameter are used to construct the Wigner-Seitz cells shown by the overlaid polygons, defining the corresponding PBZs (with thresholds $\eps=0.28$ and $0.04$ generating the smaller and larger PBZs, respectively).
    Decreasing the threshold parameter $\eps$ retains only stronger diffraction vectors and generates a hierarchy of increasingly larger PBZs, providing systematically improvable reciprocal-space quadrature domains.}
    \label{fig:quasi-structure}
\end{figure}

A conventional construction selects a set of reciprocal-space vectors with strong diffraction amplitudes and defines the PBZ as the origin-centered Wigner-Seitz cell generated by these vectors \cite{GambaudoVignolo2014,WangLiuHuang2022}. 
However, a single such cell does not provide a systematic refinement toward the full reciprocal-space average in \eqref{eq:dos_limit_def}.
We therefore vary the selection threshold according to the normalized structure factor $F(\vG)$ defined in \eqref{structure_factor}, thereby generating a hierarchy of PBZs.
The quantity $|F(\vG)|\leq 1$ measures the overlap between two reciprocal-space scattering channels separated by $\vG$. 
A value close to one indicates that the Fourier states at $\vk$ and $\vk+\vG$ become nearly linearly dependent, implying that they represent approximately equivalent reciprocal-space descriptions and determine candidate boundaries of the PBZ.
Let $\eps>0$ be a threshold parameter, we define the set of strong diffraction vectors as
\begin{equation*}
\sds := \left\{ \vG\in\Lambda^*\setminus\{\vzero\} :~ |F(\vG)|\ge 1-\eps \right\}.
\end{equation*}
The PBZ associated with the threshold $\eps$ is defined as the origin-centered Wigner-Seitz cell generated by the selected strong diffraction vectors,
\begin{equation*}
\pbz := \big\{\vk\in\R^d :~ |\vk|\le |\vk-\vG|, ~~
\forall ~\vG\in\sds \big\} .
\end{equation*}
Geometrically, the facets of PBZ lie on the perpendicular bisectors between the origin and the selected strong diffraction vectors in $\sds$.
Although $\sds$ may contain vectors from several diffraction shells, only those whose bisectors form active constraints contribute to the PBZ boundary. For the rotationally symmetric Penrose and Ammann-Beenker examples, these active vectors belong to the innermost symmetry-related shell and generate regular polygonal cells, as shown in Figure \ref{fig:quasi-structure}.
These boundaries separate momentum points according to their nearest high-overlap representatives, thereby reducing repeated sampling of nearly equivalent channel states.
Therefore, integration over a single PBZ samples one representative branch of the reciprocal-space quasiperiodic structure while avoiding redundant contributions.

Using this diffraction-guided domain, the bulk observable can be approximated by an averaged integration of the local quantities \eqref{eq:ldos_def} over the PBZ
\begin{equation*}
\dos(\kg) \approx {\dos}_\eps(\kg)
:= \frac{1}{|\pbz|} \int_{\pbz} \ldos(\vk;\kg)\,\diff \vk.
\end{equation*}
The threshold parameter provides a systematic refinement procedure. 
A smaller $\eps$ imposes a stricter criterion on the channel equivalence and retains only diffraction vectors with structure factors closer to 1. 
Since exact reciprocal equivalence is absent in quasicrystals, these increasingly correlated scattering vectors generally appear at larger reciprocal-space distances, leading to increasingly larger PBZs. 
Therefore, reducing $\eps$ enlarges the quadrature domain and improves the approximation to the full reciprocal-space average. 
Conversely, a larger threshold includes weaker diffraction vectors and produces a smaller PBZ dominated by nearby scattering channels.
The resulting family $\{\pbz\}_{\eps\rightarrow 0}$ provides a hierarchy of quadrature domains that systematically improves the approximation to the full reciprocal-space average, with convergence demonstrated numerically in Figure \ref{fig:pbz_convergence}.

This PBZ approximation for the density of states and other bulk quantities has a direct interpretation. 
The strong diffraction vectors with $|F(\vG)|\approx1$ are associated with small internal-space components $\vG_\perp$. For a fixed $\vk\in\R^d$ and a corresponding Fourier component $\ee^{\im(\ppara\vn)\cdot\vk}$ of a momentum-resolved local quantity, duality gives 
\begin{equation*}
\ee^{\im(\ppara\vn)\cdot(\vk+\vG)}
=
\ee^{-\im(\pperp\vn)\cdot\vG_\perp}
\ee^{\im(\ppara\vn)\cdot\vk}.
\end{equation*}
When the internal-space component $\vG_\perp$ is small, the phase mismatch $\ee^{-\im(\pperp\vn)\cdot\vG_\perp}\approx 1$, so the momentum shift changes the Fourier modes weakly and the local quantities remain approximately invariant. Because opposite PBZ facets are separated by the same strong diffraction vector $\vG$, the correspondence of $\vk$ and $\vk+\vG$ carries the approximate invariance directly to the paired facets. 
This produces an additional cancellation between the boundary contributions, making the PBZ average a natural finite-domain approximation to the full reciprocal-space average. 
A convergence analysis of this mechanism is provided in Appendix \ref{app:pbz-mechanism}.

In summary, the diffraction-guided PBZ construction extends Brillouin-zone quadrature from exact reciprocal periodicity in crystals to approximate reciprocal invariance in quasicrystals.
$\pbz$ is not an exact fundamental domain, but a systematically refinable quadrature domain for evaluating bulk observables from momentum-resolved local quantities.

\subsection{Local-to-global formulation of current-current correlations}
\label{subsec:current_current_local_global}

The local-to-global reciprocal-space framework established in Section \ref{subsec:kspace_hamiltonian_overlap} and \ref{subsec:pbz_quadrature} extends naturally from spectral observables to the two-energy current-current correlation functional introduced in \eqref{eq:current_correlation}. 
At each reference momentum $\vk$, we first construct a local correlation tensor in the coupled scattering-channel representation. 
Averaging this local quantity over the full reciprocal space recovers the corresponding bulk current-current correlation, while the diffraction-guided PBZ hierarchy provides a sequence of finite-domain approximations.
This extension allows transport coefficients and topological responses to be treated within the same reciprocal-space framework as the density of states.

For a fixed momentum $\vk\in\R^d$, we follow the definitions in Section \ref{subsec:kspace_hamiltonian_overlap}, denoting the reciprocal-space Hamiltonian by $\hat{\ham}_L(\vk)$, the overlap matrix by $\hat{S}_L$ and the orthonormalized Hamiltonian by $\HeffL(\vk)$ at the finite scattering-channel cutoff $L$.
At the same cutoff, the reciprocal-space representation of the current operator is
\begin{equation*}
(\hat J_L)_i(\vk):=
(e/\hbar)\partial_{k_i}\hat{\ham}_L(\vk),
\end{equation*}
which is obtained from its real-space definition in Section \ref{sec:tight_binding_real_space} and a direct differentiation of the phase factors entering $\hat{\ham}_L(\vk)$.
In the orthonormalized scattering-channel representation, the current operator becomes
\begin{equation*}
(\JeffL)_{i}(\vk)
:= \hat S_L^{-1/2} (\hat J_L)_i(\vk) \hat S_L^{-1/2}.
\end{equation*}
Both $\HeffL(\vk)$ and $(\JeffL)_i(\vk)$ act on the same family of coupled scattering channels $\vk+\vG$ with $\vG\in\Lambda_L^*$.
As in Section \ref{subsec:kspace_hamiltonian_overlap}, these operators are understood on the retained nonsingular subspace of the truncated overlap matrix.

Let $\Pspec{\vk,L}(\diff E):=\Pspec{\HeffL(\vk)}(\diff E)$ denote the projection-valued spectral measure of the orthonormalized Hamiltonian at cutoff $L$. Then for the correlation kernel $\kG:\R\times\R\rightarrow\R$, the momentum-resolved local current-current correlation tensor is defined by
\begin{multline}
\label{eq:local_correlation_def}
\lccc(\vk;\kG)_{ij}
:= \lim_{R\to\infty} \sum_{\alpha=1}^{\Nb} \Big\langle \psi^R_{\vk\alpha} \Big| \iint_{\R^2} \kG(E,E')\, \Pspec{\ham}(\diff E)\, J_i\, \Pspec{\ham}(\diff E')\, J_j \Big| \psi^
R_{\vk\alpha} \Big\rangle 
\\
= \lim_{L\to\infty} \sum_{\alpha=1}^{\Nb} \bigg[ \hat S_L^{1/2} \iint_{\R^2} \kG(E,E')\, \Pspec{\vk,L}(\diff E)\, (\JeffL)_i(\vk)\, \Pspec{\vk,L}(\diff E')\, (\JeffL)_j(\vk)\, \hat S_L^{1/2} \bigg]_{\vzero\alpha,\vzero\alpha}
\end{multline}
for $1\leq i,j\leq d$.
The last equality follows in the same way as \eqref{eq:ldos_def}, by applying the L\"owdin transformation at each finite channel cutoff and then taking $L\to\infty$. Expressing the resulting correlation matrix element in the original nonorthogonal $\vG=\vzero$ channel introduces the two factors $\hat S_L^{1/2}$.
The kernel $\kG(E,E')$ determines how correlations between spectral components near the two energies $E$ and $E'$ contribute to the observable. 
Different kernels and different tensor components therefore yield different physical quantities. 
In particular, the symmetric or diagonal components enter dissipative transport coefficients, whereas antisymmetric combinations can encode Hall and topological responses.                                
Because momentum shifts by strong Fourier-module vectors generate nearly redundant scattering-channel representations, the local correlation tensor exhibits the same approximate invariance under these shifts as the local spectral quantity. Numerical examples of these momentum-space structures are presented in Figure \ref{fig:local_responses}.

Like the spectral function in \eqref{eq:dos_limit_def}, averaging the local correlation tensor over the full reciprocal space recovers the real-space bulk correlation functional \cite{MassattCarrLuskin2020}, that is
\begin{equation*}
\ccc_{ij}(\kG) =
\lim_{R\to\infty}\frac{1}{|B_R|}\int_{B_R} \lccc(\vk;\kG)_{ij}\,\diff\vk.
\end{equation*}
This full-space average can again be replaced by quadrature over the diffraction-guided PBZs.
For a threshold $\eps>0$, we construct the approximation of $\ccc_{ij}(\kG)$ by
\begin{align*}
{\ccc}_{\eps,ij}(\kG)
:= \frac{1}{|\pbz|} \int_{\pbz} \lccc(\vk;\kG)_{ij}\,\diff\vk .
\end{align*}
The direction of refinement is the same as in Section \ref{subsec:pbz_quadrature}.
Varying the threshold $\eps$ generates a hierarchy $\{\pbz\}_{\eps\rightarrow 0}$, which provides a systematically refined approximation to the full reciprocal-space average. 

Consequently, both single-energy spectral observables and two-energy current-current correlations can be evaluated using the same diffraction-guided PBZ hierarchy. 
The reciprocal-space formulation separates the calculation into a momentum-resolved local problem determined by the coupled scattering-channel Hamiltonian, and a global quadrature problem determined solely by the quasiperiodic diffraction geometry. 
This common local-to-global structure provides a unified route to spectral, transport, and topological observables in quasicrystals.

\subsection{Comparison with existing methods}
\label{subsec:comparison_models_systematic}

A principal feature of the proposed framework is its systematic refinement from momentum-resolved reciprocal-space calculations to bulk observables of the original quasicrystal. 
This refinement is organized at two complementary levels: at the local level, the momentum-dependent Hamiltonian is approximated by enlarging finite families of scattering channels generated from the Fourier module; 
at the global level, the reciprocal-space average is approximated by an expanding hierarchy of diffraction-guided PBZs as $\eps\rightarrow0$. 
Together, these two refinements recover the real-space bulk spectral functional $\dos(\kg)$ and current-current correlation functional $\ccc(\kG)$, providing a common basis for comparison with existing real-space, periodic-approximant, higher-dimensional, multilayer, and low-energy formulations.

\vskip 0.1cm
\noindent
{\bf Real-space and supercell methods.}
Real-space calculations evaluate bulk observables using increasingly large finite or periodic representations of the quasiperiodic structure, but lack a direct momentum-resolved formulation \cite{Trambly2014,MassattLuskinOrtner2017,EtterMassattLuskinOrtner2020,JohnstoneColbrookNielsenOhbergDuncan2022}.
The present framework instead performs the approximation through scattering-channel truncation in reciprocal space, providing the local density of states and local current-current correlation function without introducing a periodic approximation of the original quasicrystal.

\vskip 0.1cm
\noindent
{\bf Higher-dimensional projection methods.}
Projection methods exploit the higher-dimensional periodic representation underlying quasicrystals, at the cost of increasing the dimensionality of the computational problem \cite{JiangZhang2014,JiangLiZhang2024}.
In the present formulation, the scattering channels are generated from the Fourier module, while the momentum $\vk$ remains in the physical reciprocal space.
This separation enables a momentum-resolved description of the full spectrum and provides band-structure information directly in physical momentum space.

\vskip 0.1cm
\noindent
{\bf Layer-based reciprocal-space methods.}
Existing reciprocal-space formulations for incommensurate multilayers exploit the periodicity of the constituent layers to organize the scattering channels \cite{MassattCarrLuskinOrtner2018,ZhouChenZhou2019,MassattCarrLuskin2020,WangChenZhouZhouMassatt2025}.
The present framework instead constructs the scattering channels from the Fourier module, allowing the same reciprocal-space formulation to be applied to a broad class of quasicrystals, including the Penrose and Ammann-Beenker tilings, without requiring a periodic layer decomposition.

\vskip 0.1cm
\noindent
{\bf Low-energy $k\cdot p$ methods.} 
PBZ-centered $k\cdot p$ models employ effective Hamiltonians to describe low-energy electronic structures and their associated topological properties \cite{WangLiuHuang2022}. 
The present formulation instead retains a truncated scattering-channel family from the Fourier module at every physical momentum and is therefore not restricted to the low-energy regime. 
This distinction is demonstrated numerically in Figure \ref{fig:dos_spectral_comparison}, where the two approaches agree in the low-energy regime while the present framework resolves the spectrum of the original tight-binding Hamiltonian beyond this regime.

The present framework combines full-spectrum access to the original quasicrystalline model with momentum resolution and a systematically refinable PBZ quadrature. This enables a unified reciprocal-space treatment of quasibands, bulk gaps, and pseudogaps, as well as transport and topological properties derived from current-current correlations.

\section{Applications}
\label{sec:applications_simulations}

To assess the reciprocal-space framework on geometrically distinct quasicrystalline systems, we consider two canonical two-dimensional examples: the rhombic Penrose and Ammann-Beenker tilings, with their structures denoted by $\Lambda_{\rm{Pen}}$ and $\Lambda_{\rm{AB}}$ respectively.
Their cut-and-project parameters, including the physical and internal projections, acceptance windows, and Fourier modules, are summarized in Appendix \ref{app:real_reciprocal_geometry}.
Each vertex hosts three atomic orbitals $\alpha\in\{s,p_x,p_y\}$, and two spin states $\sigma\in\{\uparrow,\downarrow\}$.
Let $\vs=(s_x,s_y,s_z)$ denote the Pauli matrices acting on the spin degree of freedom, we use the following spinful tight-binding Hamiltonian
\begin{align*}
\ham
&= \sum_{\vr,\alpha,\sigma} E_\alpha |\phi_{\vr \alpha\sigma}\rangle \langle \phi_{\vr \alpha\sigma}| + \sum_{\langle \vr \alpha,\vr'\alpha'\rangle,\sigma} t_{\vr \alpha,\vr'\alpha'} |\phi_{\vr \alpha\sigma}\rangle \langle \phi_{\vr'\alpha'\sigma}|\\
&\quad + \im\lambda \sum_{\vr,\sigma,\sigma'} (s_z)_{\sigma\sigma'} \Big( |\phi_{\vr p_y\sigma}\rangle \langle \phi_{\vr p_x\sigma'}| - |\phi_{\vr p_x\sigma}\rangle \langle \phi_{\vr p_y\sigma'}| \Big) \\
&\quad + \sum_{\vr,\alpha,\sigma,\sigma'} \delta_M (\vm\cdot\vs)_{\sigma\sigma'} |\phi_{\vr \alpha\sigma}\rangle \langle \phi_{\vr \alpha\sigma'}| .
\end{align*}

The first two terms describe the onsite orbital energies and spin-independent hopping, respectively.
Hopping is included between sites connected by tile edges or by the shorter diagonals of the thin rhombi and is parametrized using the two-center Slater--Koster form \cite{SlaterKoster1954}. 
The hopping matrix element $t_{\vr\alpha,\vr'\alpha'}$ is expressed in terms of the Slater--Koster integrals $V_{\mu}(|\vr-\vr'|)$ with bonding type $\mu\in\{ss\sigma,sp\sigma,pp\sigma,pp\pi\}$.
We take their distance dependence as $V_{\mu}(d)=(d_0/d)^2 V_{\mu,0}$ with $d_0$ the reference bond length. 
The third term is an onsite atomic spin–orbit coupling in the $(p_x,p_y)$ orbital subspace, while the last term is an orbital-independent Zeeman coupling of strength $\delta_M$ along direction $\vm$. 
To isolate the effects of quasicrystalline geometry and to facilitate comparison with the low-energy model of \cite{WangLiuHuang2022}, we use the same parameters for both tilings: the onsite energies are $E_s=0.7$ and $E_{p_x}=E_{p_y}=-2.3$, the reference Slater--Koster integrals are $V_{ss\sigma,0}=-V_{sp\sigma,0}=-0.17$ and $V_{pp\sigma,0}=V_{pp\pi,0}=0.34$ with $d_0=1$, and the spin--orbit coupling strength is $\lambda=1.0$. All energies are expressed in electronvolts.

The numerical studies follow the local-to-global framework developed in Section \ref{sec:tight_binding_reciprocal_space}.
We first examine the momentum-resolved spectral and current-current correlation maps and their approximate invariance under shifts by strong diffraction vectors, and then assess the convergence of PBZ quadrature. 
The full-band formulation is further used to analyze gaps and pseudogaps, compare with a $k\cdot p$ model, track Zeeman-driven Chern transitions, and test phason invariance. 
Unless otherwise stated, the reciprocal-space calculations use the mixed scattering-channel cutoff $(\rpara,\rperp)=(15,150)$, with channel-truncation tests and additional numerical details provided in Appendix \ref{app:reciprocal_truncation}.

\subsection{Momentum-resolved local observables and PBZ quadrature}
\label{subsec:local_quantities}

\begin{figure}[htbp!]
    \centering
    \begin{subfigure}[t]{0.47\textwidth}
        \centering
    \includegraphics[width=\textwidth]{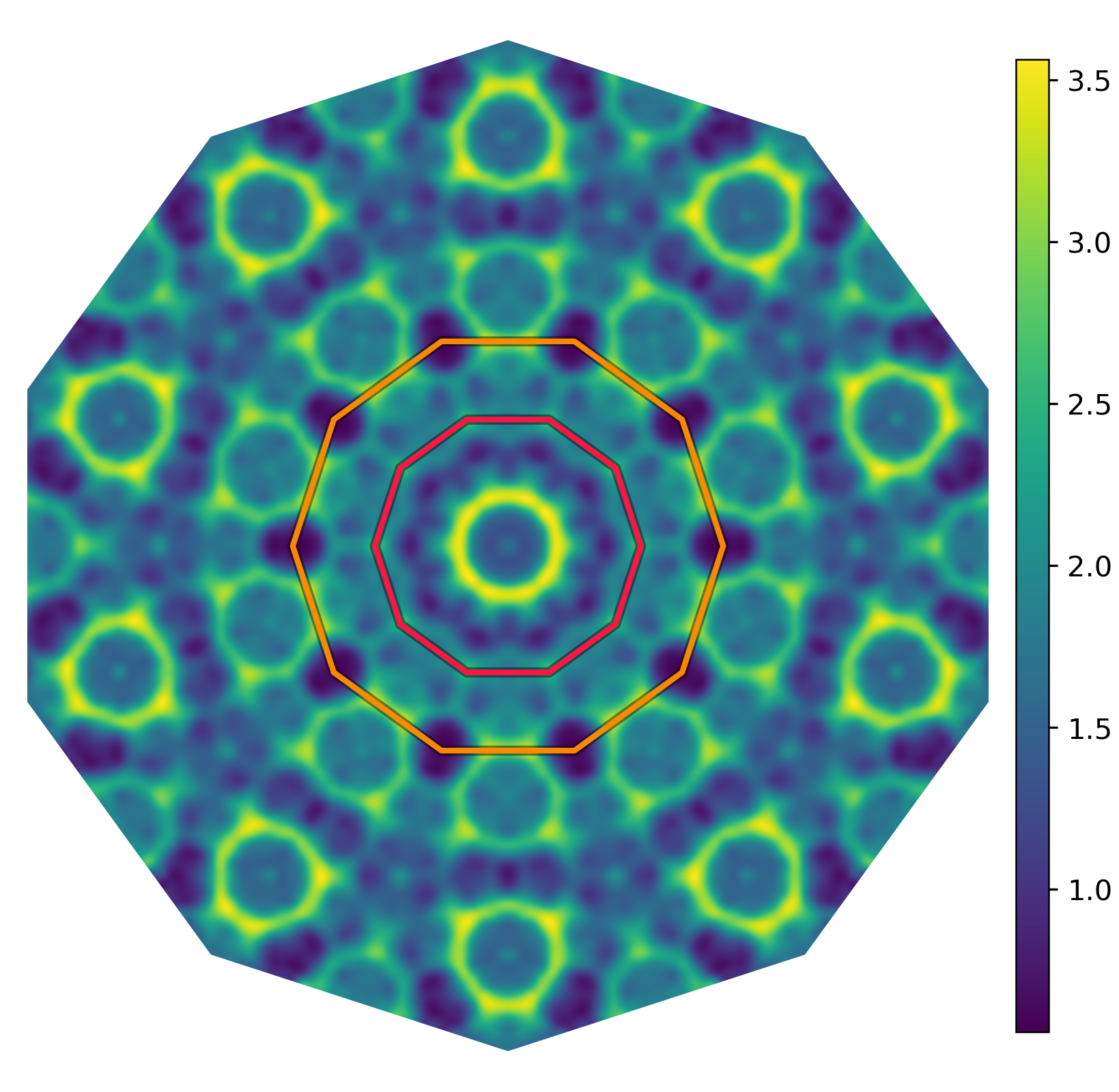}
        \caption{Penrose: local density of states}
        \label{fig:penrose_local_ldos}
    \end{subfigure}
    \hskip 0.2cm
    \begin{subfigure}[t]{0.47\textwidth}
        \centering
    \includegraphics[width=\textwidth]{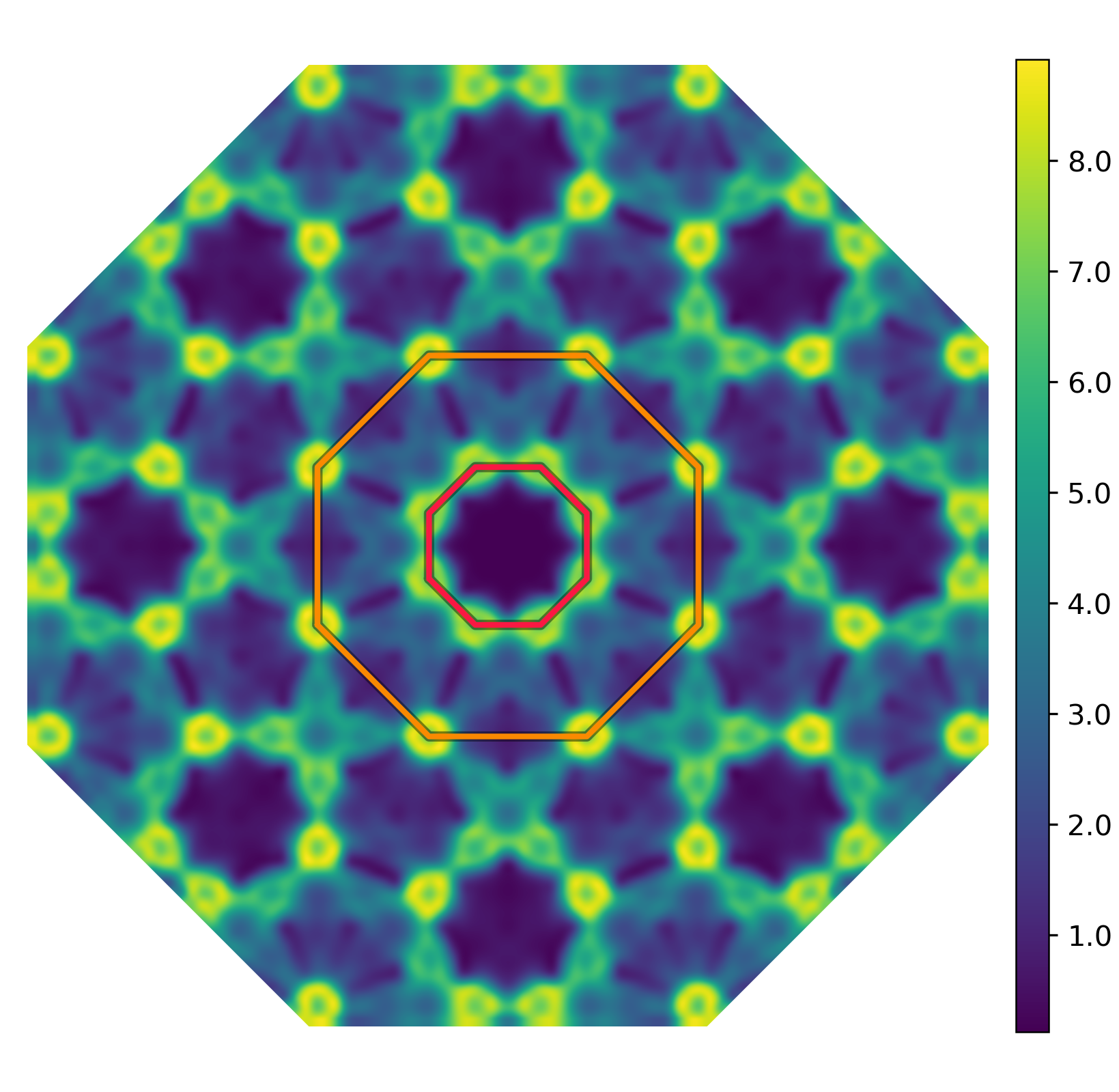}
        \caption{Ammann-Beenker: local density of states}
        \label{fig:ab_local_ldos}
    \end{subfigure}
    \medskip
    \begin{subfigure}[t]{0.47\textwidth}
        \centering
    \includegraphics[width=\textwidth]{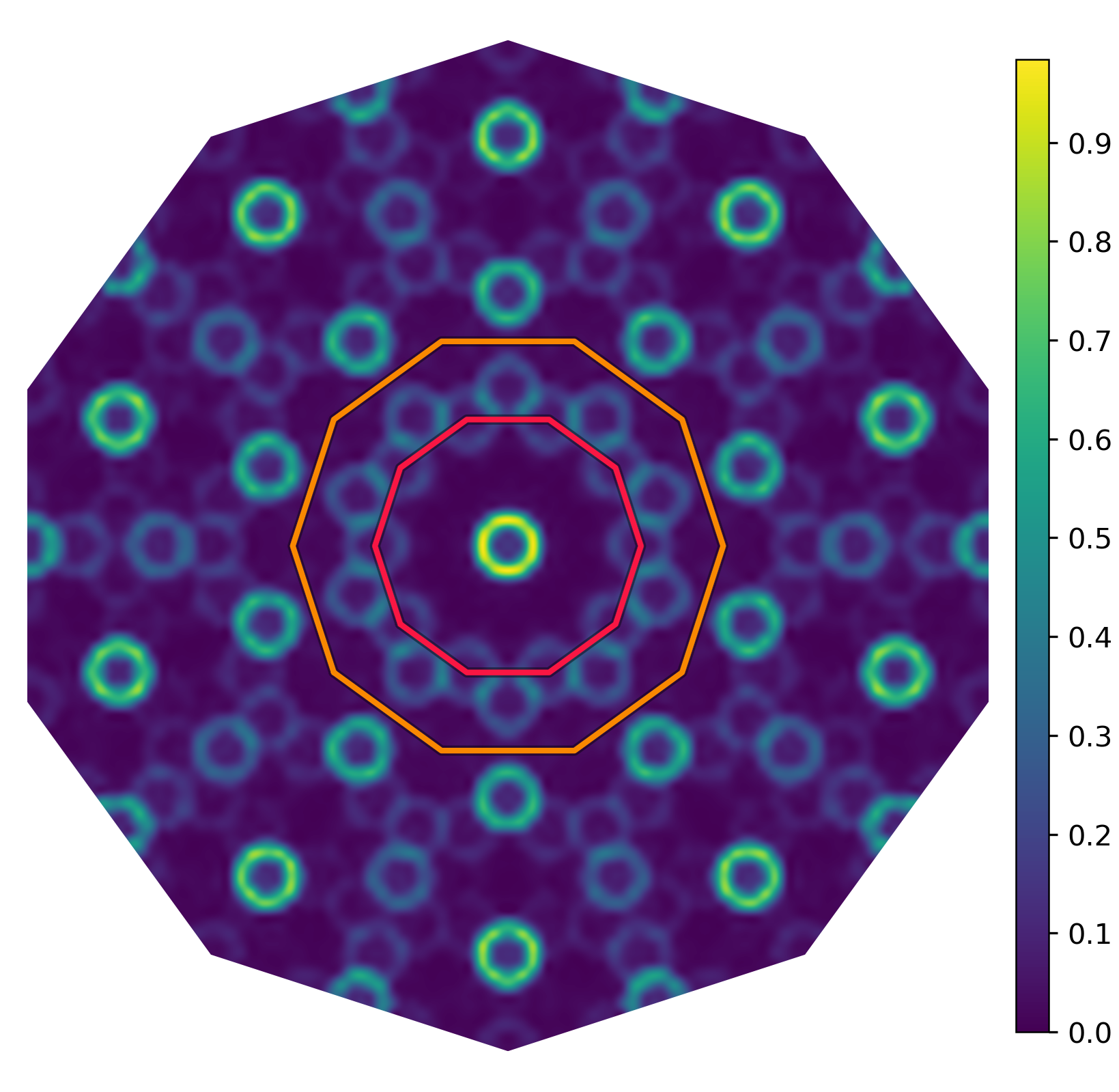}
        \caption{Penrose: local correlation function}
        \label{fig:penrose_local_lcf}
    \end{subfigure}
    \hskip 0.2cm
    \begin{subfigure}[t]{0.47\textwidth}
        \centering
     \includegraphics[width=\textwidth]{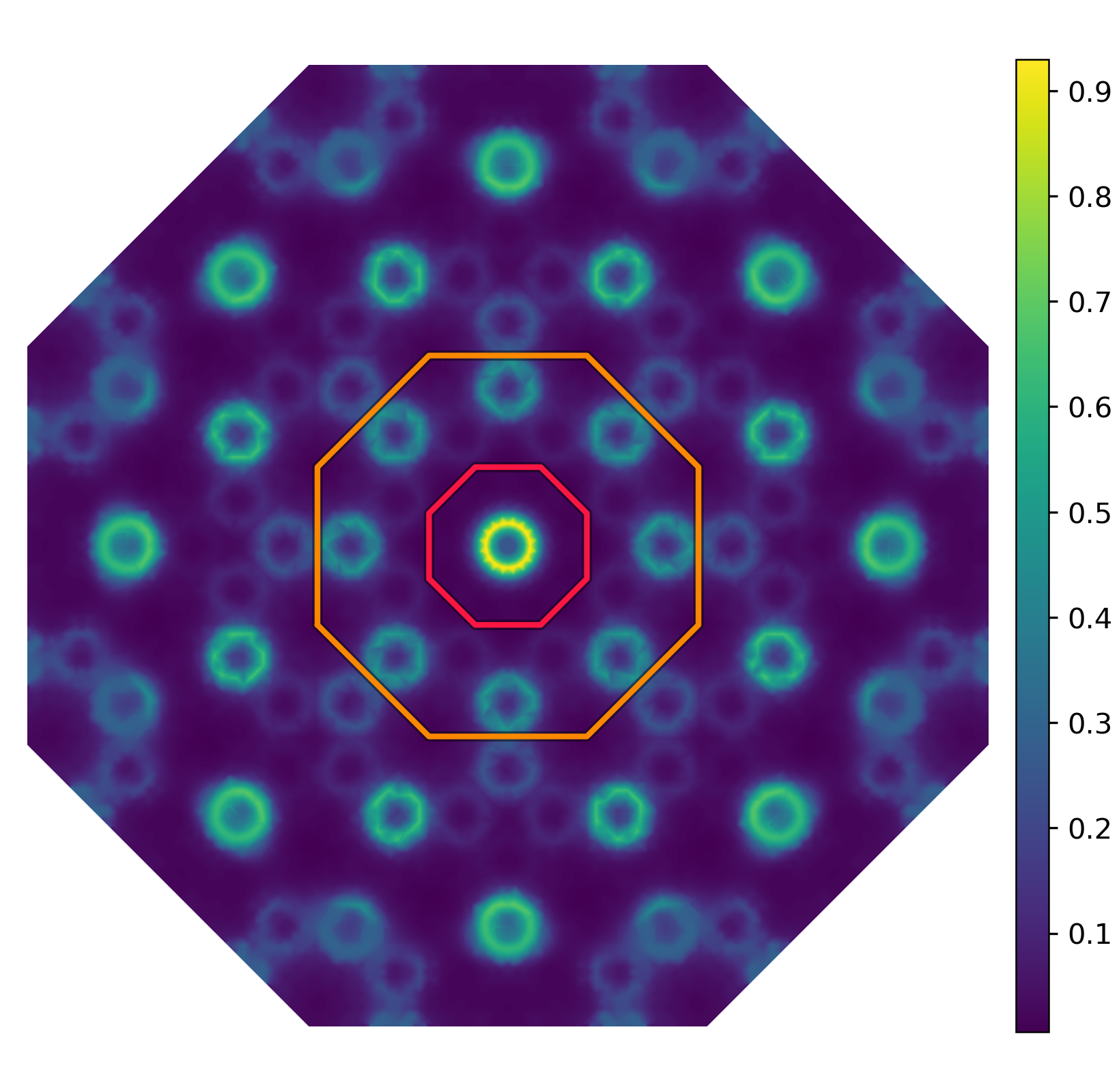}
        \caption{Ammann-Beenker: local correlation function}
        \label{fig:ab_local_lcf}
    \end{subfigure}
    \caption{
    Momentum-resolved local quantities for the Penrose and Ammann-Beenker models. 
    Panels (a) and (b) show Gaussian-smeared local density of states $\ldos(\vk;\kg)$, with $\kg$ the Gaussian of width $\eta=0.1$ centered at $E_0=0.77$ for the Penrose model and $E_0=1.16$ for the Ammann-Beenker model. 
    Panels (c) and (d) show the local correlation component $|\lccc(\vk;\GFermi)_{xy}|$, where the Fermi level $E_{\rm{F}}$ is selected inside the gap at $5/6$-filling.
    The PBZs marked in the pictures are obtained by using $\eps=0.28$ and $\eps=0.04$, respectively.
    }
    \label{fig:local_responses}
\end{figure}

To evaluate the momentum-resolved quantities in \eqref{eq:ldos_def} and \eqref{eq:local_correlation_def},
we use a normalized Gaussian spectral test function for the density of states, and a zero-temperature Fermi kernel for the current-current correlation.
When $E_{\rm{F}}$ lies in a bulk gap, this kernel selects occupied–unoccupied spectral pairs across the Fermi level, and the antisymmetric combination of the corresponding PBZ averaged correlation tensor is used to evaluate the Chern number later (see Section \ref{subsec:topological}).
We present in Figure \ref{fig:local_responses} the local density of states and the $xy$-component magnitude of the local correlation function over a reciprocal-space domain, together with the associated PBZ polygons.
Both local quantities inherit the rotational symmetry of their diffraction patterns: the Penrose maps exhibit decagonal symmetry, whereas the Ammann-Beenker maps exhibit octagonal symmetry.

We next examine the numerical convergence of the diffraction-guided PBZ hierarchy introduced in Section \ref{subsec:pbz_quadrature}.
For the numerical integration, each PBZ is decomposed into nonoverlapping triangles, with the refinement level adjusted according to the PBZ size to maintain a comparable reciprocal-space resolution across the hierarchy.
The integral is then evaluated using a piecewise-linear triangular quadrature rule \cite{LeeShishidouFreeman2002}.
The scattering-channel cutoff is held fixed at the value specified above; convergence with respect to the channel cutoff is documented separately in Appendix \ref{app:reciprocal_truncation}.
To quantify the convergence, we compute the real-space benchmark on extremely large clusters containing $N_{\rm ref}= 18936$ sites for the Penrose tiling and $N_{\rm ref}=18593$ sites for the Ammann-Beenker tiling. 

\begin{figure}[!htb]
    \centering
    \begin{subfigure}[t]{0.49\textwidth}
        \centering
        \includegraphics[width=\linewidth]{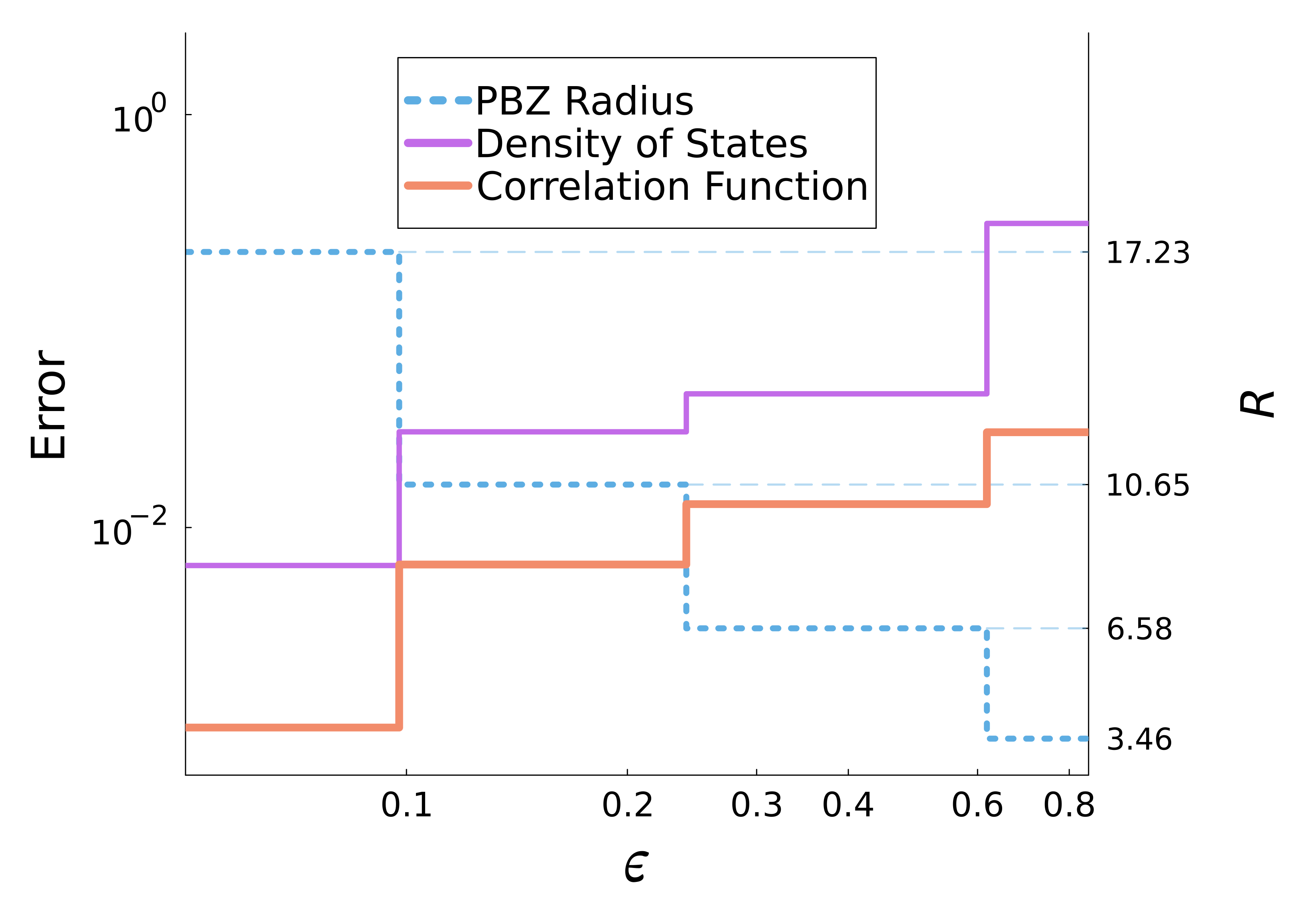}
        \caption{Penrose: error against threshold $\eps$}
        \label{fig:penrose_structure_factor_error}
    \end{subfigure}
    \hfill
    \begin{subfigure}[t]{0.49\textwidth}
        \centering
        \includegraphics[width=\linewidth]{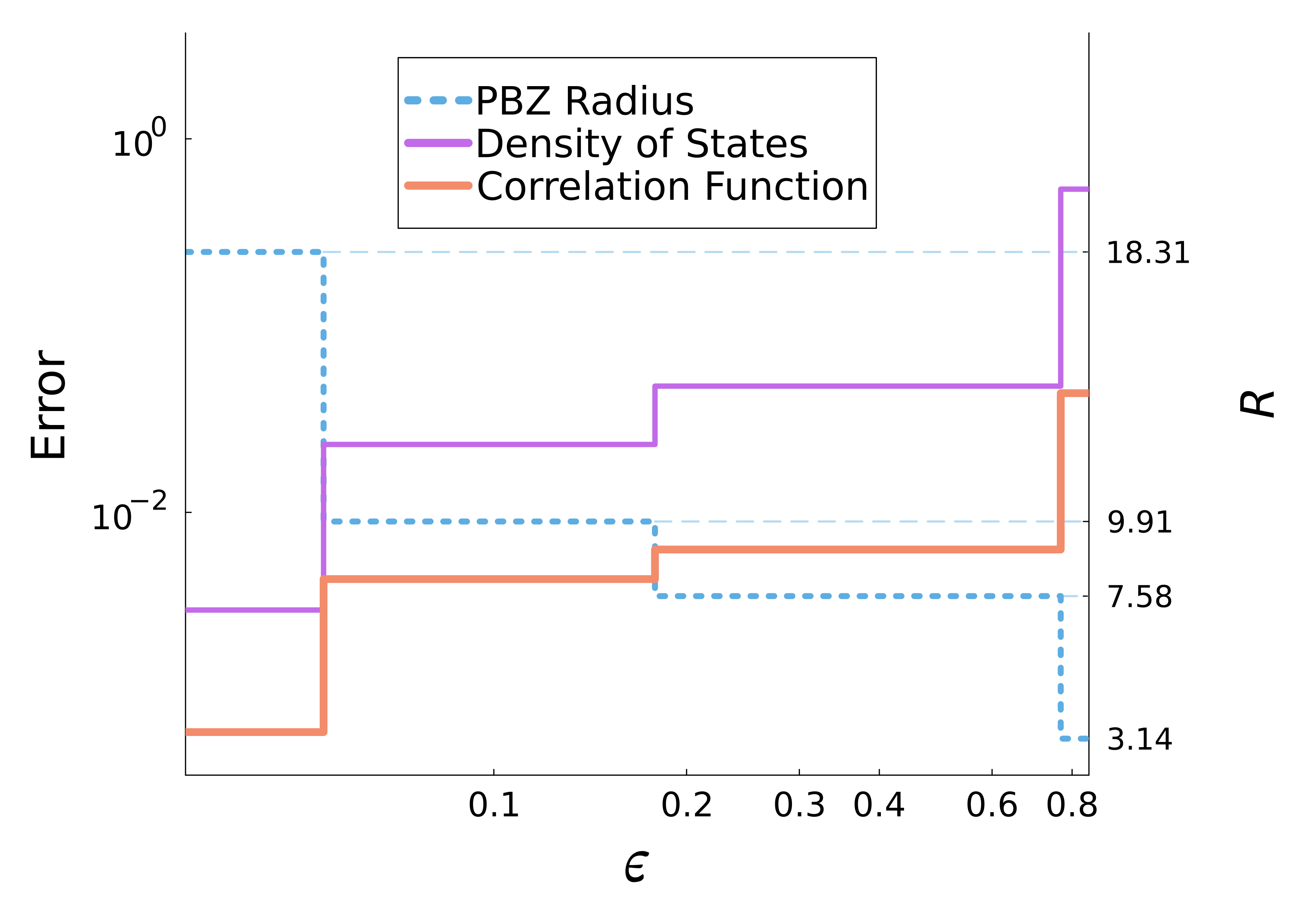}
        \caption{Ammann-Beenker: error against threshold $\eps$}
        \label{fig:ab_structure_factor_error}
    \end{subfigure}
    \vspace{0.4em}
    \begin{subfigure}[t]{0.48\textwidth}
        \centering
        \includegraphics[width=\linewidth]{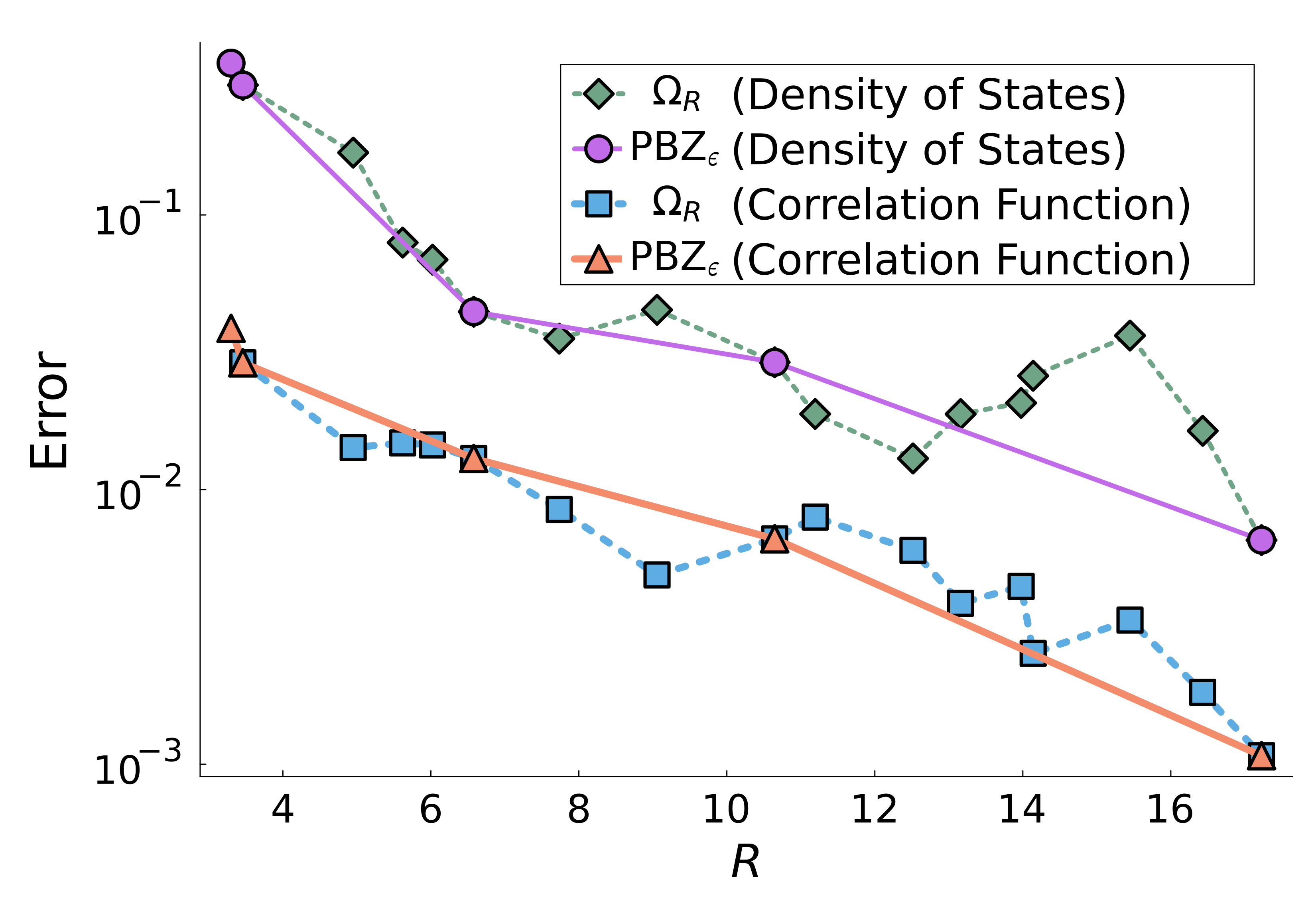}
        \caption{Penrose: error against radius $R$}
        \label{fig:penrose_pbz_radius_error}
    \end{subfigure}
    \hfill
    \begin{subfigure}[t]{0.48\textwidth}
        \centering
        \includegraphics[width=\linewidth]{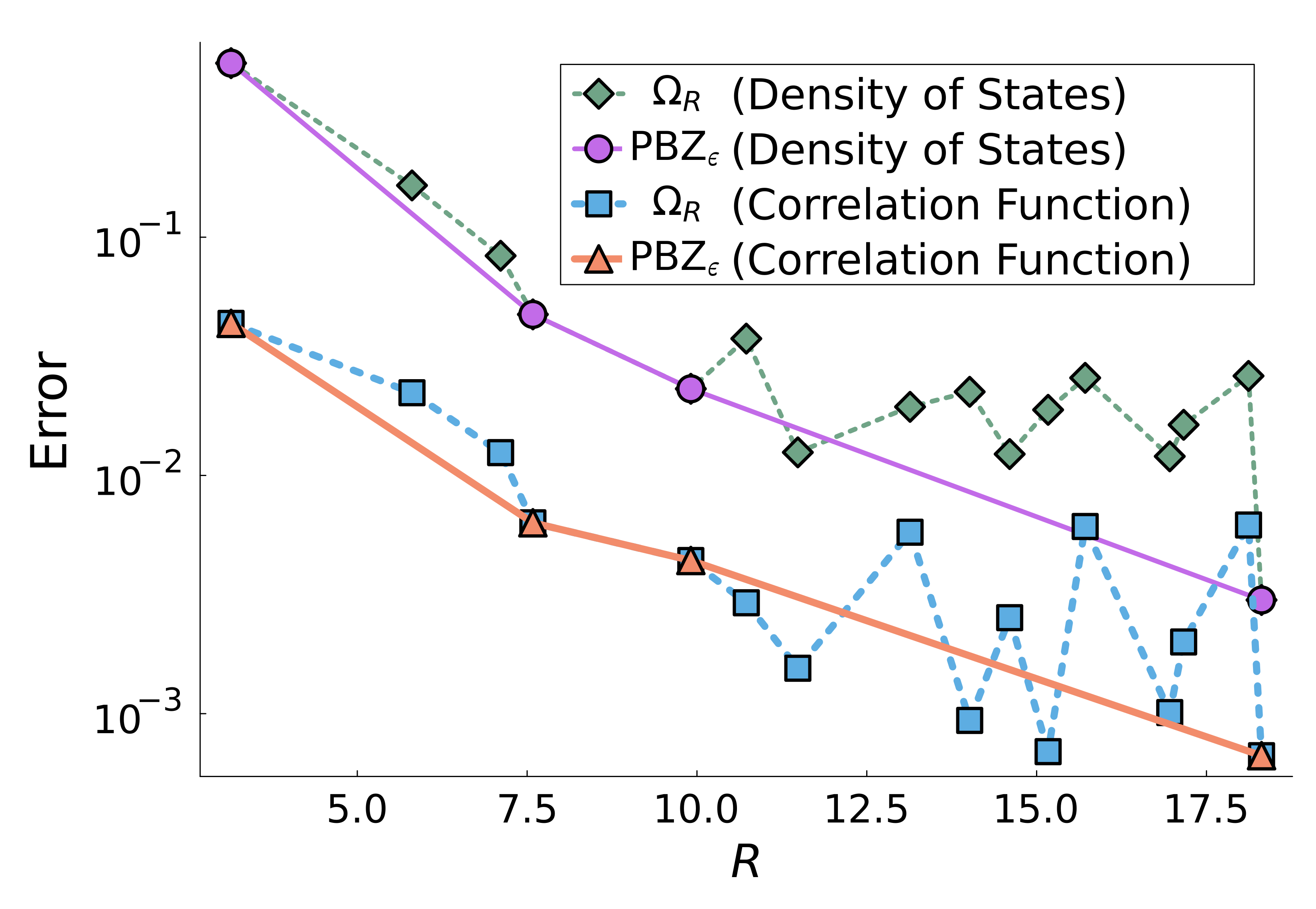}
        \caption{Ammann-Beenker: error against radius $R$}
        \label{fig:ab_pbz_radius_error}
    \end{subfigure}
    \caption{Convergence of the diffraction-guided PBZ quadrature for the Penrose and Ammann-Beenker models.
    Panels (a) and (b) show the errors of spectral quantities (purple) and correlation functions (orange) along the PBZ hierarchy as the threshold parameter $\eps$ is varied, together with the corresponding PBZ radius $R$.
    Panels (c) and (d) compare the errors along the PBZ sequence with those obtained from regular polygon regions $\Omega_R$ (with arbitrary domain radius $R$).
    }
    \label{fig:pbz_convergence}
\end{figure}

We show in Figure \ref{fig:pbz_convergence} (a)-(b) the PBZ radius and the error of the density of states and current-current correlation function with respect to $\eps$.
The radius and error vary stepwise because the selected set ${\sds}$ changes only when the threshold $1-\eps$ crosses the amplitude of a diffraction peak.
We then present in Figure \ref{fig:pbz_convergence} (c)-(d) the errors against the domain radius.
We observe that the numerical error decreases correspondingly along the hierarchy, demonstrating that the PBZ family yields increasingly accurate reciprocal-space averages.
To distinguish the effect of diffraction-guided geometry from simple domain enlargement, we further compare the PBZ hierarchy with a sequence of origin-centered regular polygonal domains $\Omega_R$ constructed without reference to the diffraction pattern.
Along the PBZ sequence, we observe that the error exhibits a regular and monotone decrease as the domain radius $R$ grows. 
In contrast, the errors associated with the generic polygonal domains show pronounced nonmonotonic oscillations, even though their sizes increase.
This comparison indicates that the improved accuracy is not a consequence of enlarging the integration domain alone. 
The diffraction-guided boundaries are adapted to high-overlap Fourier-module shifts and therefore reduce repeated sampling of approximately equivalent momentum regions.

\subsection{Full-band spectral structure and quasiperiodic pseudogaps}
\label{subsec:bands}

Having established the numerical reliability of the reciprocal-space formulation, we now use it to investigate quasicrystalline electronic structures over the full tight-binding bandwidth. 
This regime is particularly relevant to quasicrystals, where a large family of Fourier-module channels can contribute at widely separated energies and generate spectral structures that are not retained by an effective theory constructed around a single low-energy momentum. 
We focus on two questions: whether the diffraction-guided PBZ formulation recovers the full-band spectrum of the original tight-binding model, and how its momentum-resolved representation reveals the origin of quasiperiodic gaps and pseudogaps.

\begin{figure}[!htb]
    \centering
    \begin{subfigure}[t]{0.48\textwidth}
        \centering
    \includegraphics[width=\linewidth]{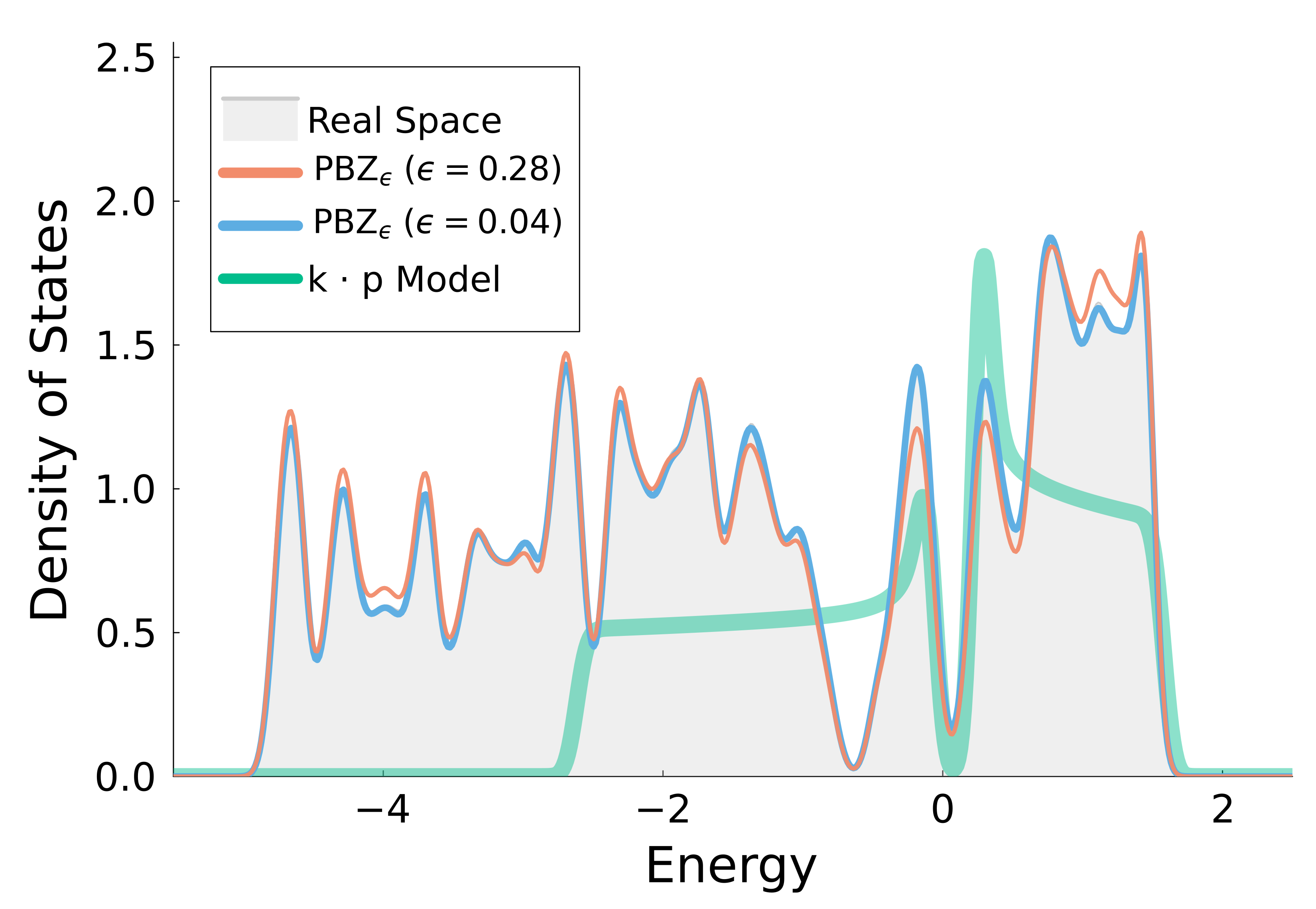}
        \caption{Penrose}
        \label{fig:penrose_dos_pbz}
    \end{subfigure}
    \hfill
    \begin{subfigure}[t]{0.48\textwidth}
        \centering
    \includegraphics[width=\linewidth]{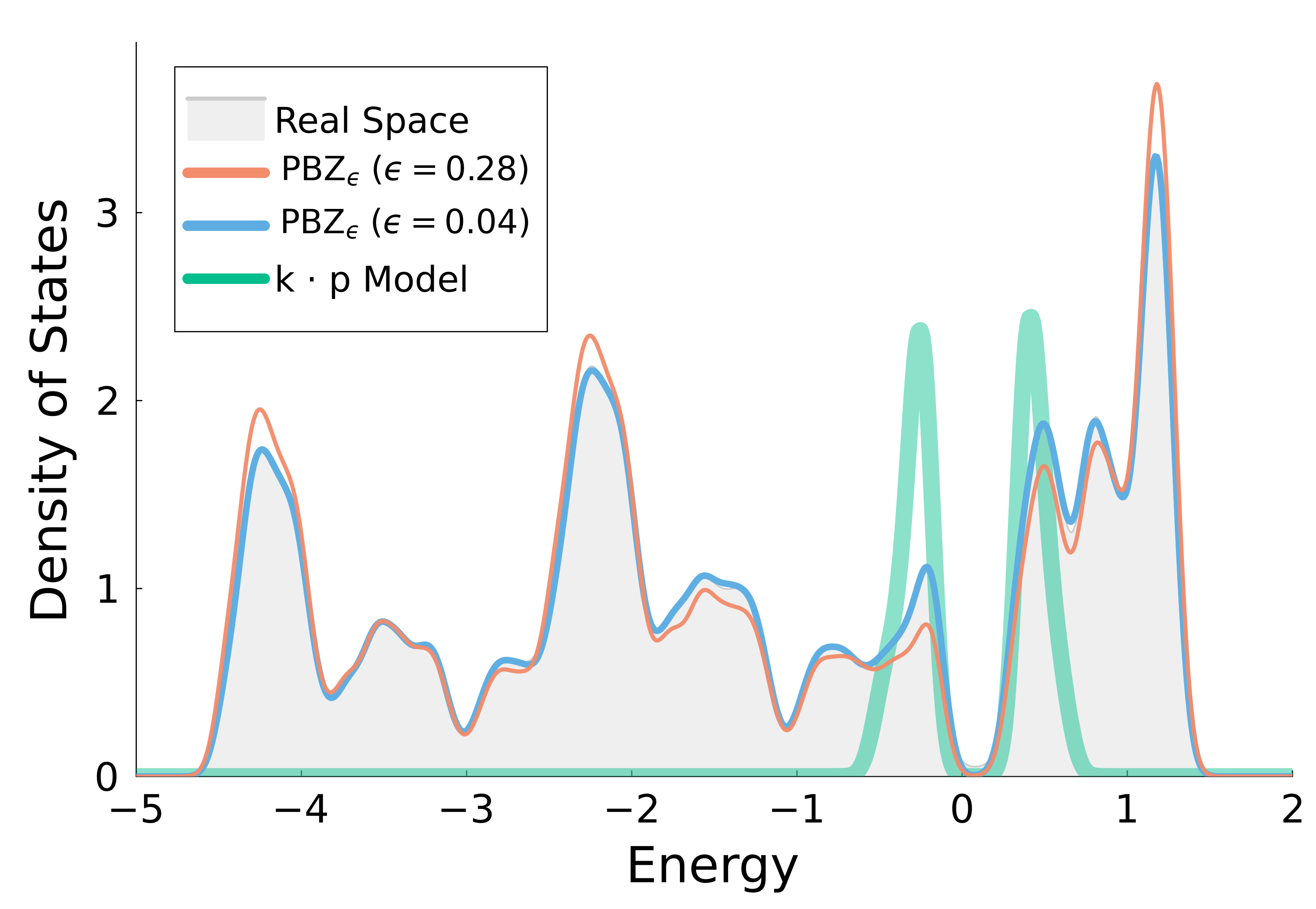}
        \caption{Ammann-Beenker}
        \label{fig:ab_dos_pbz}
    \end{subfigure}
    \caption{
    Gaussian-smeared density of states for the (a) Penrose and (b) Ammann-Beenker models. Each panel compares the real-space reference (shaded) with the reciprocal-space approximation evaluated over the small and large diffraction-guided PBZs, as well as the results from a PBZ-centered low-energy $k\cdot p$ effective model.
    }
    \label{fig:dos_spectral_comparison}
\end{figure}

In Figure \ref{fig:dos_spectral_comparison}, we compare the Gaussian-smeared density of states obtained from the reciprocal-space formulation with the large-cluster real-space reference above. 
The smaller PBZ with $\eps=0.28$ reproduces the principal spectral structure but retains visible errors in several peak amplitudes and in parts of the higher-energy spectrum. 
Reducing the threshold to $\eps=0.04$ enlarges the PBZ and substantially improves the agreement throughout the full bandwidth. 
We further compare the results with the low-energy $k\cdot p$ model \cite{WangLiuHuang2022}, whose effective Hamiltonian is constructed by expanding around the PBZ center $\Gamma$ and retaining a reduced set of low-energy degrees of freedom. 
The $k\cdot p$ model therefore provides an efficient description of the spectral structure within its intended low-energy window, but it is not designed to reproduce remote bands or spectral features generated by scattering channels outside the retained subspace.
By contrast, the present formulation resolves the electronic structure across the full bandwidth, as increasing the scattering-channel cutoff systematically recovers the spectrum of the original tight-binding Hamiltonian. 

\begin{figure}[!htb]
    \centering
    \begin{subfigure}[t]{0.48\textwidth}
        \centering
    \includegraphics[width=\linewidth]{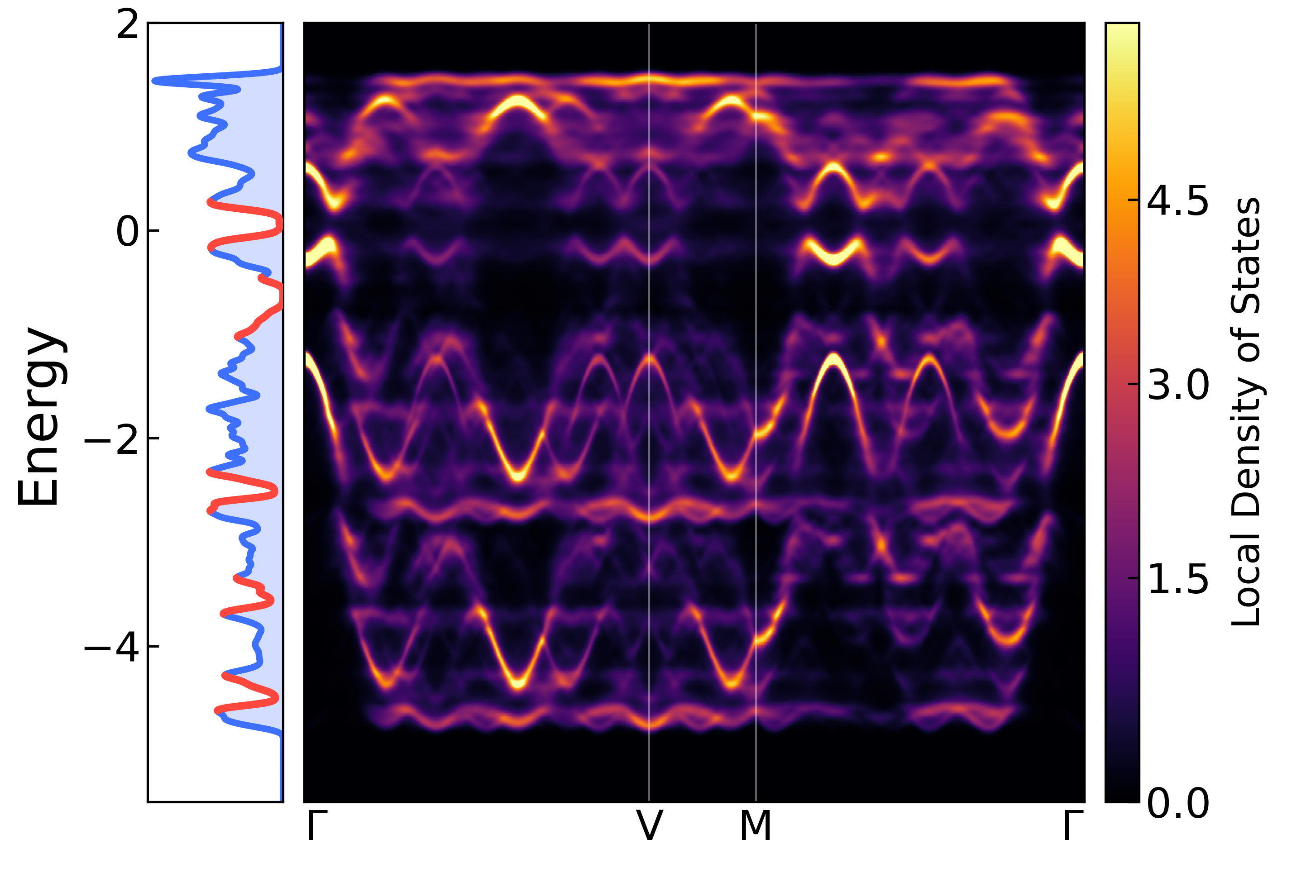}
        \caption{Penrose}
        \label{fig:Penrose_band}
    \end{subfigure}
    \hfill
    \begin{subfigure}[t]{0.48\textwidth}
        \centering
    \includegraphics[width=\linewidth]{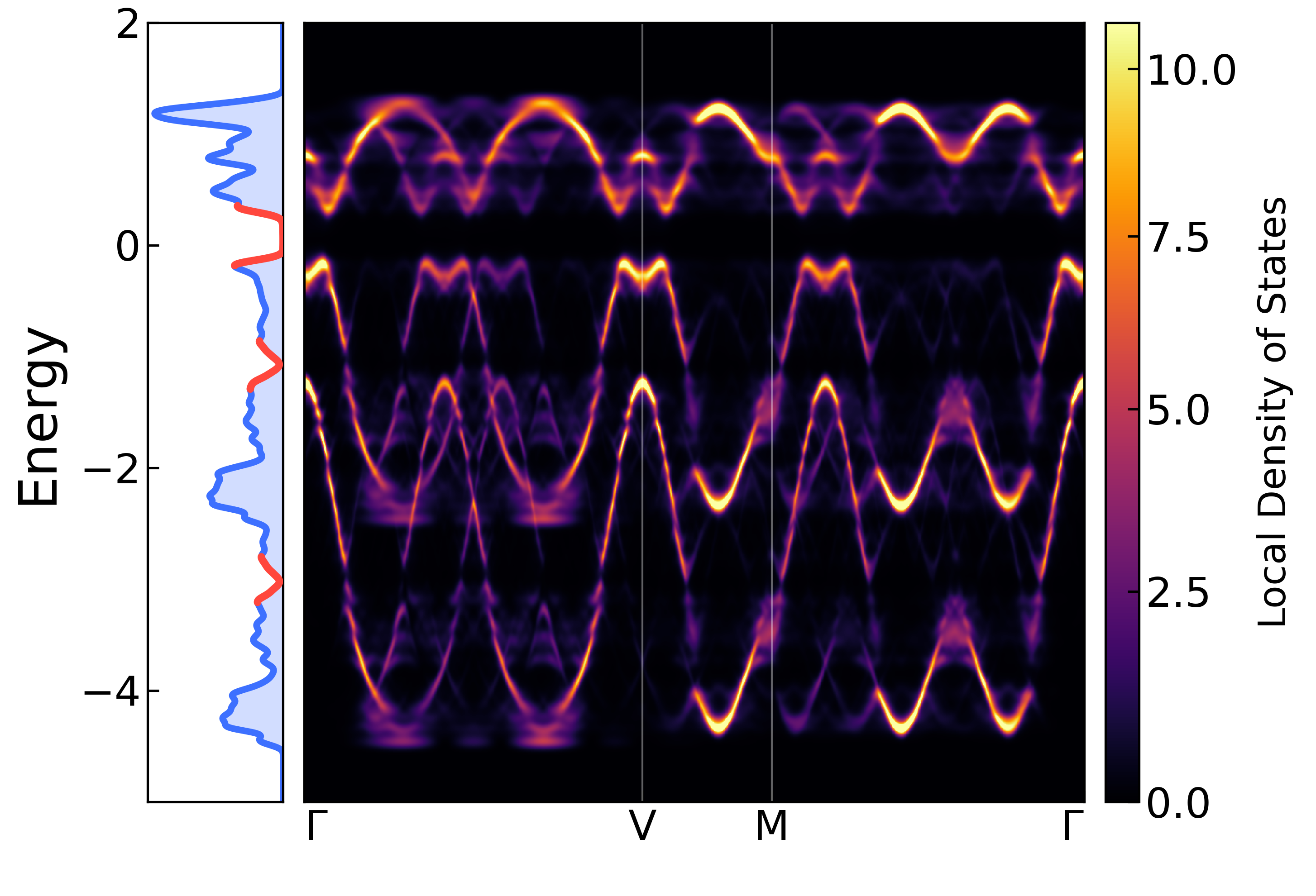}
        \caption{Ammann-Beenker}
        \label{fig:ab_band}
    \end{subfigure}
    \caption{Full-band spectral structure for quasicrystals: local density of states $\ldos(\vk;\kg)$ along the high-symmetry path of the $\pbz$ with $\eps=0.04$ for the Penrose and Ammann-Beenker models. 
    We use a smaller $\eta=0.05$ to better resolve the spectral features and quasiband dispersions.
    The side panels show the corresponding density of states, with the bulk gaps and representative pseudogap regions highlighted in red. For the Penrose model, these regions are centered near $E=0.05,-0.74,-2.51,-3.51$, and $-4.45$, while for the Ammann-Beenker model they occur near $E=-0.06,-1.30$, and $-3.37$. At these energies, the quasiband maps exhibit pronounced suppression in spectral weight, consistent with the gap and pseudogap features shown in Figure \ref{fig:dos_spectral_comparison}.
    }
    \label{fig:band}
\end{figure}

To expose the momentum-space organization underlying the bulk density of states, we present in Figure \ref{fig:band} the Gaussian-smeared local spectral quantity $\ldos(\vk;\kg)$ as a function of energy and momentum along representative high-symmetry paths of the PBZs. 
The bright ridges form quasiband-like branches rather than conventional Bloch bands, with their intensities measuring the spectral weight carried by the central scattering channel. 
A bulk gap corresponds to an energy interval in which the density of states vanishes within the chosen broadening resolution, whereas a pseudogap exhibits strong but incomplete spectral suppression.
The scattering-channel representation provides a natural interpretation of this depletion. 
At each reference momentum $\vk$, the reciprocal-space Hamiltonian $\hat{\ham}(\vk)$ couples the central channel to channels $\vk+\vG$, with $\vG\in\Lambda^*$. 
Their hybridization, governed by the off-diagonal matrix elements of $\hat{\ham}(\vk)$, produces avoided crossings and redistributes spectral weight among quasiband branches. 
When this redistribution occurs only over part of reciprocal space, it suppresses the density of states over a finite energy interval without opening a complete bulk gap, thereby generating a quasiperiodic pseudogap.

Several of the resulting pseudogaps occur outside the energy range accurately represented by the PBZ-centered $k\cdot p$ model. 
Their appearance in the full-band calculation therefore reflects scattering processes involving channels omitted from the low-energy reduction. 
This multichannel mechanism is consistent with the Hume–Rothery picture, in which quasiperiodic Bragg scattering reconstructs the electronic spectrum and produces partial depletion near resonant energies \cite{SmithAshcroft1987,FujiwaraYokokawa1991,Fujiwara1993ElectronicStructures,SatoTakeuchiMizutani2001,RogalevEtAl2015}. 
The present formulation extends this interpretation by resolving the associated spectral-weight redistribution directly in the momentum-dependent full-band tight-binding problem.
These results demonstrate that our reciprocal-space framework can support a multichannel interpretation of the gaps and pseudogaps that are obscured in purely real-space averages and absent from a fixed low-energy effective model.

\subsection{Zeeman-driven gap evolution and Chern transitions}
\label{subsec:topological}

\begin{figure}[!htb]
    \centering
    \begin{subfigure}[t]{0.48\textwidth}
        \centering
        \includegraphics[width=\textwidth]{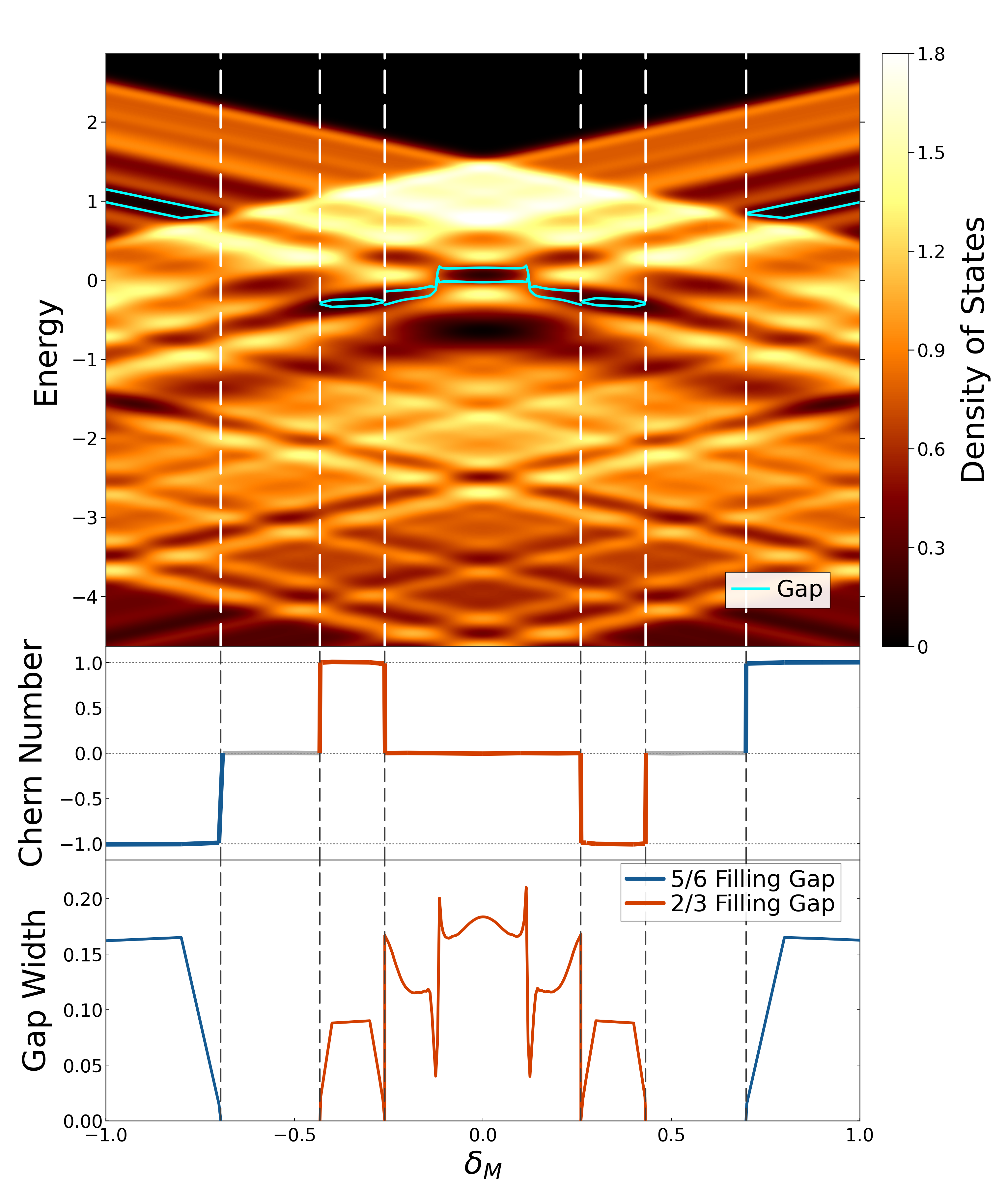}
        \caption{Penrose}
        \label{fig:penrose_phase}
    \end{subfigure}
    \hfill
    \begin{subfigure}[t]{0.48\textwidth}
        \centering
        \includegraphics[width=\textwidth]{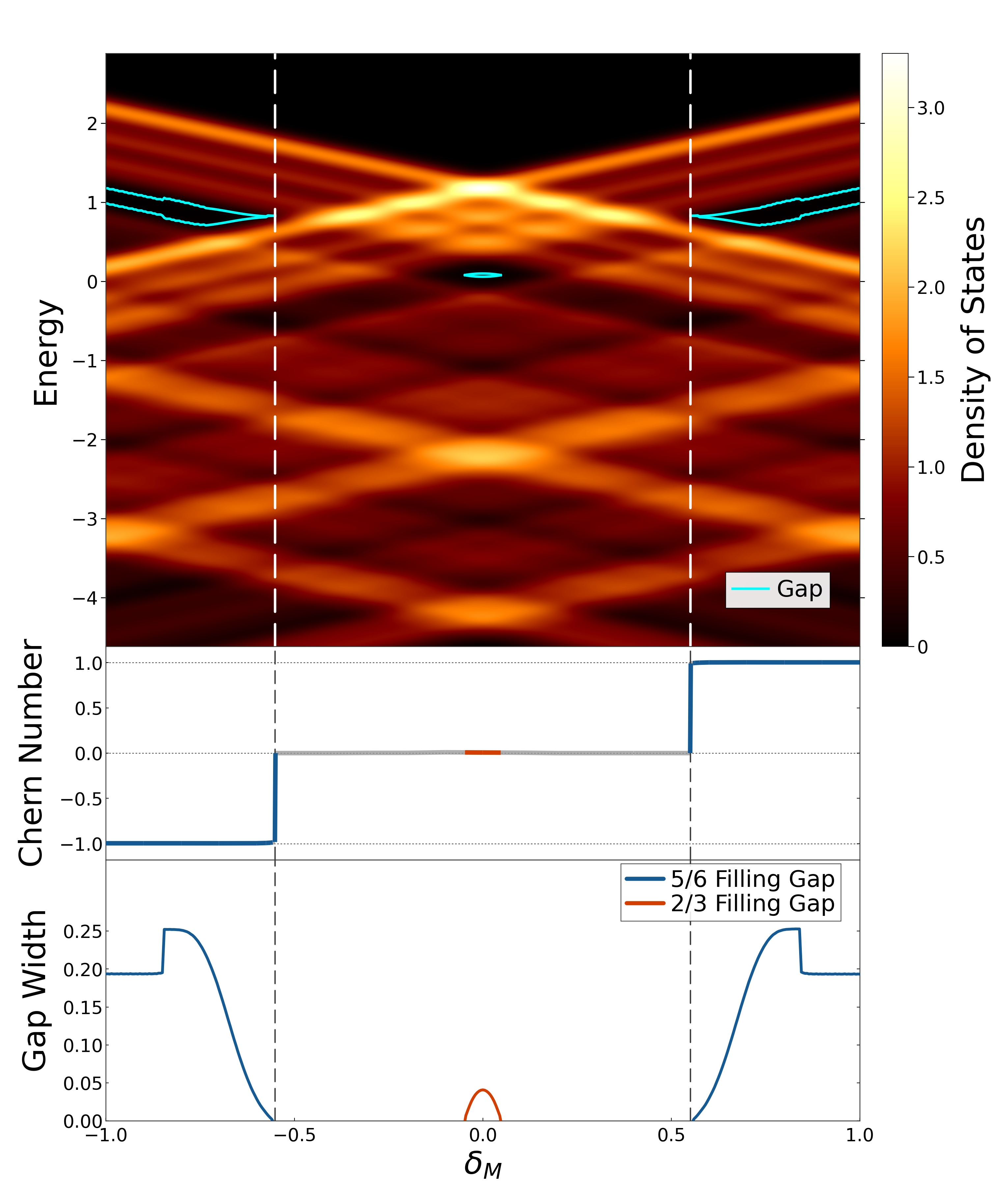}
        \caption{Ammann-Beenker}
        \label{fig:ab_phase}
    \end{subfigure}
    \caption{ Zeeman-driven spectral and topological evolution of the Penrose and the Ammann-Beenker models. Each panel displays the Gaussian-smeared density of states (top), Chern number (middle), and gap width (bottom) as a function of the out-of-plane coupling $\delta_M$. 
    The tracked gap edges for the indicated fillings ($\nu=2/3$ and $\nu=5/6$) are overlaid on the spectra, and the Chern numbers are evaluated with $E_{\rm{F}}$ inside the open gaps.
    }
    \label{fig:topology_phase_diagrams}
\end{figure}

We next apply the current-current correlation formulation to Zeeman-driven topological transitions. 
A uniform out-of-plane Zeeman coupling is introduced by setting $\vm=\hat{\vz}$ and varying its strength $\delta_M$, thereby breaking time-reversal symmetry. 
For a Fermi level $E_{\rm{F}}$ inside an open bulk gap, the Chern number $C$ is estimated from the antisymmetric correlation tensor
\begin{equation*} 
C = \frac{2\pi \im\hbar^2}{e^2} \rho_\Lambda \big( \ccc_{yx}(\GFermi)-\ccc_{xy}(\GFermi) \big),
\end{equation*}
where $\rho_\Lambda = \lim_{R\to\infty}|\Lambda_R|/|B_R|$ is the point set density, which converts the trace per site into a trace per unit area.
We use the larger $\pbz$ constructed with $\eps=0.04$, place $E_{\rm{F}}$ at the center of the tracked gap, and report a Chern plateau only when the gap remains open.
For each $\delta_M$, we track the gaps at the selected filling fractions $\nu=2/3$ and $\nu=5/6$, where $\nu$ denotes the fraction of occupied single-particle states.
We present the Zeeman-driven spectral and topological evolution in  Figure \ref{fig:topology_phase_diagrams}.
The top panels show the Gaussian-smeared density of states as a function of $\delta_M$, with the tracked gaps marked at the target fillings.
We use a small smearing width $\eta=0.01$ to clearly resolve the corresponding gap edges and widths.
We further show the Chern number transitions with respect to $\delta_M$, with the corresponding critical values of $\delta_M$ subsequently refined by a bisection search.

In all cases, changes in the Chern number coincide with gap closing and reopening. 
For the Penrose model, the reopened $2/3$-filling gap has $|C|=1$ for $0.26<|\delta_M|<0.43$, while the $5/6$-filling gap becomes topological for $|\delta_M|>0.70$.
For the Ammann-Beenker model, the nontrivial phase occurs in the $5/6$-filling gap for $|\delta_M|>0.55$, whereas no nontrivial $2/3$-filling plateau is found in the scanned range.
Time-reversal symmetry of the zero-field model implies that the density of states and gap widths are even in $\delta_M$.
A reopened gap with $C=\pm1$ realizes a quantum anomalous Hall phase with $\sigma_{xy}=Ce^2/h$ \cite{ThoulessKohmotoNightingaleNijs1982,Haldane1988,Bellissard1994}. Reversing the sign of $\delta_M$ reverses the Chern number and hence the Hall response and the chirality of the associated edge modes through bulk--boundary correspondence~\cite{Hatsugai1993Chern}.
These results demonstrate that the full-band reciprocal-space framework resolves the gap closures and Chern transitions of the original quasicrystalline Hamiltonian. 

\subsection{Invariance of bulk observables under uniform phason shifts}
\label{subsec:phason_invariance}

We finally examine whether the bulk observables obtained from the reciprocal-space formulation depend on a uniform phason shift.
In the cut-and-project construction, a uniform translation of the acceptance window in internal space generates a phason-shifted configuration \cite{deBruijn1981I,deBruijn1981II,Socolar1986Phason,Cui2026Phason}.
We parameterize the shift by $\bm{\gamma}=(\gamma_1,\ldots,\gamma_D)^{\rm T}\in\R^D$ and denote its internal-space component by $\bm{\gamma}_\perp:=\pperp\bm{\gamma}$. The shifted acceptance window and the corresponding quasicrystalline point set are
\begin{equation*}
W_{\bm{\gamma}} = W+\bm{\gamma}_{\perp}, \qquad
\Lambda_{\bm{\gamma}} = \left\{ \ppara\vn :~ \vn\in\Z^D,~ \pperp\vn\in W_{\bm{\gamma}}\right\}.
\end{equation*}
As $\bm{\gamma}_\perp$ varies, the boundary of the shifted window crosses internal-space projections $\pperp\vn$, adding or removing the corresponding real-space sites $\ppara\vn$ and producing local rearrangements of the tiling \cite{Socolar1986Phason,Lifshitz2011,JagannathanDuneau2024}.
For nonsingular shifts, no projected lattice point lies on the window boundary. These configurations belong to the same quasicrystalline hull and share the same asymptotic patch frequencies \cite{BaakeGrimm2013}.
We provide representative phason-shifted configurations for both Penrose and Ammann-Beenker models in Appendix \ref{app:real_reciprocal_geometry}.

\begin{figure}[!htb]
    \centering
    \begin{subfigure}[t]{0.49\linewidth}
        \centering
        \includegraphics[width=\linewidth]{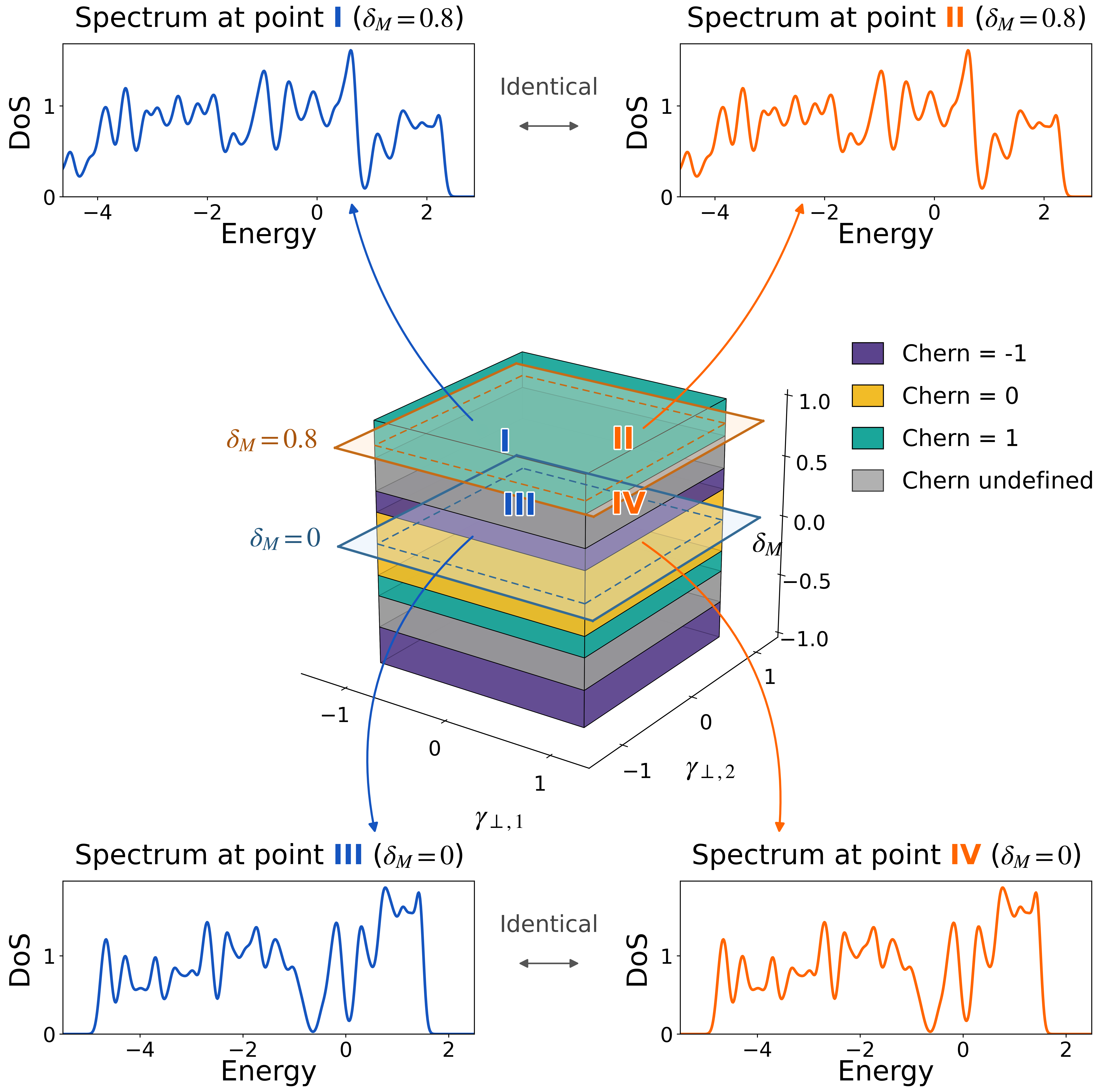}
        \caption{Penrose}
        \label{fig:penrose_phason}
    \end{subfigure}
    \hfill
    \begin{subfigure}[t]{0.49\linewidth}
        \centering
        \includegraphics[width=\linewidth]{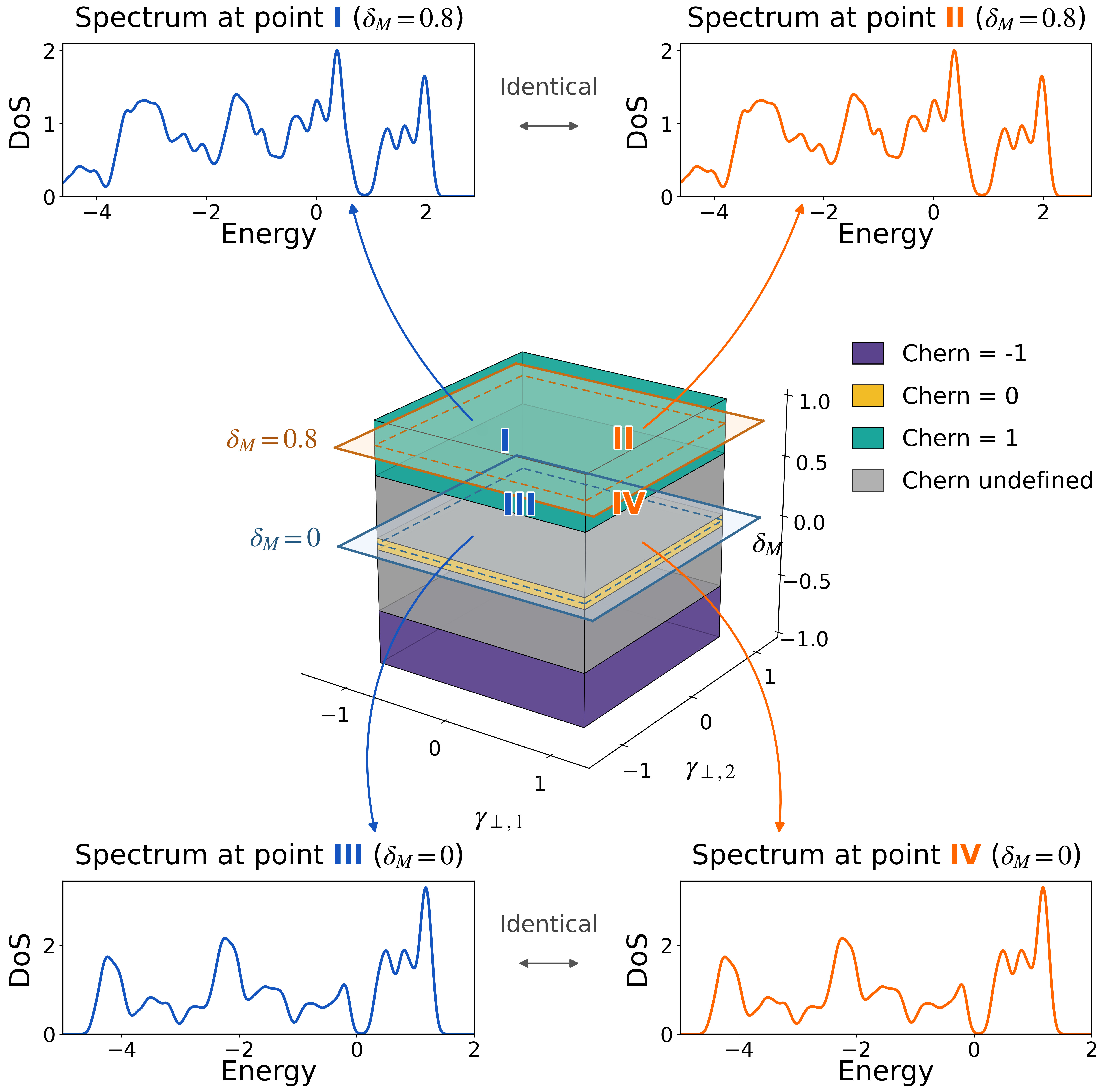}
        \caption{Ammann-Beenker}
        \label{fig:ab_phason}
    \end{subfigure}
    \caption{Phason invariance of the bulk quantities for the Penrose and the Ammann-Beenker models. 
    Panels (a) and (b) show the Penrose and Ammann-Beenker results, respectively. 
    In each central panel, the color indicates the Chern number of the tracked bulk gap at the filling fractions $\nu=2/3$ or $\nu=5/6$. This Chern number is presented as a function of the internal-space shift coordinates $(\gamma_{\perp,1},\gamma_{\perp,2})$ and the Zeeman coupling $\delta_M$.
    Gray regions indicate that the tracked gap is closed and the Chern number is not assigned.
    The resulting plateaus are uniform along the phason directions and change only when $\delta_M$ drives a bulk-gap closing. 
    The upper spectra compare the density of states at the distinct phason configurations I and II for $\delta_M=0.8$, while the lower spectra compare configurations III and IV for $\delta_M=0$. 
    Each pair coincides within plotting resolution, demonstrating the phason invariance of the bulk density of states.
    }
    \label{fig:phason_invariance}
\end{figure}

A window translation changes the Fourier amplitudes only through phason-dependent phase factors $\ee^{\im\theta_{\vG}(\bm{\gamma}_\perp)} F(\vG)$, and therefore leaves $|F(\vG)|$ unchanged. 
The diffraction-guided PBZ hierarchy can consequently be used without modification for all phason-shifted configurations. 
Although their local real-space environments may differ, their bulk spectral observables are expected to be invariant under uniform phason shifts. 
The Chern number is likewise invariant provided that the corresponding bulk gap remains open.
We test these properties for the Penrose and Ammann-Beenker models by sampling $\bm{\gamma}_\perp$ and present the resulting density of states and Chern numbers in Figure \ref{fig:phason_invariance}. 
The density of states obtained from distinct phason configurations coincides at both $\delta_M=0$ and $\delta_M=0.8$. 
The Chern-number plateaus are also constant along the phason directions and vary only with the Zeeman coupling, indicating no phason-induced change as long as the tracked gap remains open.
These results show that the computed density of states and Chern number are bulk properties of the quasicrystalline hull rather than of a particular real-space realization. 
They also confirm that the local-to-global formulation reproduces the same bulk observables across phason-shifted configurations.

\section{Conclusions}
\label{sec:conclusion}

We have developed a systematically convergent reciprocal-space tight-binding framework for a broad class of cut-and-project quasicrystals. 
Fourier-module shifts generate the coupled scattering channels at each physical momentum, enabling momentum-resolved local spectral and current-current correlation quantities to be constructed without a Bloch decomposition. 
Their full reciprocal-space means are equivalent to the real-space thermodynamic observables, while finite calculations converge through two complementary refinements: enlargement of the scattering-channel space and expansion of the diffraction-guided PBZ hierarchy. 

For the Penrose and Ammann-Beenker models, the framework reproduces large-cluster densities of states across the full bandwidth and resolves quasiband structures and multichannel pseudogaps beyond the range of the PBZ-centered $k\cdot p$ description. 
The current-current formulation further captures Zeeman-driven gap closings and reopenings together with quantized Chern plateaus, while the density of states and Chern numbers remain invariant under uniform phason shifts. 
These results demonstrate that the same local-to-global construction provides a unified route to full-band spectral, transport, and topological observables in quasicrystals.

An important perspective is to extend the present local-to-global construction to incommensurate moir\'{e} heterostructures, including twisted bilayer and multilayer graphene. 
While widely used continuum models primarily describe the small-angle, low-energy regime, the present framework provides a route to construct more general coupled-channel Hamiltonians and systematically refinable quadrature domains of PBZs.
Such an extension would enable full-band spectral and topological calculations for incommensurate moir\'{e} materials.

\section*{Acknowledgment}

This work was supported by the National Key R\&D Program of China (No. 2025YFA1016600).
XL was also supported by the National Natural Science Foundation of China (No. 12301548).
HC was also supported by the National Natural Science Foundation of China (No. 12371431).
DZ was also supported by the National Natural Science Foundation of China (No. 124B2020).

\section*{Appendix}
\appendix

\makeatletter
\@addtoreset{figure}{section}
\@addtoreset{equation}{section} 
\makeatother

\section{Quasicrystalline geometry}
\label{app:real_reciprocal_geometry}

This appendix presents further details of the cut-and-project scheme and provides the explicit parameters for two typical quasicrystals: the Penrose and Ammann-Beenker tilings \cite{DuneauKatz1985,BaakeGrimm2013}.

The quasicrystalline point set is obtained by projecting selected points of the $D$-dimensional integer lattice onto the $d$-dimensional real space ($D>d$), with the selection controlled by an acceptance window $W$ in the internal space
\cite{DuneauKatz1985},
\begin{equation*}
\Lambda = 
\left\{ \ppara\vn :~ \vn\in\Z^D,~ \pperp\vn\in W \right\} \subset\R^d.
\end{equation*}
All acceptance windows are chosen to be compact sets with nonempty interior and boundary of zero Haar measure in the internal space. Let $\ppara : \R^D\to\R^d$ and $\pperp : \R^D\to\R^{D-d}$ denote the physical and internal projections chosen such that $\ppara^{\rm T}\ppara+\pperp^{\rm T}\pperp=c_{\rm lat} I_{D}$, where $c_{\rm lat}>0$ is a normalization factor determined by the choice and overall scale of the projection matrices. These projections naturally form a matrix $Q = \begin{pmatrix} \ppara\\ \pperp \end{pmatrix}\in \R^{D\times D}$, which generates the associated higher-dimensional embedding lattice
\begin{equation*}
\mathcal L:= Q \Z^D\subset\R^D.
\end{equation*}
The corresponding dual high-dimensional lattice is defined by
\begin{align*}
\mathcal L^* 
&:= \left\{ \vy\in\R^D :~ \vy^{\rm T}\vx\in 2\pi\Z ,~~ \forall~\vx\in\mathcal L \right\} \\
&= 2\pi Q^{-T}\Z^D = (2\pi/c_{\rm lat}) Q\Z^D = (2\pi/c_{\rm lat}) 
\begin{pmatrix} 
\ppara \\[2pt]  
\pperp 
\end{pmatrix} 
\Z^D \subset\R^D,
\end{align*}
yielding the natural definitions for the physical and internal reciprocal coordinate maps,
\begin{equation*}
\ppara^* :=
(2\pi/c_{\rm lat}) \ppara, \qquad \pperp^*:=(2\pi/c_{\rm lat}) \pperp . 
\end{equation*}
Projecting $\mathcal L^*$ onto the reciprocal space then yields the Fourier module,
\begin{equation*}
\Lambda^* = \left\{ \ppara^*\vn :~ \vn\in\Z^D \right\} \subset \R^d.
\end{equation*}
For the projections considered here, the resulting Fourier modules are dense in reciprocal space $\R^d$, and their elements label the scattering channels used in Section \ref{sec:tight_binding_reciprocal_space}.

In general, a quasicrystalline lattice generated via the cut-and-project scheme is specified by two projections, $\ppara$ and $\pperp$, and the acceptance window $W$.  For the Penrose and Ammann-Beenker tilings, these three characterizing quantities and their corresponding Fourier modules are explicitly provided in the following subsections.

\begin{figure}[!htb]
    \centering
    \begin{subfigure}[t]{0.32\textwidth}
        \centering
    \includegraphics[width=\textwidth]{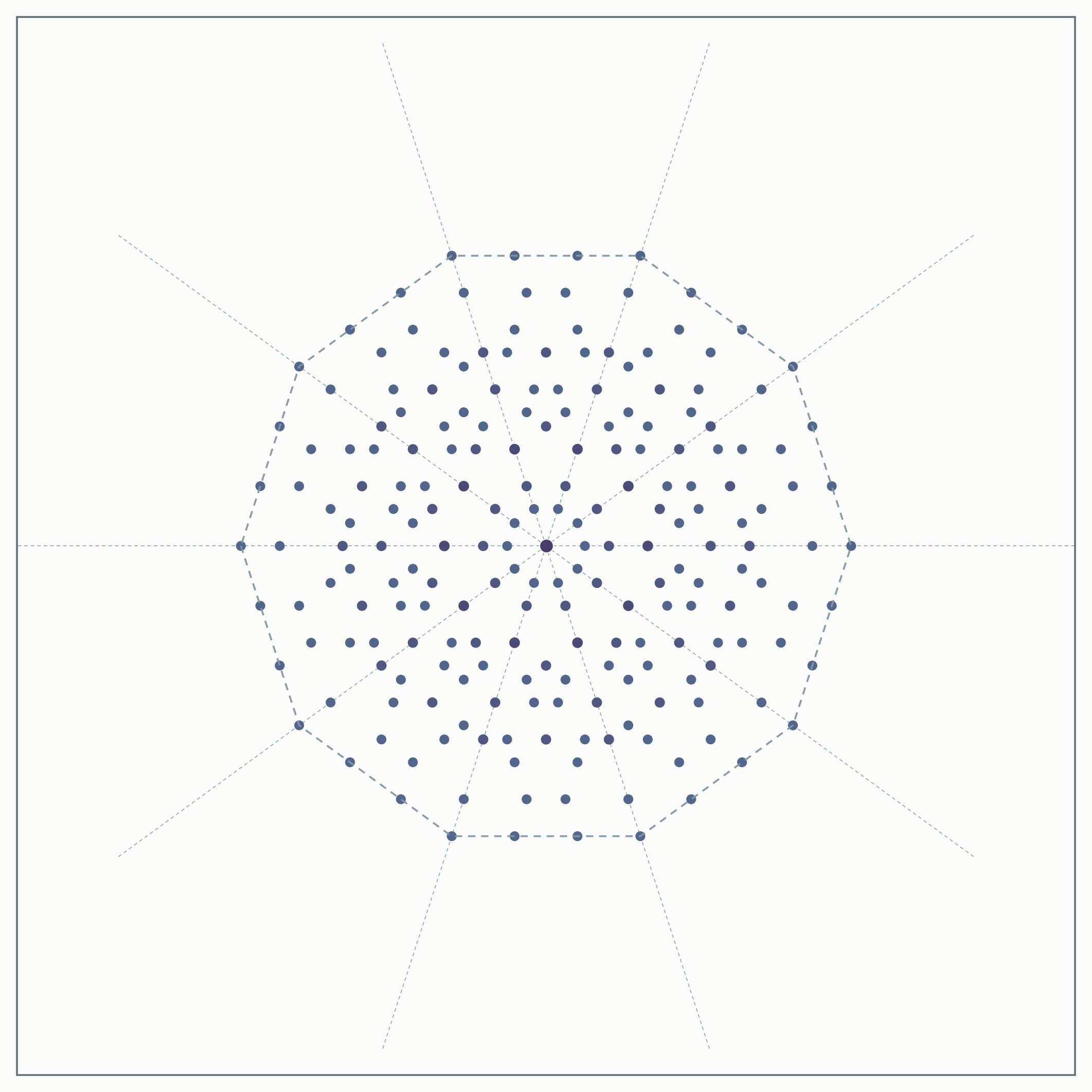}
        \caption{Penrose: $\rhigh=3$}
    \label{fig:penrose_reciprocal_L3}
    \end{subfigure}
    \hfill
    \begin{subfigure}[t]{0.32\textwidth}
        \centering
    \includegraphics[width=\textwidth]{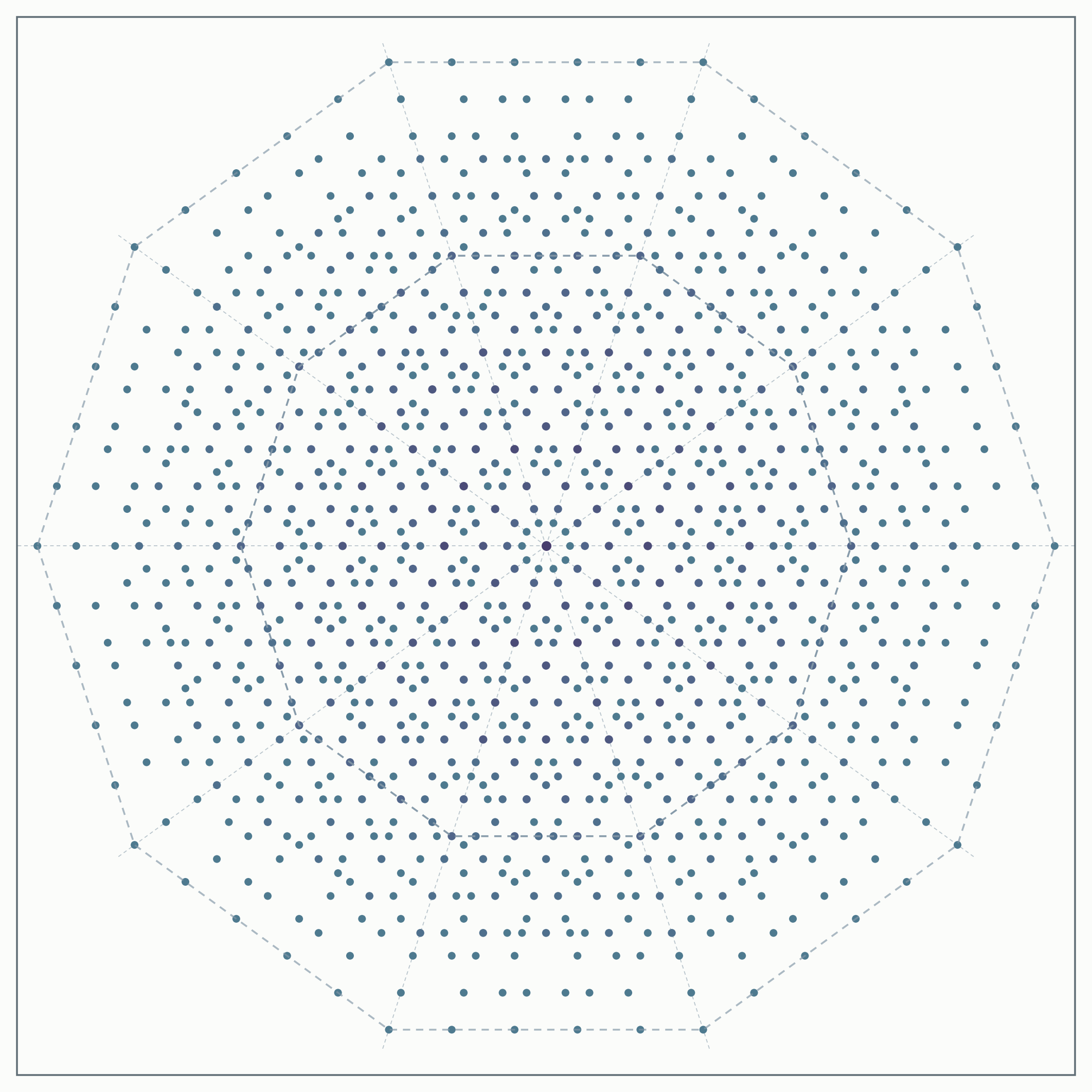}
        \caption{Penrose: $\rhigh=5$}
    \label{fig:penrose_reciprocal_L5}
    \end{subfigure}
    \hfill
    \begin{subfigure}[t]{0.32\textwidth}
        \centering
    \includegraphics[width=\textwidth]{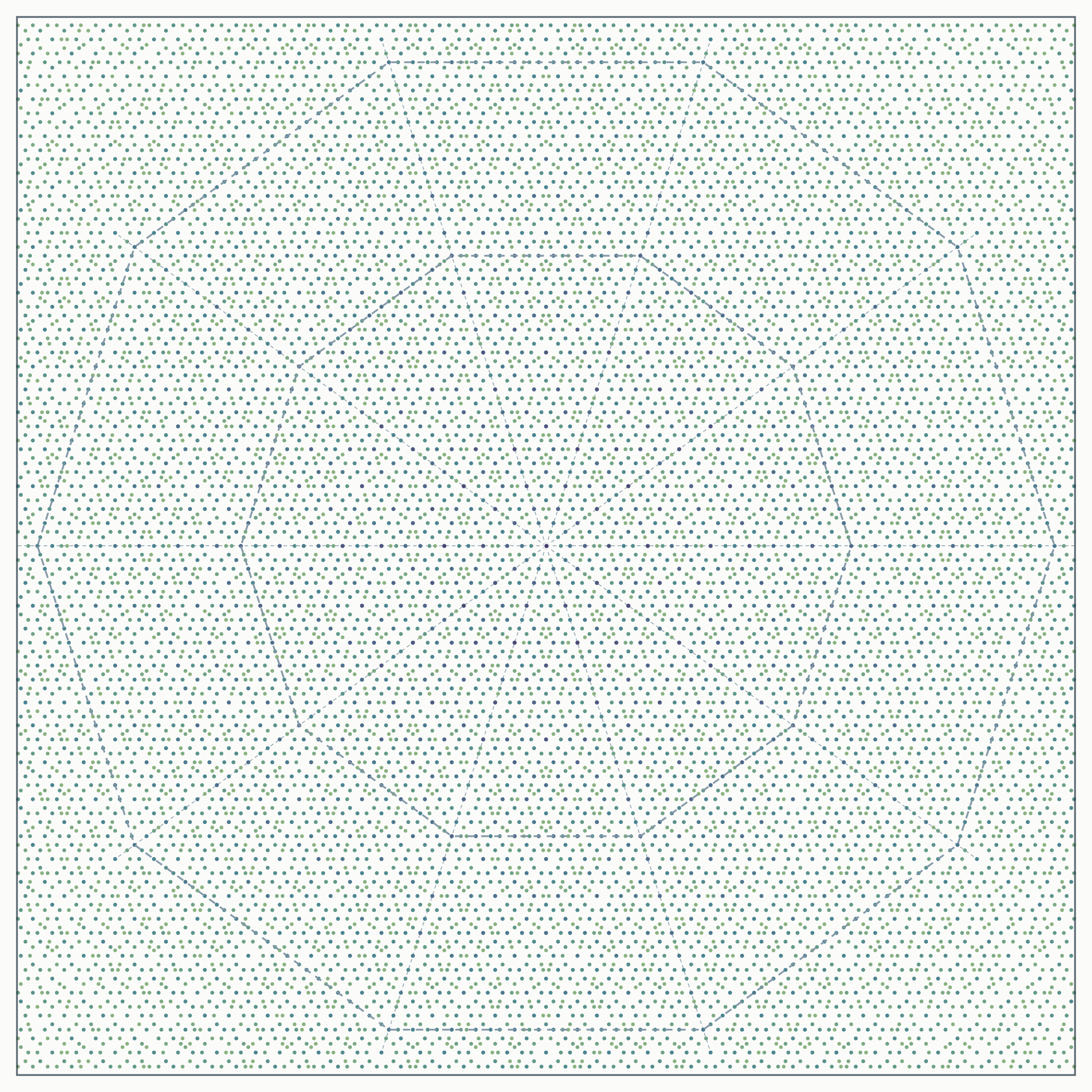}
        \caption{Penrose: $\rhigh=12$}
    \label{fig:penrose_reciprocal_L12}
    \end{subfigure}

    \vspace{0.8em}
    \begin{subfigure}[t]{0.32\textwidth}
        \centering
    \includegraphics[width=\textwidth]{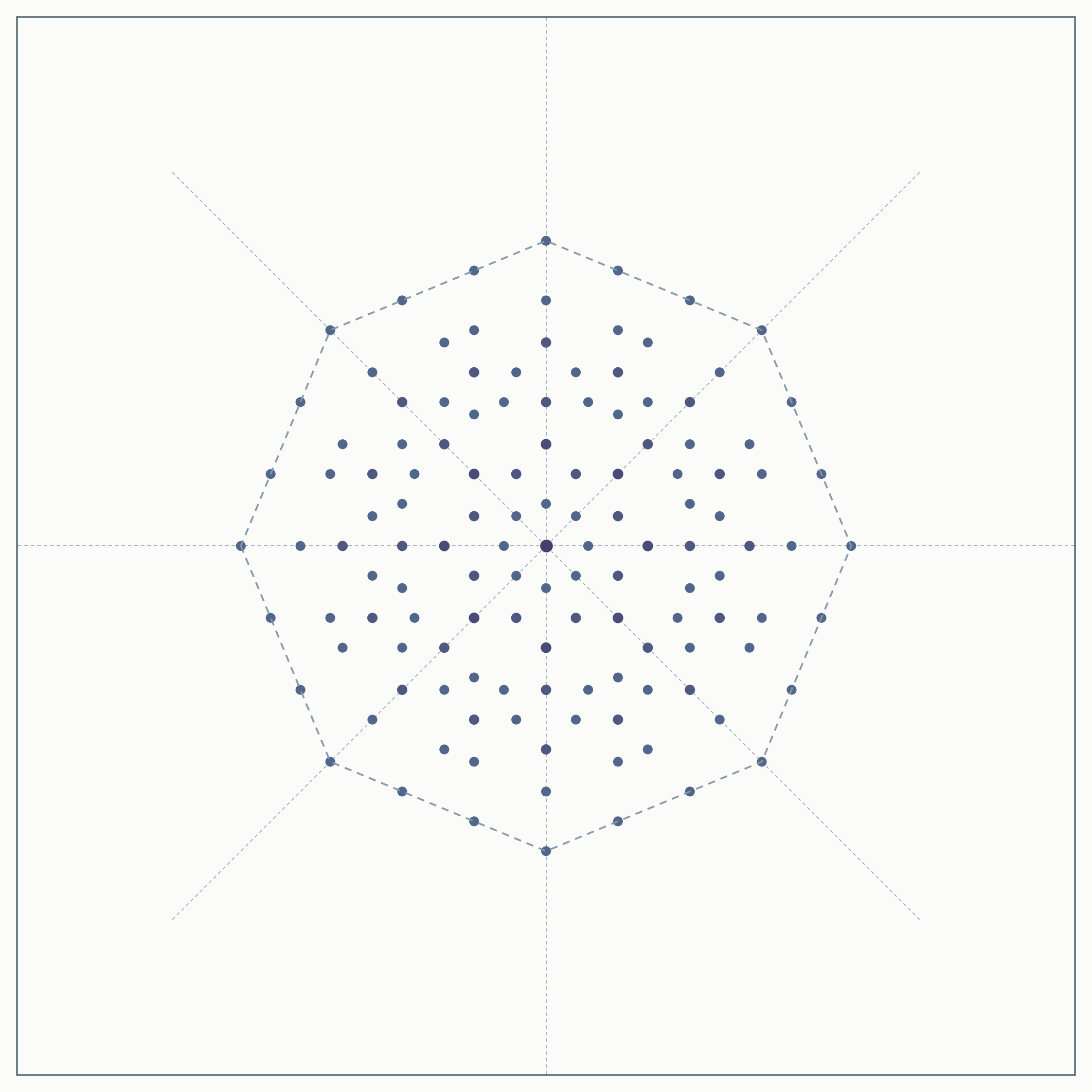}
        \caption{\footnotesize Ammann-Beenker: $\rhigh=3$}
        \label{fig:ab_reciprocal_L3}
    \end{subfigure}
    \hfill
    \begin{subfigure}[t]{0.32\textwidth}
        \centering
    \includegraphics[width=\textwidth]{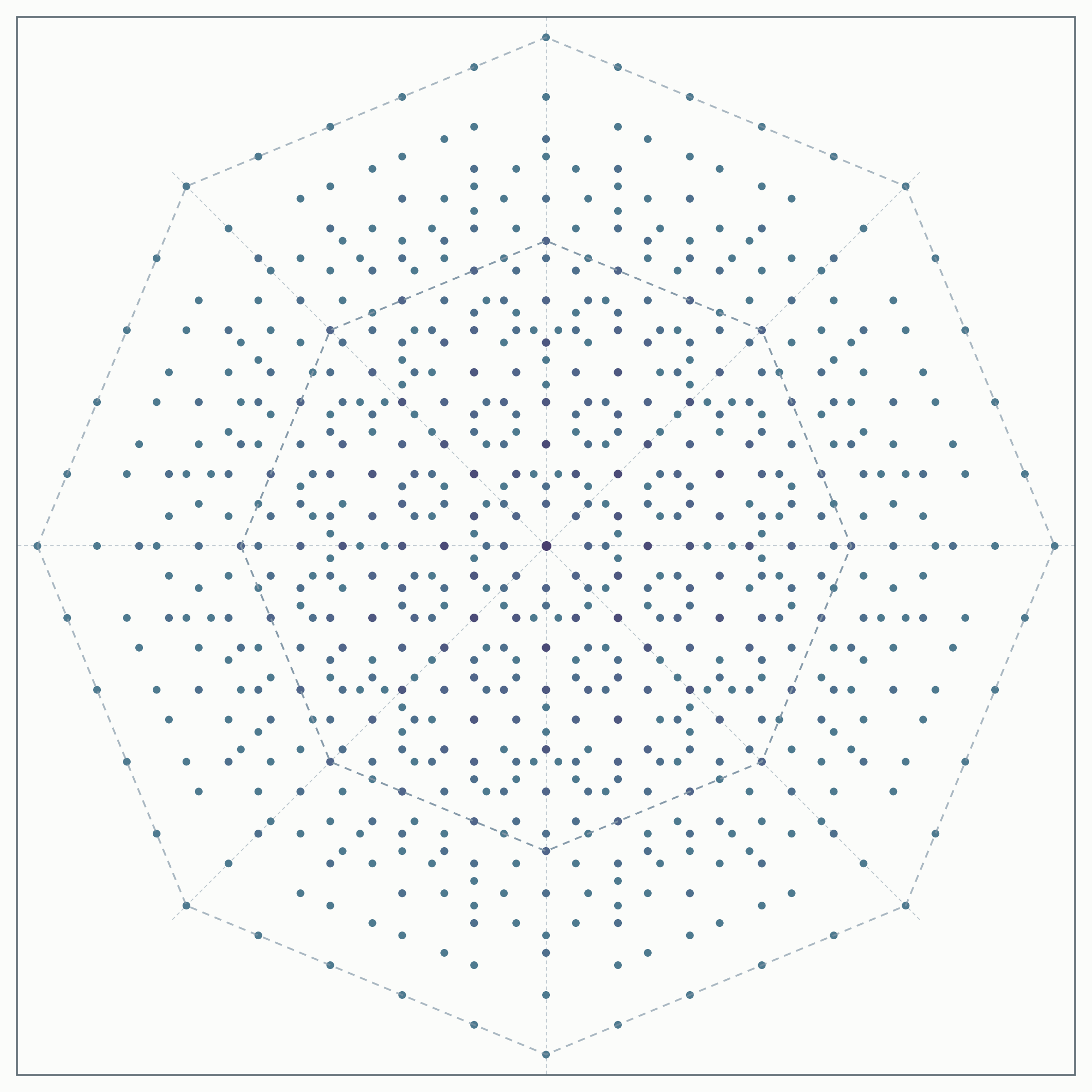}
        \caption{\footnotesize Ammann-Beenker: $\rhigh=5$}
        \label{fig:ab_reciprocal_L5}
    \end{subfigure}
    \hfill
    \begin{subfigure}[t]{0.32\textwidth}
        \centering
    \includegraphics[width=\textwidth]{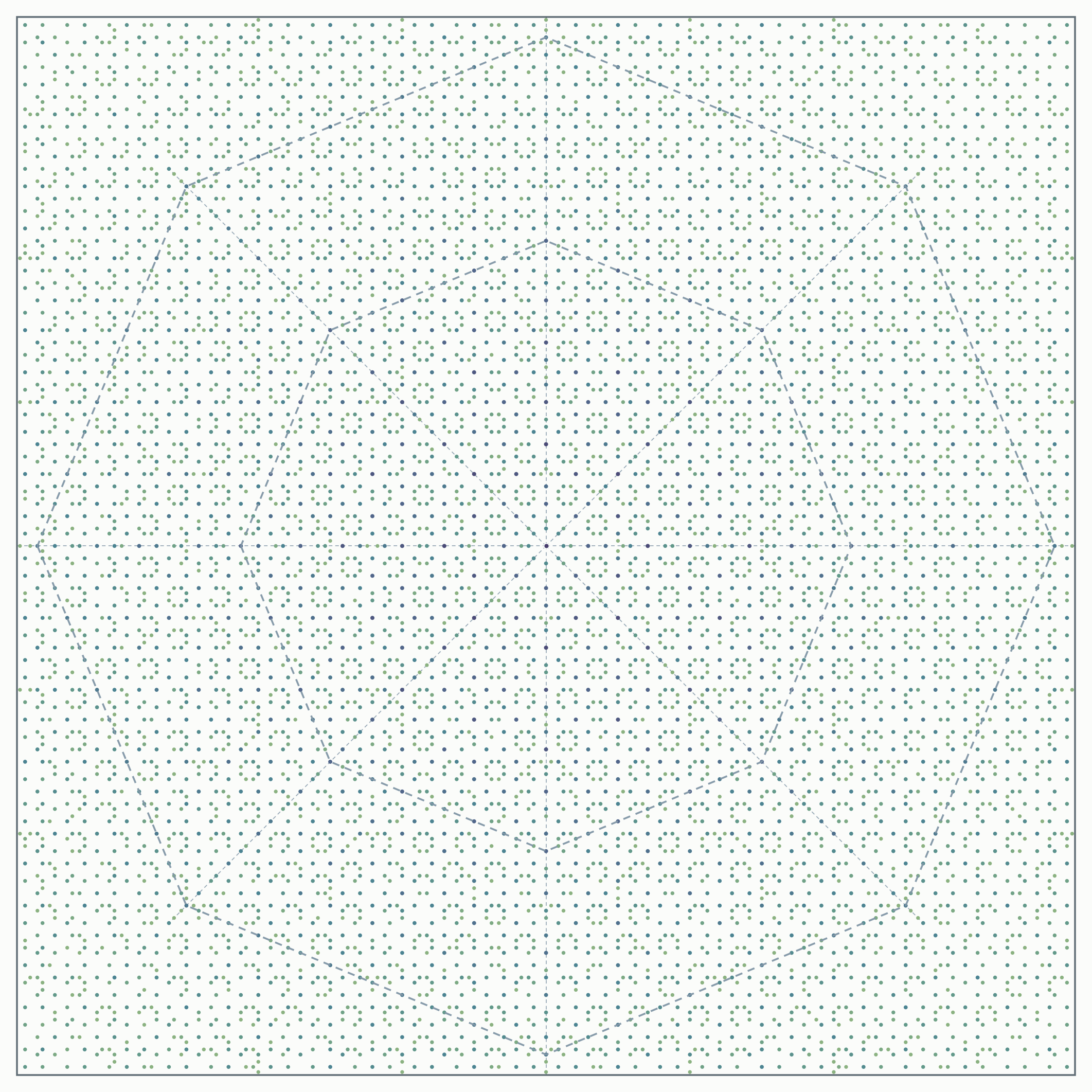}
        \caption{\footnotesize Ammann-Beenker: $\rhigh=12$}
        \label{fig:ab_reciprocal_L12}
    \end{subfigure}
    \caption{
    Reciprocal space high-dimensional truncation (given in \eqref{eq:high_dim_truncation}) of the Fourier modules for the Penrose and Ammann-Beenker  models. The Fourier modules naturally exhibit rotational symmetry, as observed at small cutoffs $\rhigh$. 
    For both models, increasing $\rhigh$ reveals the progressively denser structure of the Fourier module.
    }
\label{fig:reciprocal_module_density}
\end{figure}

\subsection{Rhombic Penrose model}
\label{sec:rhombic-penrose}

For the rhombic Penrose model \cite{BaakeGrimm2013,deBruijn1981I,deBruijn1981II}, we use the standard five-dimensional representation and define the five-fold rotation angles $\theta_j=2\pi j/5$, $j=0,\ldots,4$. The physical projection $\ppara^{\rm{Pen}}$ and the internal projection $\pperp^{\rm{Pen}}$ are
\begin{equation*}
\ppara^{\rm{Pen}} = 
\begin{pmatrix}
\cos\theta_0 & \cos\theta_1 & \cos\theta_2 & \cos\theta_3 & \cos\theta_4\\
\sin\theta_0 & \sin\theta_1 & \sin\theta_2 & \sin\theta_3 & \sin\theta_4
\end{pmatrix},
\end{equation*}
\begin{equation*}
\pperp^{\rm{Pen}} =
\begin{pmatrix}
\cos 2\theta_0
& \cos 2\theta_1
& \cos 2\theta_2
& \cos 2\theta_3
& \cos 2\theta_4\\[4pt]
\sin 2\theta_0
& \sin 2\theta_1
& \sin 2\theta_2
& \sin 2\theta_3
& \sin 2\theta_4\\[4pt]
\frac{1}{\sqrt2}
& \frac{1}{\sqrt2}
& \frac{1}{\sqrt2}
& \frac{1}{\sqrt2}
& \frac{1}{\sqrt2}
\end{pmatrix}.
\end{equation*}
We should point out that in this five-dimensional representation, the Penrose internal space is $\R^2\times(\Z/5\Z)$, represented within the three-dimensional coordinates \cite{BaakeGrimm2013,SingWelberry2006}.
The third coordinate labels the discrete $\Z/5\Z$ component rather than an additional continuous internal direction. The corresponding four acceptance window components are
\begin{equation*}
W_{\ell}^{\rm{Pen}} = 
\left\{ \pperp^{\rm{Pen}}\vx :~ \vx =(s_0,\ldots,s_4)^{\rm T}\in[0,1]^5,~ \sum_{j=0}^{4}s_j=\ell \right\} \subset\R^3, \qquad \ell=1,\ldots,4 .
\end{equation*}
Because of $\sum_{j=0}^{4}s_j=\ell$, the third coordinate of every point in $W_\ell^{\rm Pen}$ is fixed at $\ell/\sqrt2$. Each $W_\ell^{\rm Pen}$ is represented as a two-dimensional polygonal slice in the three-dimensional coordinate representation of $\R^2\times(\Z/5\Z)$. The four components $\ell=1,\ldots,4$ together define the Penrose acceptance window.
The quasicrystalline Penrose lattice is then given by
\begin{equation*}
\Lambda_{\rm{Pen}} = 
\left\{ \ppara^{\rm{Pen}}\vn :~ \vn\in\Z^5,~ \pperp^{\rm{Pen}}\vn \in  \bigcup_{\ell=1}^{4}W_{\ell}^{\rm{Pen}} \right\} \subset\R^2 .
\end{equation*}
The physical projection satisfies
$\ppara^{\rm Pen}(1,1,1,1,1)^{\rm T}=\vzero$, so that the five-dimensional integer representation contains the redundancy $\vn\sim\vn+m(1,1,1,1,1)^{\rm T}$ for $m\in\Z$. However, the four acceptance window components select lattice points with $\sum_{j=0}^{4}n_j\in\{1,2,3,4\}$. Consequently, each site of $\Lambda_{\rm Pen}$ has a unique five-dimensional representative satisfying the acceptance-window condition. 

A minimal rank-four representation is obtained by quotienting this redundant direction, $\Z^5/\Z(1,1,1,1,1)^{\rm T}\simeq\Z^4$, which is equivalent to the standard four-dimensional $A_4$-based description of the Penrose tiling
\cite{BaakeGrimm2013,Mazac2023}. In this minimal representation, the restrictions of the physical and continuous internal projections to the rank-four indexing lattice are injective, and the internal coordinate is two-dimensional. The $\Z/5\Z$ label carried by the third coordinate in the five-dimensional construction is encoded by the corresponding window component rather than by an additional internal coordinate. 

The four-dimensional and five-dimensional constructions therefore provide equivalent descriptions of the same Penrose tiling and Fourier module. The analytical derivations in Appendices \ref{app:coupling_rule} and \ref{app:pbz-mechanism} use the minimal-rank representation described above.
For numerical calculations, we retain the five-dimensional representatives to keep the rotational symmetry explicit. The mixed scattering-channel truncation used for bulk observables is constructed from the physical reciprocal projection and the first two continuous components of the internal reciprocal projection, as detailed in Appendix \ref{app:reciprocal_truncation}.

For the phason-shifted Penrose configurations considered in Section \ref{subsec:phason_invariance}, the four acceptance windows $\{W_{\ell,\bm{\gamma}}^{\rm Pen}\}_{\ell=1,\cdots,4}$ are translated by the phason shift $\pperp^{\rm Pen}\bm{\gamma}$. Following the de Bruijn parameterization, the parameters $\bm{\gamma}\in\R^5$ are restricted to $\sum_{j=0}^{4}\gamma_j=0$ \cite{deBruijn1981I,deBruijn1981II}. Under this constraint, the third component of $\bm{\gamma}_{\perp}^{\rm Pen}$ vanishes, so the Penrose phason shift is parameterized by two independent internal-space coordinates. The top row of Figure \ref{fig:phason_configuration_comparison} compares two representative Penrose configurations and their direct overlay in the same physical coordinates.

Since the phason shift leaves the projection maps unchanged, the Fourier module is independent of $\bm{\gamma}$, given by
\begin{equation*}
\Lambda_{\rm{Pen}}^* =  \left\{ (\ppara^{\rm{Pen}})^*\vn :~ \vn\in\Z^5 \right\}.
\end{equation*}
As shown in Figures \ref{fig:quasi-structure} and \ref{fig:reciprocal_module_density}, the Penrose tiling exhibits five-fold rotational symmetry in real space and decagonal symmetry in its Fourier module. 

\subsection{Ammann-Beenker model}
\label{subsec:AB}

\begin{figure}[!htb] 
    \centering 
    \begin{subfigure}[t]{0.32\textwidth} 
    \centering 
    \captionsetup{justification=centering}
    \includegraphics[width=\textwidth]{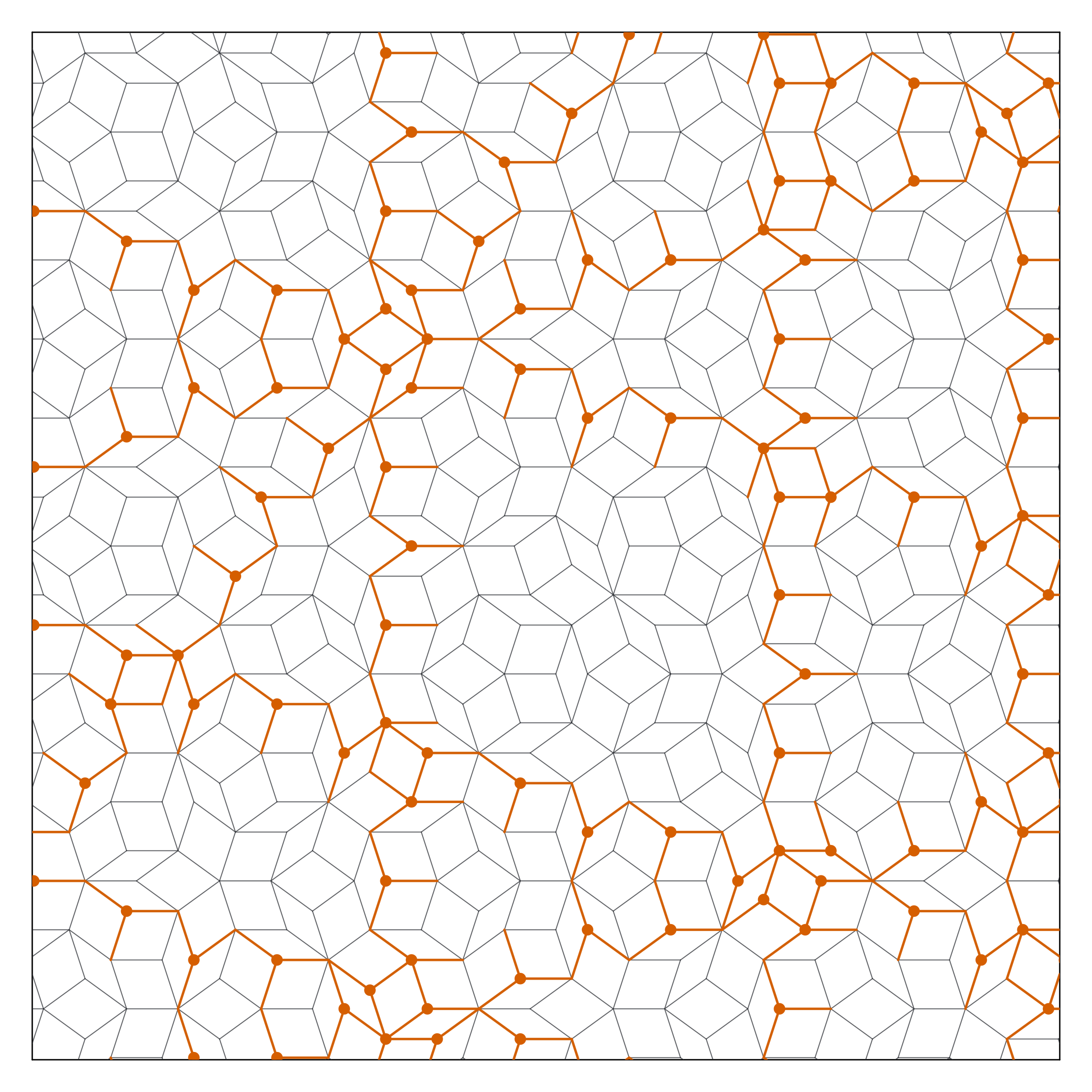} 
    \caption{Penrose:\\ {\footnotesize $\bm{\gamma} =(0.08,0.16,0.24,-0.16,-0.32)$}} \label{fig:penrose_phason_config1} 
    \end{subfigure} 
    \hfill 
    \begin{subfigure}[t]{0.32\textwidth} 
    \centering
    \captionsetup{justification=centering}
    \includegraphics[width=\textwidth]{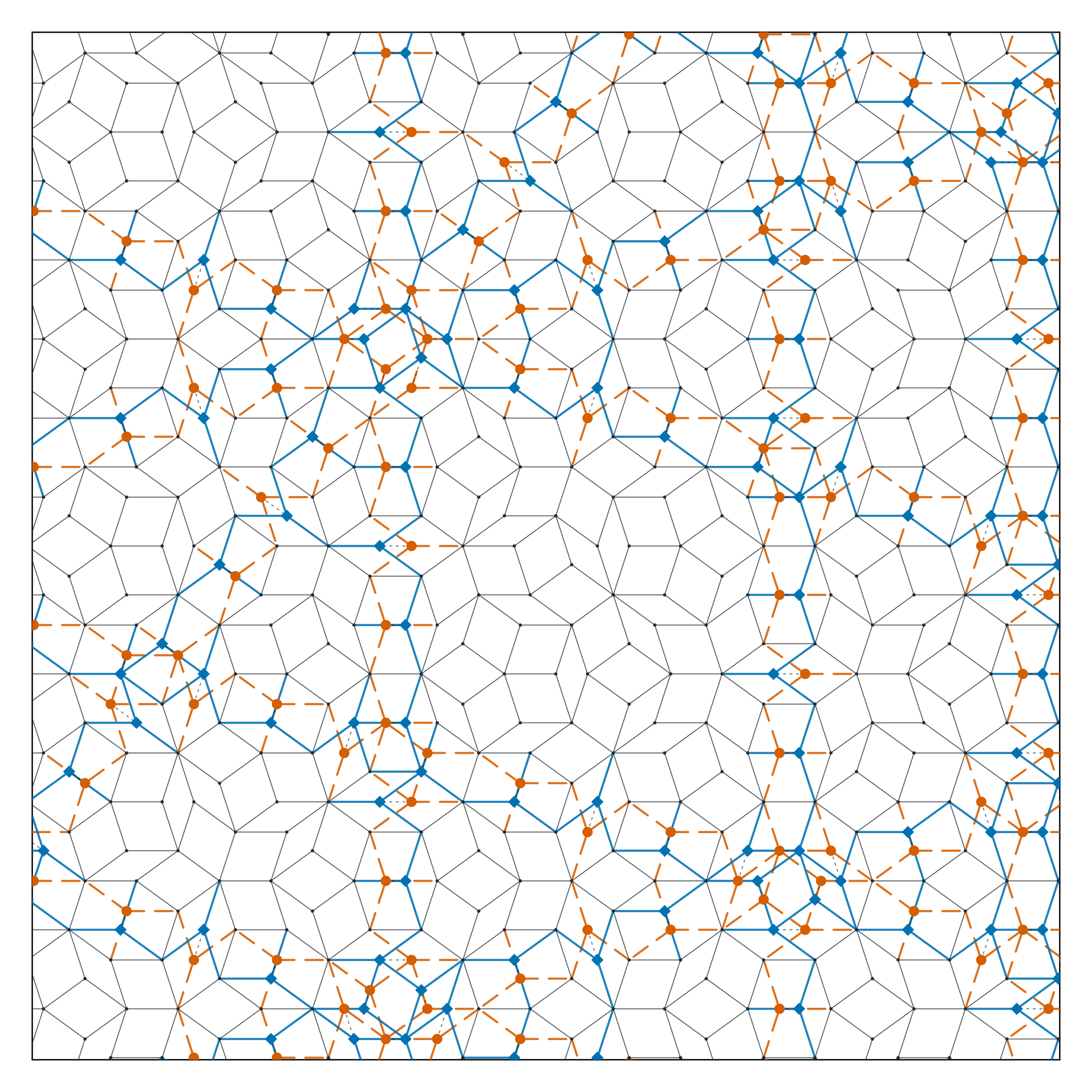} 
    \caption{Penrose: overlay} \label{fig:penrose_phason_overlay} 
    \end{subfigure} 
    \hfill 
    \begin{subfigure}[t]{0.32\textwidth} 
    \centering 
    \captionsetup{justification=centering}
    \includegraphics[width=\textwidth]{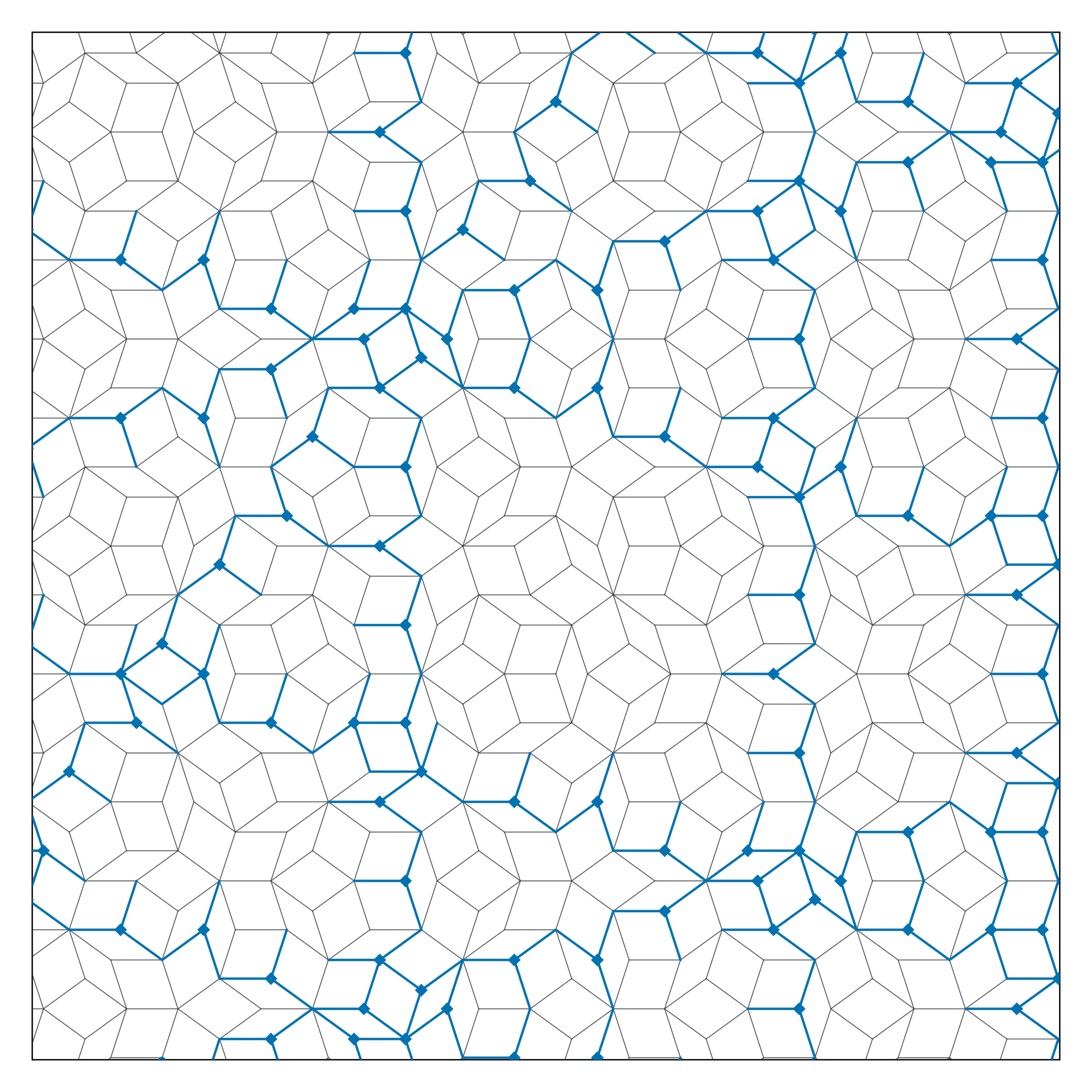} 
    \caption{Penrose:\\ {\footnotesize $\bm{\gamma}=(0.20,0.40,0.60,-0.40,-0.80)$}} \label{fig:penrose_phason_config2} 
    \end{subfigure} 
    \vspace{0.8em} 
    \begin{subfigure}[t]{0.32\textwidth} 
    \centering 
    \captionsetup{justification=centering}
    \includegraphics[width=\textwidth]{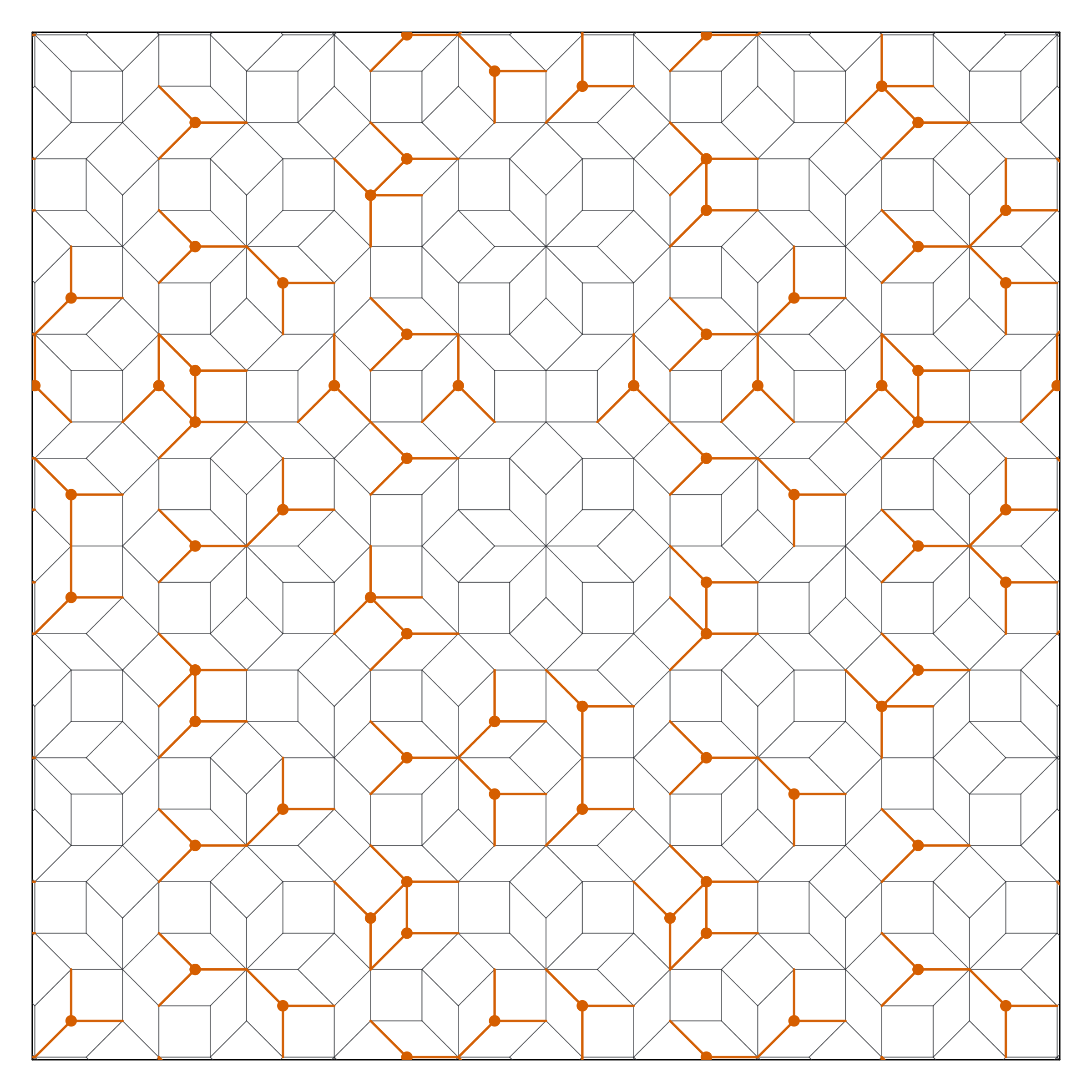} 
    \caption{Ammann-Beenker: \\{\footnotesize $\bm{\gamma} =(-0.08,0.04,0.03,-0.07)$}}
    \label{fig:ab_phason_config1} 
    \end{subfigure} 
    \hfill 
    \begin{subfigure}[t]{0.32\textwidth} 
    \centering 
    \captionsetup{justification=centering}
    \includegraphics[width=\textwidth]{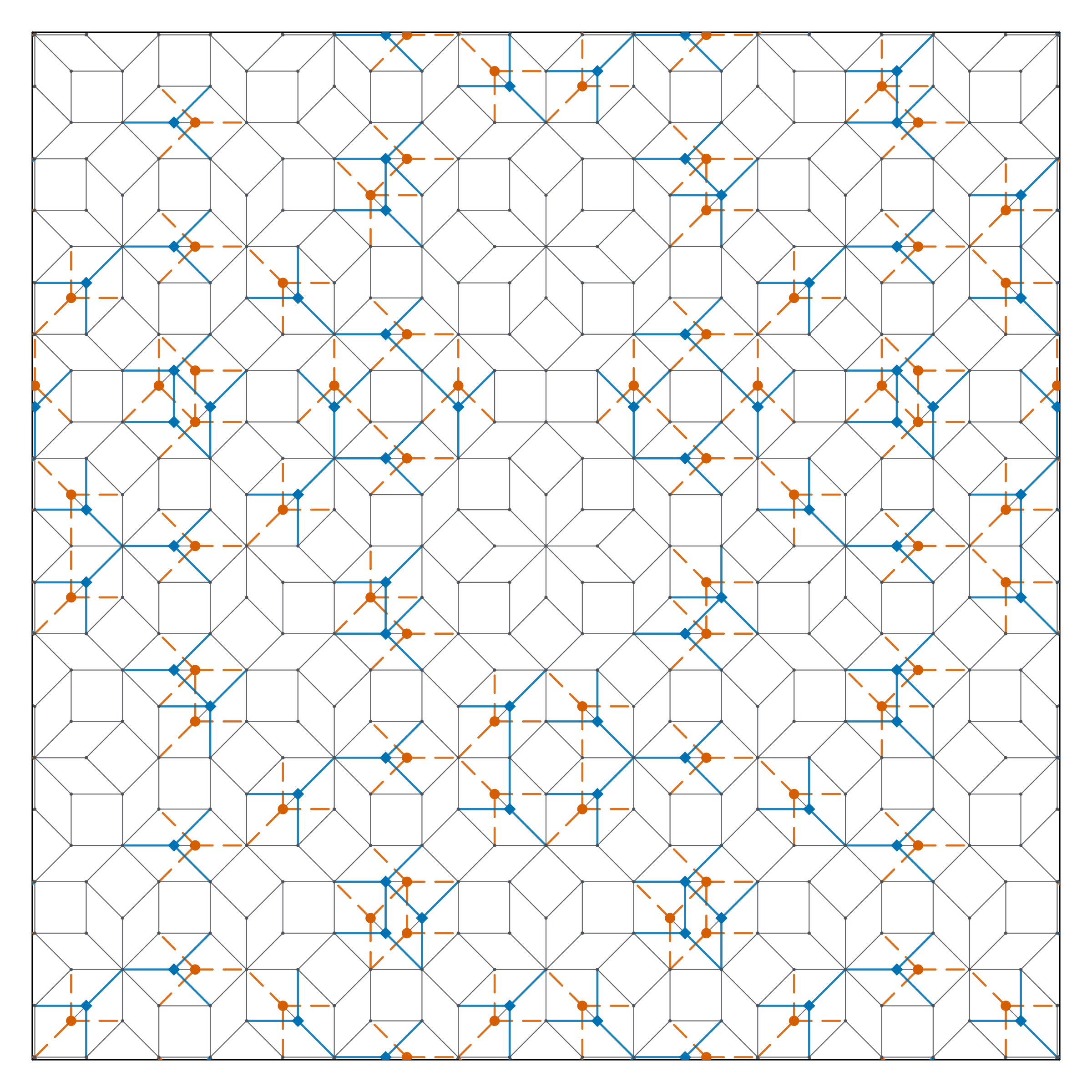} 
    \caption{Ammann-Beenker: overlay} \label{fig:ab_phason_overlay} 
    \end{subfigure} 
    \hfill 
    \begin{subfigure}[t]{0.32\textwidth} 
    \centering 
    \captionsetup{justification=centering}
    \includegraphics[width=\textwidth]{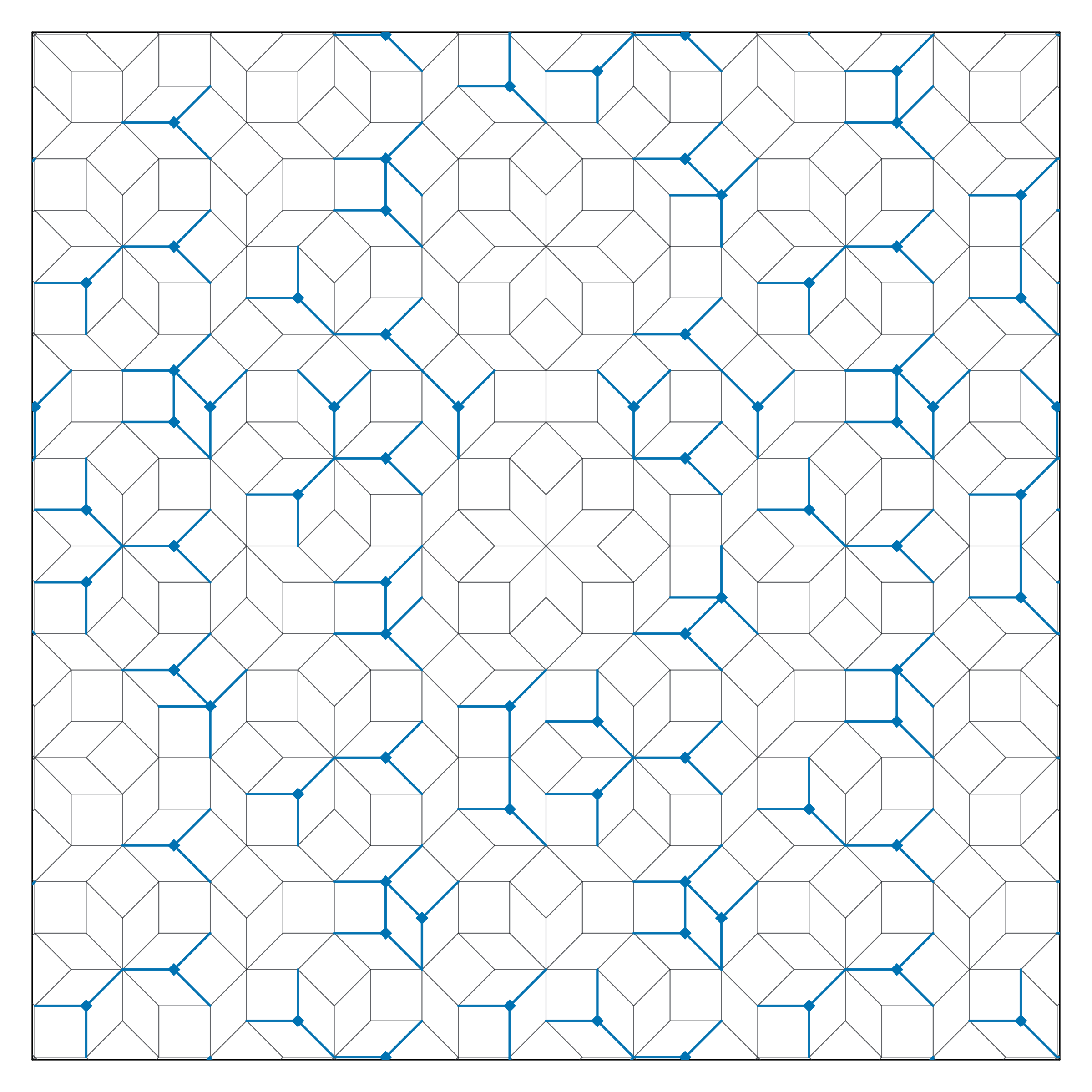} 
    \caption{Ammann-Beenker: \\ {\footnotesize $\bm{\gamma}=(0.10,-0.10,0.05,0.04)$}} 
    \label{fig:ab_phason_config2} 
    \end{subfigure} 
    \caption{Representative phason-shifted configurations of the Penrose (upper row) and Ammann-Beenker (lower row) tilings. The left and right panels show configurations generated by two different phason shifts, while the middle panels directly overlay the two configurations in the same physical coordinates. Black sites and edges are common to both configurations, orange sites and edges occur only in the left configuration, and blue sites and edges occur only in the right configuration. 
    The mismatched regions arise from sites entering or leaving the shifted acceptance window and correspond to local rearrangements of the tiling.} 
    \label{fig:phason_configuration_comparison} 
\end{figure}

For the Ammann-Beenker model, let the eight-fold rotation angles be $\theta_j=\pi j/4$ ($j=0,\ldots,3$). The physical space projection $\ppara^{\rm{AB}}$ and the corresponding internal space projection $\pperp^{\rm{AB}}$ \cite{BaakeGrimm2013} are
\begin{equation*}
\ppara^{\rm{AB}} =  
\begin{pmatrix}
\cos\theta_0 & \cos\theta_1 & \cos\theta_2 & \cos\theta_3\\
\sin\theta_0 & \sin\theta_1 & \sin\theta_2 & \sin\theta_3
\end{pmatrix},
\end{equation*}
\begin{equation*}
\pperp^{\rm{AB}} =  
\begin{pmatrix}
\cos 3\theta_0 & \cos 3\theta_1 & \cos 3\theta_2 & \cos 3\theta_3\\
\sin 3\theta_0 & \sin 3\theta_1 & \sin 3\theta_2 & \sin 3\theta_3
\end{pmatrix}.
\end{equation*}
The acceptance window is the centered regular octagon obtained by projecting the four-dimensional unit cube onto internal space \cite{BaakeGrimm2013},
\begin{equation*}
W^{\rm{AB}} = 
\left\{ \pperp^{\rm{AB}}\vx :~ \vx\in \left[-\frac12,\frac12\right]^4 \right\} \subset\R^2 .
\end{equation*}
The Ammann-Beenker point set is
\begin{equation*}
\Lambda_{\rm{AB}} = 
\left\{ \ppara^{\rm{AB}}\vn :~ \vn\in\Z^4,~ \pperp^{\rm{AB}}\vn \in W^{\rm{AB}} \right\} \subset\R^2 .
\end{equation*}
Distinct Ammann-Beenker configurations are produced by varying the phason shift $\bm{\gamma}_{\perp}^{\rm AB}=\pperp^{\rm AB}\bm{\gamma}$. The bottom row of Figure \ref{fig:phason_configuration_comparison} shows the corresponding comparison for two representative Ammann-Beenker configurations. The Fourier module is likewise independent of $\bm{\gamma}$, given by
\begin{equation*}
\Lambda_{\rm{AB}}^* = \left\{ (\ppara^{\rm{AB}})^*\vn :~ \vn\in\Z^4 \right\}.
\end{equation*}
As shown in Figures \ref{fig:quasi-structure} and \ref{fig:reciprocal_module_density}, the Ammann-Beenker tiling and its Fourier module both exhibit octagonal symmetry. 

\section{Coupling rule for scattering channels}
\label{app:coupling_rule}

In this appendix, we derive the fundamental reciprocal space coupling rule for the Hamiltonian matrix elements.

\medskip

\noindent\textbf{Coupling Rule.}
Two scattering channels can couple only if their momentum transfer belongs to the Fourier module, yielding the condition
\begin{equation}
\label{eq:app_coupling_rule_statement}
\lim_{R\to\infty} \langle\psi^R_{\vk\alpha}|\ham|\psi^R_{\vk'\alpha'}\rangle \ne0 \quad\Longrightarrow\quad \vk'-\vk\in\Lambda^* .
\end{equation}

Throughout the derivation, we employ the standard internal space notation for model sets: for any projected point $\vr=\ppara\vn\in\Lambda$, its corresponding internal space projection is denoted by $\vr_\perp=\pperp\vn$. 
Following the tight-binding formulation of Section \ref{sec:tight_binding_real_space}, we introduce two standard locality assumptions on the matrix elements: a finite hopping range and local dependence on the atomic configuration \cite{Goringe1997,LenzStollmann2003,Kellendonk2003}.

In the orthonormal localized basis, the Hamiltonian matrix elements retain their locality, possibly with small tails that decay rapidly with intersite distance.
For simplicity, we make the finite-range assumption \cite{LenzStollmann2003}: there exists $R_{\ham}>0$ such that $t_{\vr\alpha,\vr'\alpha'}$ vanishes when $|\vr-\vr'|>R_{\ham}$.
Consequently, a nonzero matrix element requires $\vr-\vr'\in\Omega_{R_{\ham}} := (\Lambda-\Lambda)\cap\overline{B_{R_{\ham}}(\vzero)}$. 
Since $\Lambda$ is also a Meyer set and possesses finite local complexity \cite{BaakeGrimm2013,BalkovaMasakovaPelantova2004}, its difference set $\Lambda-\Lambda$ is uniformly discrete \cite{Moody2000}, which naturally ensures that $\Omega_{R_{\ham}}$ is a finite set.
As a second assumption, we assume that the hopping elements depend on the local atomic configuration \cite{LenzStollmann2003,Kellendonk2003}.
Specifically, for a fixed displacement $\vz\in\R^d$, the matrix element $t_{\vr\alpha,(\vr+\vz)\alpha'}$ is completely determined by the local configuration of atoms within a radius $R_{\rm{env}}+|\vz|$ around the site $\vr$. Consequently, if two ordered site pairs, $(\vr_1,\vr_1+\vz)$ and $(\vr_2,\vr_2+\vz)$, share identical local configurations,
\begin{equation*}
(\Lambda-\vr_1)\cap\overline{B_{R_{\rm{env}}+|\vz|}(\vzero)}
= (\Lambda-\vr_2)\cap\overline{B_{R_{\rm{env}}+|\vz|}(\vzero)},
\end{equation*}
then their matrix elements coincide, yielding $t_{\vr_1 \alpha,(\vr_1+\vz)\alpha'} = t_{\vr_2 \alpha,(\vr_2+\vz)\alpha'}$.
These assumptions are widely applicable, as tight-binding models fundamentally rely on the strong localization of orbitals at atomic sites \cite{Kaxiras2003}. 

For each $\vz\in\Omega_{R_{\ham}}$, we define the set $
\Lambda_{\vz} := \left\{ \vr\in\Lambda :~ \vr+\vz\in\Lambda \right\}$. In the cut-and-project scheme, a point $\vr$ belongs to $\Lambda_{\vz}$ if and only if its internal-space projection  satisfies $\vr_\perp\in W_{\vz} := W\cap(W-\vz_\perp)$. Consequently, this set can be explicitly rewritten as
\begin{equation*}
\Lambda_{\vz} = 
\left\{ \vr=\ppara\vn :~ \vn\in\Z^D,~ \vr_\perp\in W_{\vz} \right\},
\end{equation*}
which demonstrates that $\Lambda_{\vz}$ is a model set generated by the same cut-and-project scheme, with the acceptance window $W_{\vz}$. We then define the weighted Fourier-Bohr coefficient
\begin{equation}
\label{eq:app_weighted_hopping_coefficient}
\mathcal T_{\vz}^{\alpha\alpha'}(\vq) := 
\lim_{R\to\infty} \frac{1}{|\Lambda_{\vz,R}|} \sum_{\vr\in\Lambda_{\vz,R}} t_{\vr \alpha,(\vr+\vz)\alpha'} \ee^{-\im\vq\cdot\vr},
\end{equation}
where $\Lambda_{\vz,R}=\Lambda_\vz \cap B_R(\vzero)$. Let $\vz=\vr'-\vr$. The finite-range assumption restricts nonzero contributions to displacements $\vz\in\Omega_{R_{\ham}}$, such that the sum over displacements remains finite. 
The reciprocal-space matrix element can thereby be derived as
\begin{align*}
\lim_{R\to\infty} \langle\psi^R_{\vk\alpha}|\ham|\psi^R_{\vk'\alpha'}\rangle
&= \lim_{R\to\infty} \frac{1}{|\Lambda_R|} \sum_{\vr,\vr'\in\Lambda_R} \ee^{\im\vk\cdot\vr}t_{\vr\alpha,\vr'\alpha'} \ee^{-\im\vk'\cdot\vr'},\\
&= \sum_{\vz\in\Omega_{R_{\ham}}} \ee^{-\im\vk'\cdot\vz} \bigg(\lim_{R\to\infty} \frac{1}{|\Lambda_R|} \sum_{\substack{ \vr\in\Lambda_R\\\vr+\vz\in\Lambda}} t_{\vr\alpha,(\vr+\vz)\alpha'} \ee^{-\im(\vk'-\vk)\cdot\vr}\bigg), \\
&= \sum_{\vz\in\Omega_{R_{\ham}}} \ee^{-\im\vk'\cdot\vz} \bigg(\frac{\rho_{\Lambda_\vz}}{\rho_\Lambda}\mathcal T_{\vz}^{\alpha\alpha'}(\vk'-\vk)\bigg),
\end{align*}
where $\rho_{\Lambda}$ and $\rho_{\Lambda_\vz}$ are the densities of $\Lambda$ and $\Lambda_{\vz}$, respectively. For the expression in the second line, replacing the condition $\vr+\vz\in\Lambda_R$ with $\vr+\vz\in\Lambda$ alters only boundary contributions, which vanish in the thermodynamic limit.

Next, we demonstrate that the coefficient $\mathcal T_{\vz}^{\alpha\alpha'}(\vk'-\vk)$ vanishes whenever $\vk'-\vk \notin \Lambda^*$. Utilizing the local atomic configuration dependence, the value of $t_{\vr\alpha,(\vr+\vz)\alpha'}$ depends solely on the local environment in $B_{R_{\rm{env}}+|\vz|}(\vr)$. These possible relative positions form the set
\begin{equation*}
\Omega_{\vz} := (\Lambda-\Lambda) \cap \overline{B_{R_{\rm{env}}+|\vz|}(\vzero)} .
\end{equation*}
By finite local complexity of the quasicrystals, $\Omega_{\vz}$ is finite. 
In this sense, as $\vr$ varies over $\Lambda_{\vz}$, there are finitely many local environments for the pair $(\vr,\vr+\vz)$.
We may decompose $\Lambda_{\vz}$ into finite disjoint local configuration classes,
\begin{equation*}
\Lambda_{\vz} = \bigcup_{j=1}^{M_{\vz}} \Lambda_{\vz,j}.
\end{equation*}
For every $\vr\in\Lambda_{\vz,j}$, the matrix element takes a constant value, yielding $t_{\vr\alpha,(\vr+\vz)\alpha'} \equiv c_{\vz,j}^{\alpha\alpha'}$.

Moreover, for all local configuration classes $\{\Lambda_{\vz,j}\}_{j=1,\cdots, M_{\vz}}$, there are corresponding acceptance windows $\{W_{\vz,j}\}_{j=1,\cdots, M_{\vz}}$ \cite{Kaiser2025}, satisfying
\begin{equation*}
\Lambda_{\vz,j} = 
\left\{ \vr=\ppara\vn :~ \vn\in\Z^D,~ \vr_\perp\in W_{\vz,j} \right\}, \qquad 1\le j \le M_{\vz}.
\end{equation*}
These acceptance windows cover $W_{\vz}$ and possess disjoint interiors. Since the projections $\ppara$ and $\pperp$ are unchanged, each $\Lambda_{\vz,j}$ is generated by the same cut-and-project scheme and therefore has the same Fourier module $\Lambda^*$ \cite{BaakeGrimm2013}.

Using the decomposition of $\Lambda_{\vz}$ into the sets $\Lambda_{\vz,j}$, \eqref{eq:app_weighted_hopping_coefficient} becomes
\begin{equation}
\label{eq:app_weighted_hopping_decomposition}
\mathcal T_{\vz}^{\alpha\alpha'}(\vk'-\vk) = 
\sum_{j=1}^{M_{\vz}} \frac{\rho_{\Lambda_{\vz,j}}c_{\vz,j}^{\alpha\alpha'} }{\rho_{\Lambda_\vz}} \lim_{R\to\infty} \frac{1}{|\Lambda_{\vz,j,R}|} \sum_{\vr\in\Lambda_{\vz,j,R}} \ee^{-\im(\vk'-\vk)\cdot\vr},
\end{equation}
where $\Lambda_{\vz,j,R}=\Lambda_{\vz,j}\cap B_R$. 
Because each subset $\Lambda_{\vz,j}$ is generated by the same cut-and-project scheme, its Fourier--Bohr coefficients are supported on the Fourier module $\Lambda^*$~\cite{BaakeGrimm2013,BaakeMoody2004}, yielding
\begin{equation*}
\vk'-\vk \notin \Lambda^*
\quad\Longrightarrow\quad
\mathcal T_{\vz}^{\alpha\alpha'}(\vk'-\vk)=0,
\qquad
\forall\,\vz\in\Omega_{R_\ham}.
\end{equation*}
Since the reciprocal-space matrix element $\lim_{R\to\infty} \langle\psi^R_{\vk\alpha}|\ham|\psi^R_{\vk'\alpha'}\rangle$ is a finite linear combination of $\mathcal T_{\vz}^{\alpha\alpha'}(\vk'-\vk)$ over $\vz\in\Omega_{R_\ham}$, it vanishes whenever $\vk'-\vk\notin\Lambda^*$, which completes the derivation of the scattering coupling rule in \eqref{eq:app_coupling_rule_statement}.

\section{Scattering-channel cutoff}
\label{app:reciprocal_truncation}

Since the dense Fourier module $\Lambda^*$ generates an infinite number of scattering channels at each wave vector $\vk$, this appendix compares two systematic truncation schemes to facilitate numerical calculations.
The first is a direct cutoff on the $D$-dimensional integer index
\begin{equation}
\label{eq:high_dim_truncation}
\Lambda_{\rhigh}^* =
\left\{ \ppara^*\vn :~ \vn\in\Z^D,\ \|\vn\|_1\le \rhigh \right\}.
\end{equation}
This direct cutoff is natural in the $D$-dimensional index space, but is computationally less efficient because it does not distinguish between the physical and internal reciprocal directions.
We therefore introduce the following mixed truncation, which applies separate cutoffs in the two directions
\begin{equation}
\label{eq:mix_truncation}
\Lambda_{\rpara,\rperp}^* =
\left\{ \ppara^*\vn :~ \vn\in\Z^D,~ \|\ppara^*\vn\|_2\le \rpara,~ \|\pperp^*\vn\|_2\le \rperp \right\}.
\end{equation}
The mixed truncation treats the physical reciprocal-space and internal-space directions separately, allowing their convergence to be controlled independently. 
For the Penrose model, the internal reciprocal projection in \eqref{eq:mix_truncation} refers to the first two continuous components of the five-dimensional representation. Reciprocal vectors generated by equivalent five-dimensional indices are identified.

We next examine the numerical convergence of the density of states and the current-current correlation function under the two scattering-channel truncation schemes.
For the higher-dimensional cutoff tests, the reference values are obtained at $\rhigh^{\rm{ref}}=15$. For the mixed cutoff tests, we use the reference parameters $\rpara^{\rm{ref}}=15$ and $\rperp^{\rm{ref}}=150$: $\rpara$ is varied with $\rperp$ fixed at $150$, whereas $\rperp$ is varied with $\rpara$ fixed at $15$. For the convergence test of the density of states, we use $\gGaus(E)$ with $\eta=0.1$, centered at $E_0=-3.48$ for the Penrose model and $E_0=-3.29$ for the Ammann-Beenker model. For the convergence of the current-current correlation function, the zero-temperature Fermi kernel $\GFermi$ is applied, with the out-of-plane Zeeman coupling strength set to $\delta_M=0.8$ and the Fermi energy $E_{\rm{F}}$ positioned within the $5/6$-filling gap.

\begin{figure}[htbp]
    \centering
    \begin{subfigure}[t]{0.48\textwidth}
        \centering
        \includegraphics[width=\linewidth]{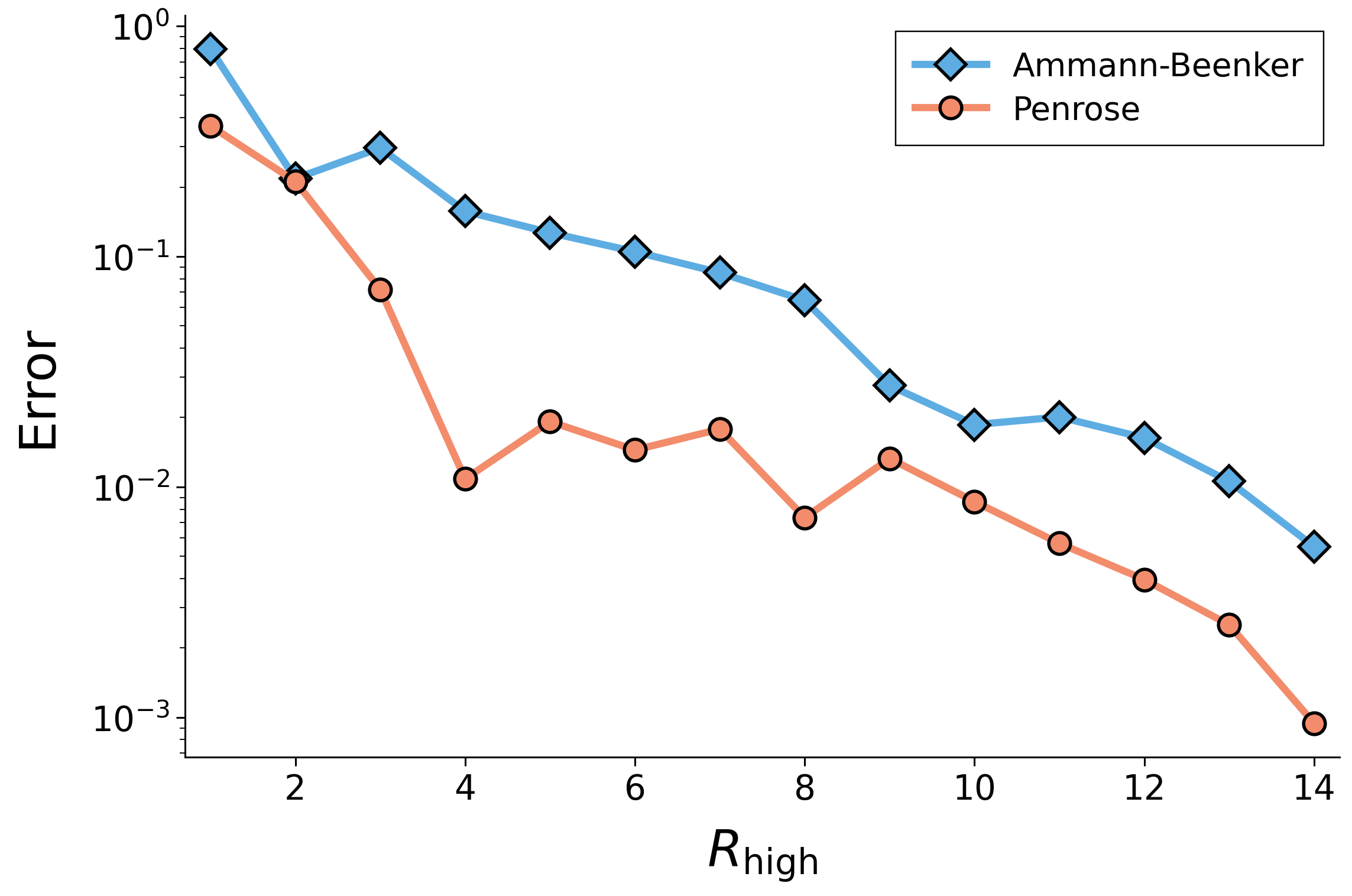}
        \caption{High-dimensional cutoff }
        \label{fig:dos_convergence_app}
    \end{subfigure}
    \begin{subfigure}[t]{0.48\textwidth}
        \centering
        \includegraphics[width=\linewidth]{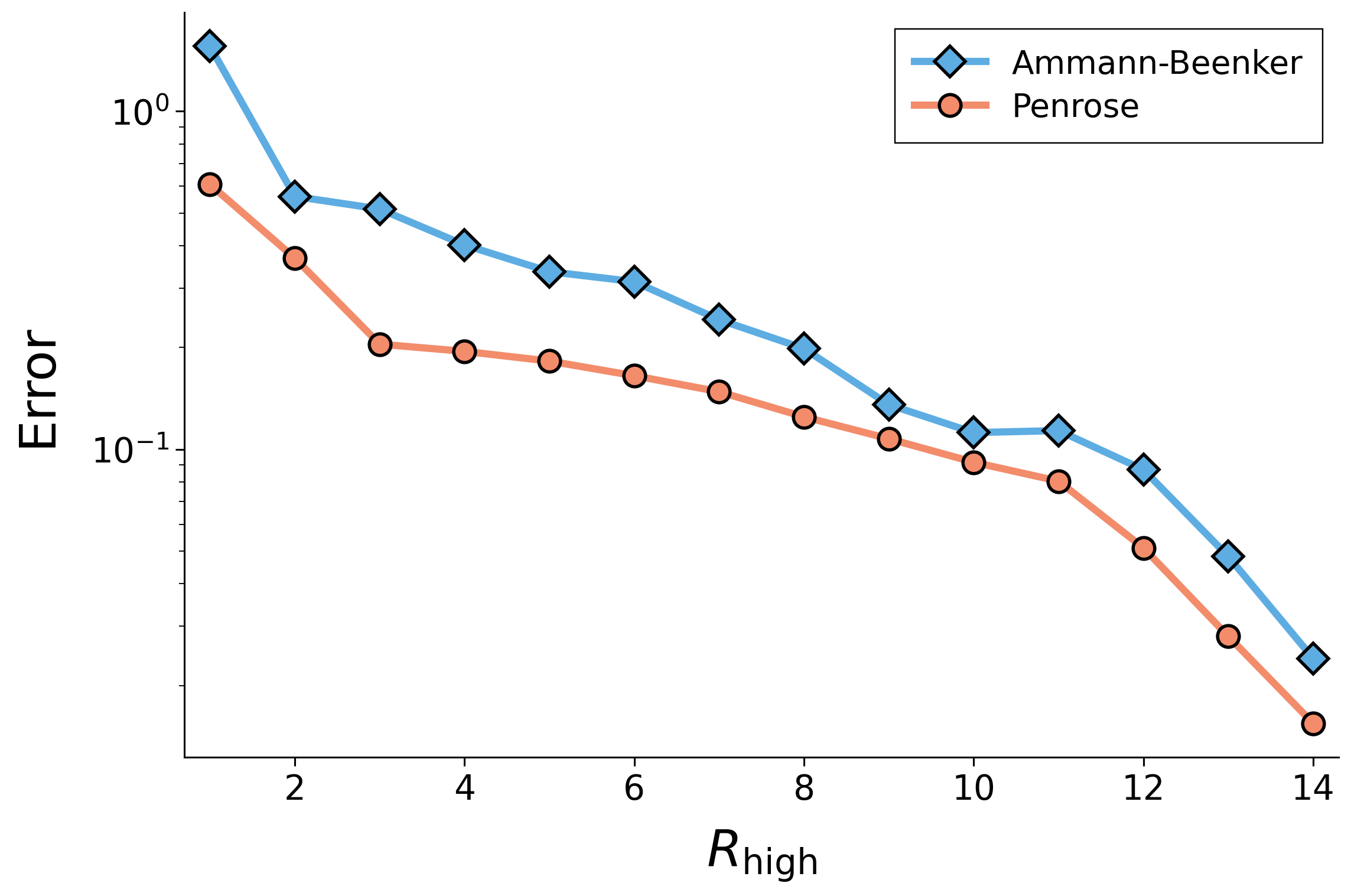}
        \caption{High-dimensional cutoff}
        \label{fig:corr_convergence_app}
    \end{subfigure}
    \hfill
    \begin{subfigure}[t]{0.48\textwidth}
        \centering
        \includegraphics[width=\linewidth]{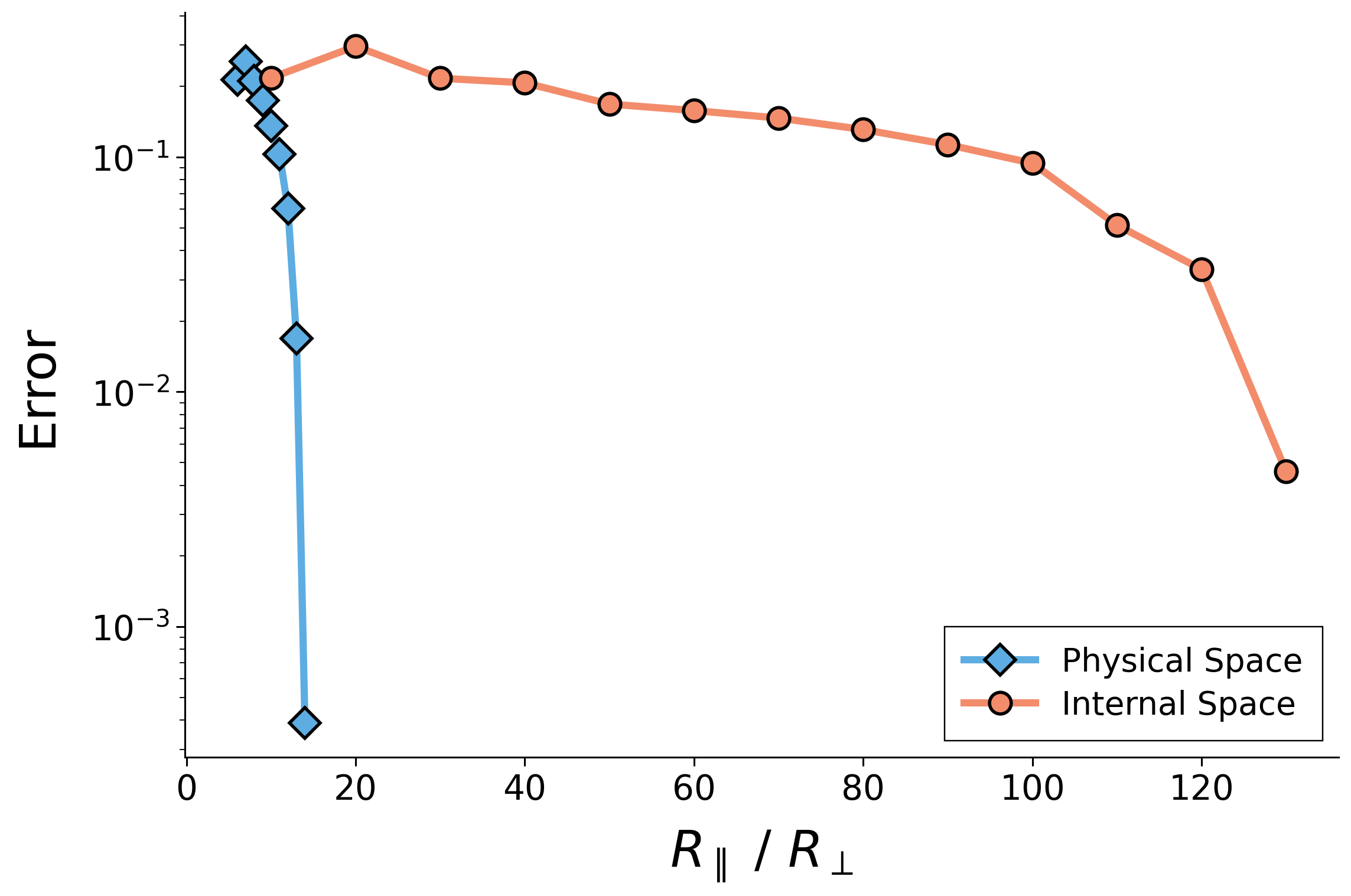}
        \caption{Mixed cutoff (Penrose)}
        \label{fig:truncation_convergence_app}
    \end{subfigure}
        \begin{subfigure}[t]{0.48\textwidth}
        \centering
        \includegraphics[width=\linewidth]{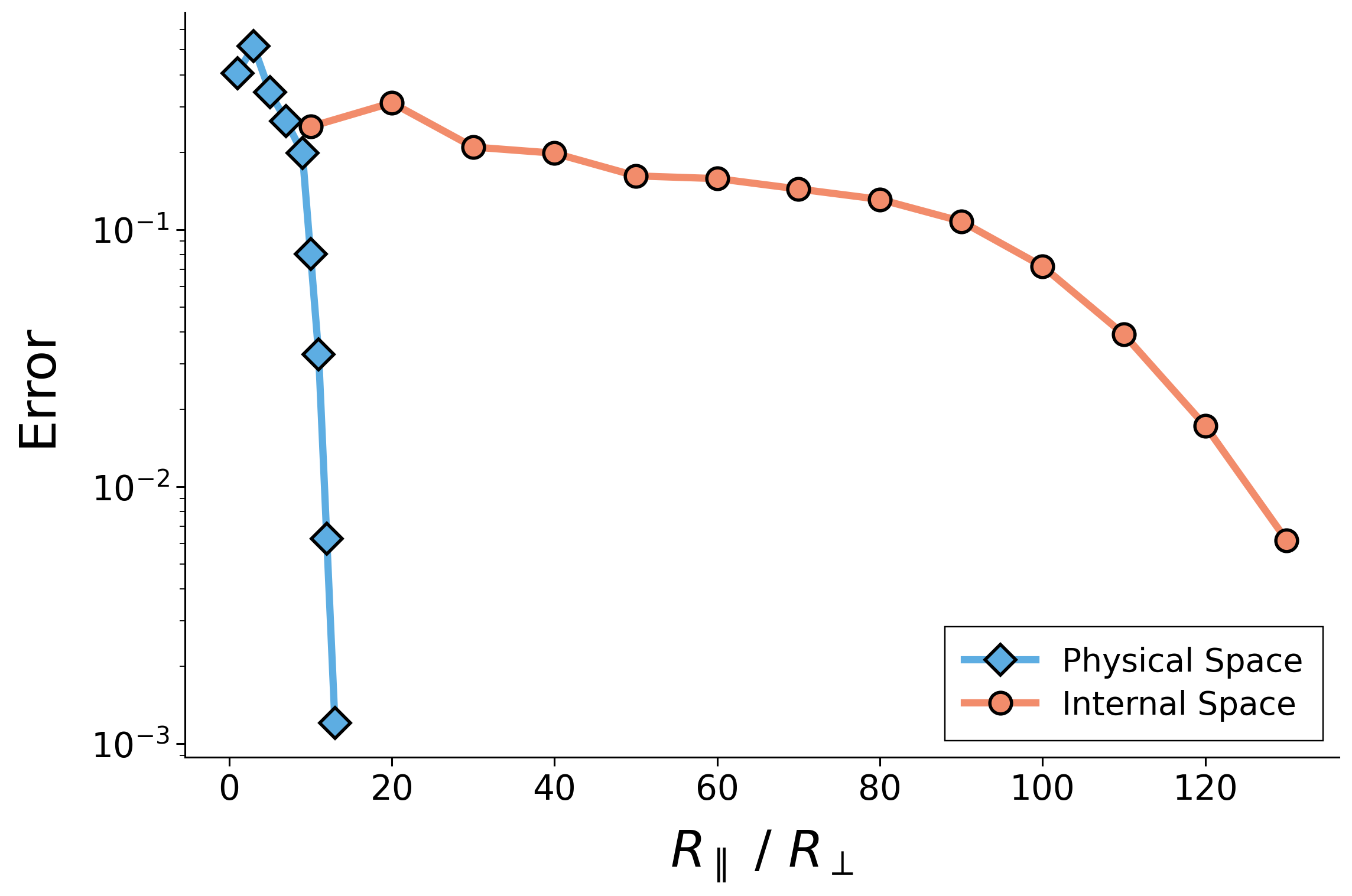}
        \caption{Mixed cutoff (Penrose)}
        \label{fig:cccf_mix_app}
    \end{subfigure}
        \hfill
    \begin{subfigure}[t]{0.48\textwidth}
        \centering
        \includegraphics[width=\linewidth]{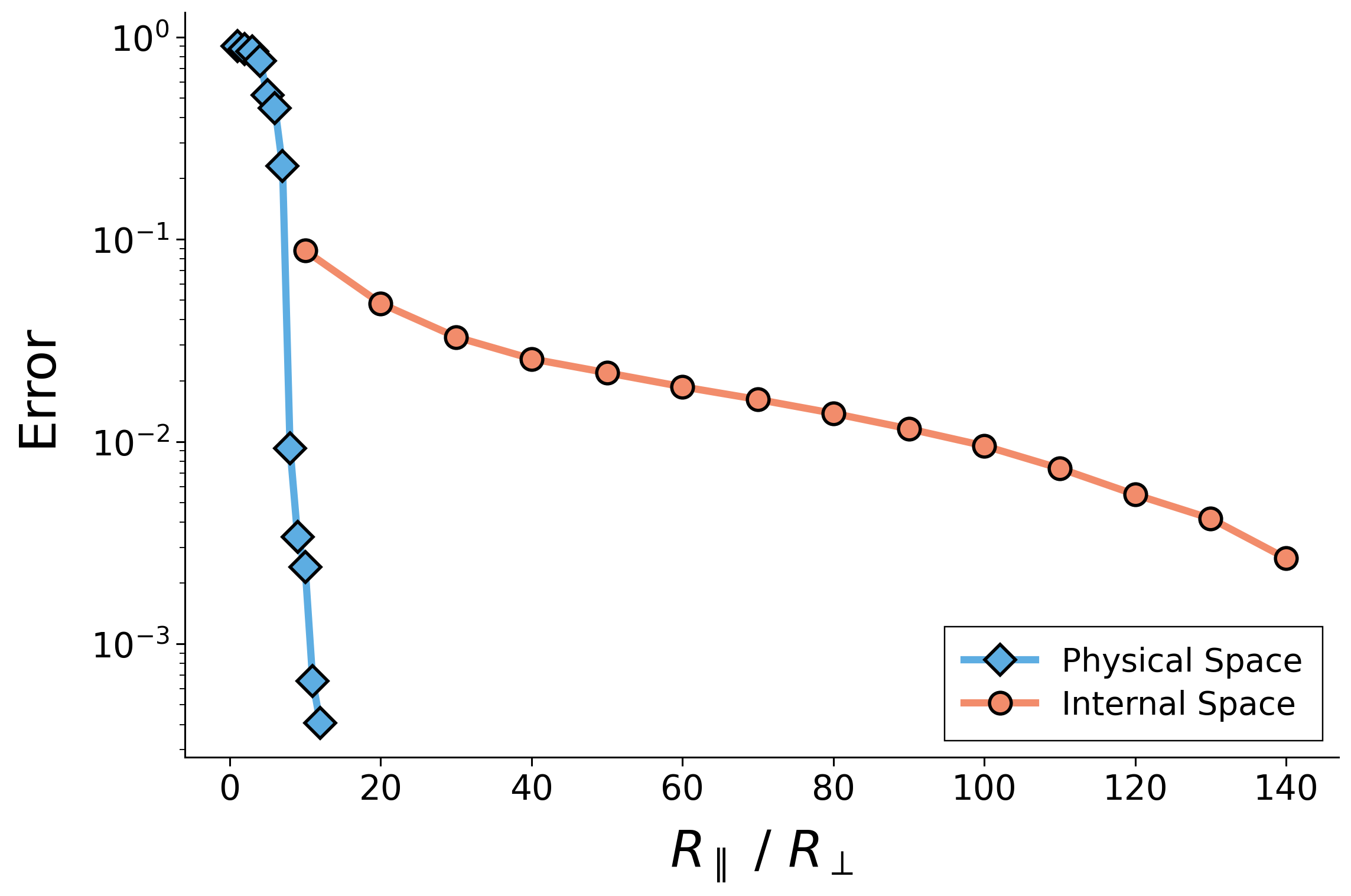}
        \caption{Mixed cutoff (Ammann-Beenker)}
        \label{fig:ab_truncation_convergence_app}
    \end{subfigure}
        \begin{subfigure}[t]{0.48\textwidth}
        \centering
        \includegraphics[width=\linewidth]{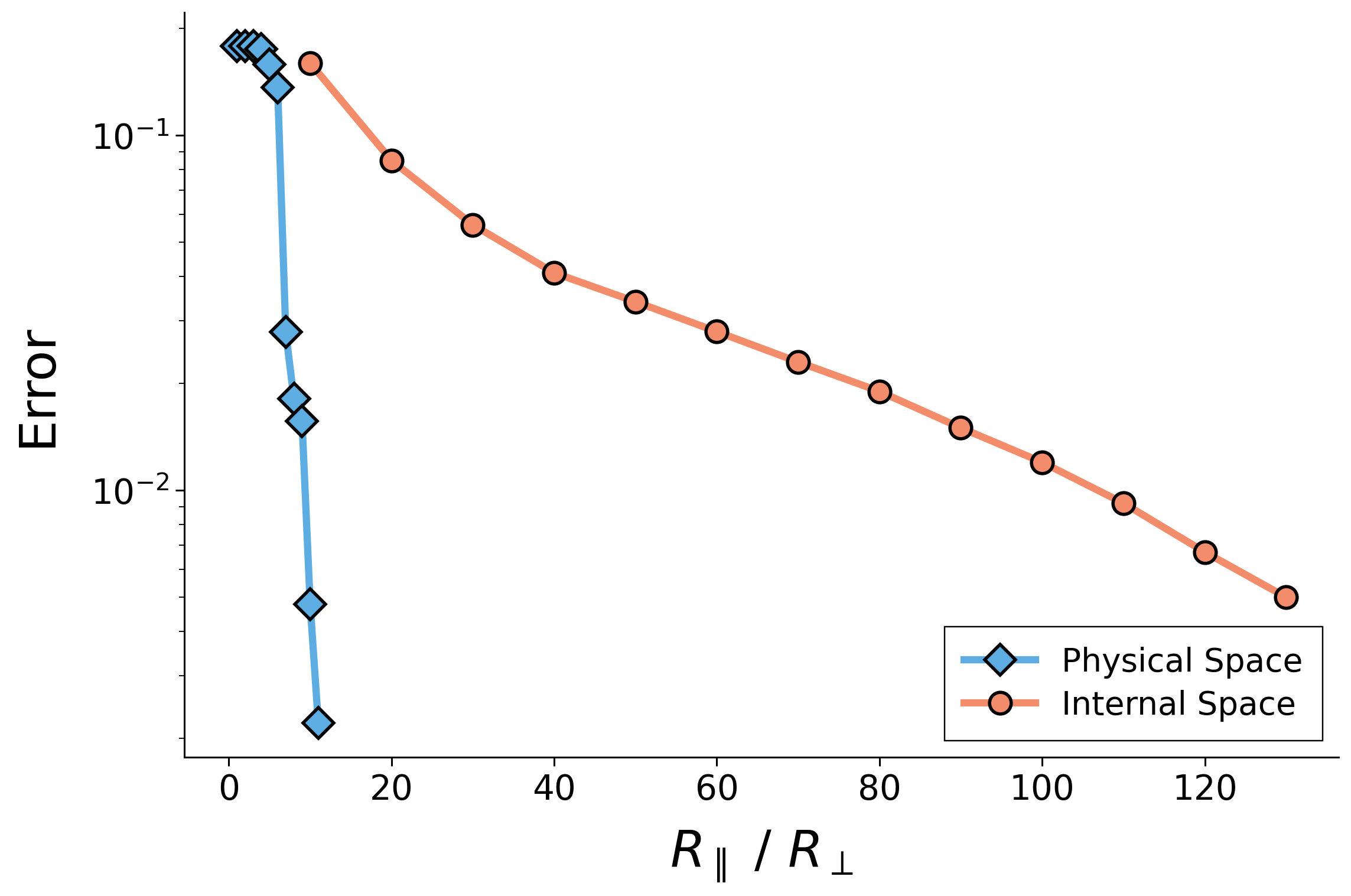}
        \caption{Mixed cutoff (Ammann-Beenker)}
        \label{fig:ab_cccf_mix_app}
    \end{subfigure}
    \caption{
    Convergence with respect to the scattering-channel cutoffs. The left and right columns show the density of states and the current-current correlation, respectively.
    Panels (a)-(b) show the higher-dimensional cutoff convergence for both the Penrose and Ammann-Beenker models. Panels (c)-(f) show the corresponding mixed cutoff convergence.
    Both cutoffs converge rapidly, while the mixed-cutoff tests show that the internal-space direction requires a larger cutoff than the physical reciprocal-space direction. 
    }
    \label{fig:reciprocal_convergence_app}
\end{figure}

For both Penrose and Ammann-Beenker models, we observe similar convergence behaviors for the density of states and the current-current correlation function from Figure \ref{fig:reciprocal_convergence_app}: their errors exhibit a rapid decay under both the higher-dimensional and mixed truncation schemes. The mixed cutoff further separates the degrees of freedom along the physical and internal directions, demonstrating that the convergence with respect to $\rpara$ is significantly faster than that with respect to $\rperp$. 
This separation motivates the use of a relatively small $\rpara$ together with a larger $\rperp$, providing an efficient mixed-truncation for calculations of the density of states and Chern numbers in Section \ref{sec:applications_simulations}.

\section{Numerical implementation}
\label{app:implementation}

This appendix describes the efficient numerical implementation of the reciprocal space framework. Our implementation uses two main techniques: truncated canonical orthogonalization of the nonorthogonal scattering channels and the KPM for evaluating the matrix functions entering the density of states and current-current correlation calculations.

\subsection{Truncated canonical orthogonalization}
\label{app:truncated_orthogonalization}

Since $\Lambda^*$ is dense in the physical reciprocal space $\R^d$, the truncated scattering-channel family becomes increasingly redundant as the cutoff increases, as shown in Figure \ref{fig:reciprocal_module_density}. This progressively dense Fourier module introduces near-linear dependencies among the scattering channels. 
For notational simplicity, throughout this section we omit the cutoff label $L$ and write $\hat{\ham}(\vk)$ and $\hat S$ for their finite-dimensional restrictions to $\Lambda_L^*$, with the same convention applied to all matrices constructed from them.
As the scattering-channel cutoff is enlarged, these near-linear dependencies give rise to increasingly small eigenvalues of $\hat S$, causing its condition number $\kappa(\hat S)$ to grow rapidly.
Consequently, forming \eqref{eq:ldos_def} and \eqref{eq:local_correlation_def} becomes numerically unstable at large scattering-channel cutoffs and can introduce substantial errors into the local quantities.

To stabilize the numerical representation on the truncated scattering-channel basis of dimension $N_G$, we introduce a canonical orthogonalization. The overlap matrix can be decomposed as $\hat S=\overlap \otimes I_{\Nb}$, where $\overlap_{\vG,\vG'}=F(\vG-\vG')$. $\overlap$ is then diagonalized as
\begin{equation*}
\overlap = U\Sigma U^\dagger, ~~ \Sigma = \operatorname{diag}(\sigma_1,\ldots,\sigma_{N_G}).
\end{equation*}
By introducing a small positive threshold $\tauorth > 0$, we restrict the eigenspace to only those eigenvectors satisfying $\sigma_\mu > \tauorth$. This restriction filters the near-zero eigenvalues that cause the ill-conditioning, effectively compressing the system into a stable, lower-dimensional approximation with $N_{\tauorth}$ retained modes. Let $U_{\tauorth} \in \C^{N_G \times N_{\tauorth}}$ and $\Sigma_{\tauorth} \in \R^{N_{\tauorth} \times N_{\tauorth}}$ denote the restricted eigenvectors and eigenvalues, we define the truncated canonical orthogonalization matrix $\hat X:=\hat{\mathcal{X}} \otimes I_{\Nb}$, where
\begin{equation*}
\hat{\mathcal{X}}:=U_{\tauorth}\Sigma_{\tauorth}^{-1/2}
\in\C^{N_G\times N_{\tauorth}}.
\end{equation*}
By construction, $\hat X^\dagger\hat S\hat X=I$, so $\hat X$ directly orthogonalizes the truncated scattering-channel subspace. The Hamiltonian and current matrices in the numerical representation are then given by
\begin{equation*}
\Htilde(\vk) := \hat X^\dagger \hat{\ham}(\vk)\hat X, \qquad
(\Jtilde)_i(\vk) := \hat X^\dagger\hat J_i(\vk)\hat X.
\end{equation*}
The threshold $\tauorth$ balances two competing numerical effects: an excessively small threshold retains near-zero eigenvalues that drive up the condition number of the overlap matrix and cause numerical instability, whereas a large threshold over-truncates the scattering channels, leading to a loss of accuracy. This construction preserves the well-conditioned subspace used to evaluate both local quantities. Using this technique, the local density of states in \eqref{eq:ldos_def} and the local correlation function in \eqref{eq:local_correlation_def} are numerically evaluated by
\begin{equation*}
\ldos(\vk;\kg)
\approx \sum_{\alpha=1}^{\Nb} \left[ \big(U_{\tauorth}\Sigma_{\tauorth}^{1/2} \otimes I_{\Nb}\big) \kg\!\left( \Htilde(\vk) \right) \big(U_{\tauorth}\Sigma_{\tauorth}^{1/2} \otimes I_{\Nb}\big)^\dagger \right]_{\vzero\alpha,\vzero\alpha},
\end{equation*}
\begin{align*}
\lccc(\vk;\kG)_{ij}
&\approx \sum_{\alpha=1}^{\Nb} \Bigg[ \big(U_{\tauorth}\Sigma_{\tauorth}^{1/2} \otimes I_{\Nb}\big) \iint_{\R^2} \kG(E,E')\, \Pspec{\Htilde(\vk)}(\diff E)\, (\Jtilde)_i(\vk) \\
&\qquad\qquad \times \Pspec{\Htilde(\vk)}(\diff E')\, (\Jtilde)_j(\vk)\, \big(U_{\tauorth}\Sigma_{\tauorth}^{1/2} \otimes I_{\Nb}\big)^\dagger \Bigg]_{\vzero\alpha,\vzero\alpha}.
\end{align*}

The truncated canonical orthogonalization not only stabilizes the calculations of the density of states and the current-current correlation function, but also reduces the dimension of the restricted orthogonal subspace and  lowers the cost of matrix-function evaluations.
Table \ref{tab:basis_truncation} compares the initial number of scattering channels $N_G$ with the dimension of the restricted orthogonal subspace $N_{\tauorth}$.
As the scattering-channel cutoff increases, the dimensional reduction becomes increasingly pronounced. Specifically, the ratio $N_{\tauorth}/N_G$ drops dramatically at larger cutoffs, indicating that a rapidly growing fraction of the overlap eigenvalues falls below the orthogonalization threshold $\tauorth$.

\begin{table}[H]
    \centering
    \renewcommand{\arraystretch}{1.08}
    \setlength{\tabcolsep}{7pt}
    \begin{tabular}{|c|ccc|ccc|}
        \hline
        \multirow{2}{*}{Cutoff}
        &
        \multicolumn{3}{c|}{Penrose}
        &
        \multicolumn{3}{c|}{Ammann-Beenker}
        \\
        \cline{2-4}\cline{5-7}
        &
        $N_G$
        &
        $N_{\tauorth}$
        &
        $N_{\tauorth}/N_G$
        &
        $N_G$
        &
        $N_{\tauorth}$
        &
        $N_{\tauorth}/N_G$
        \\
        \hline
        1  & 11    & 11   & 100.0\% & 9     & 9    & 100.0\% \\
        2  & 61    & 61   & 100.0\% & 41    & 41   & 100.0\% \\
        3  & 211   & 211  & 100.0\% & 129   & 129  & 100.0\% \\
        4  & 551   & 528  & 95.8\%  & 321   & 246  & 76.6\%  \\
        5  & 1201  & 835  & 69.5\%  & 681   & 388  & 57.0\%  \\
        6  & 2311  & 1192 & 51.6\%  & 1289  & 547  & 42.4\%  \\
        7  & 4061  & 1641 & 40.4\%  & 2241  & 727  & 32.4\%  \\
        8  & 6661  & 2159 & 32.4\%  & 3649  & 911  & 25.0\%  \\
        9  & 10351 & 2691 & 26.0\%  & 5641  & 1100 & 19.5\%  \\
        10 & 15401 & 3210 & 20.8\%  & 8361  & 1304 & 15.6\%  \\
        11 & 22111 & 3612 & 16.3\%  & 11969 & 1521 & 12.7\%  \\
        12 & 30811 & 3857 & 12.5\%  & 16641 & 1757 & 10.6\%  \\
        13 & 41861 & 4304 & 10.3\%  & 22569 & 1989 & 8.8\%   \\
        14 & 55651 & 5409 & 9.7\%   & 29961 & 2206 & 7.4\%   \\
        15 & 72601 & 6660 & 9.2\%   & 39041 & 2479 & 6.3\%   \\
        \hline
    \end{tabular}
    \caption{Reduction of the scattering channels under truncated canonical orthogonalization for the Penrose and Ammann-Beenker models. Following the high-dimensional truncation in \eqref{eq:high_dim_truncation}, $N_G$ is the initial number of scattering channels and $N_{\tauorth}$ is the dimension of restricted orthogonal subspace, both reported per internal degree of freedom. The orthogonalization threshold is fixed at $\tauorth=10^{-14}$ throughout.}
    \label{tab:basis_truncation}
\end{table}

\subsection{Kernel polynomial method}

Following the truncated canonical orthogonalization, the conventional approach to computing the density of states and current-current correlation functions relies on the exact diagonalization of the orthogonalized Hamiltonian $\Htilde(\vk)$. However, diagonalizing $\Htilde(\vk)$ at every wave vector becomes computationally expensive as the scattering-channel cutoff is enlarged.
We therefore apply the KPM, which approximates kernels by finite Chebyshev expansions and evaluates the resulting matrix functions recursively without explicit diagonalization \cite{weisse2006}.

Let $E_{\min}$ and $E_{\max}$ denote lower and upper bounds for the spectrum of the Hamiltonian. To ensure the numerical stability of the KPM, we rescale the Hamiltonian to $\Hbar(\vk)$ with spectrum contained in $[-1,1]$,
\begin{equation*}
\Hbar(\vk) = \frac{\Htilde(\vk)-bI}{a},
\end{equation*}
where $a=(E_{\max}-E_{\min})/2$, $b=(E_{\max}+E_{\min})/2$. For the local density of states, the spectral kernel $\kg$ is approximated by a one-dimensional Chebyshev expansion. For the local correlation function, the correlation kernel is approximated by a two-dimensional Chebyshev expansion. Specifically, for $E_1,E_2\in (E_{\min},E_{\max})$, the kernels are approximated by
\begin{equation*}
\kg(E_1) \approx \sum_{n=0}^{N_C} c_{n}\, \Tn{\frac{E_1-b}{a}}, \qquad
\kG(E_1,E_2)\approx \sum_{p,q=0}^{N_C} c_{pq}\, T_p(\frac{E_1-b}{a})T_q( \frac{E_2-b}{a}),
\end{equation*}
where $N_C$ is the expansion order, $c_n$ and $c_{pq}$ are the corresponding Chebyshev coefficients.
For the Chern number calculations, we first smooth the Fermi kernel within the bulk gap, where the Hamiltonian has no spectrum, so that its values for all spectral energy pairs remain unchanged \cite{ElbauGraf2002}.
We then apply the Jackson kernel to its finite Chebyshev expansion to suppress Gibbs oscillations \cite{weisse2006}.
Specifically, the two-dimensional coefficients are modified as $c_{pq}\rightarrow g_p^{\rm J}g_q^{\rm J}c_{pq}$, where $g_n^{\rm J}=\left(1-\frac{n}{N_C+2}\right)\cos\frac{\pi n}{N_C+2}+\frac{1}{N_C+2}\cot\frac{\pi}{N_C+2}\sin\frac{\pi n}{N_C+2}$ is the Jackson damping factor.
Substituting these expansions yields the following approximations to the local density of states and the local correlation function,
\begin{equation*}
\ldos^{N_C}(\vk;\kg) = 
\sum_{n=0}^{N_C} c_{n}\sum_{\alpha=1}^{\Nb} \left[ \big(U_{\tauorth}\Sigma_{\tauorth}^{1/2} \otimes I_{\Nb}\big) \Tn{\Hbar(\vk)} \big(U_{\tauorth}\Sigma_{\tauorth}^{1/2} \otimes I_{\Nb}\big)^\dagger \right]_{\vzero\alpha,\vzero\alpha},
\end{equation*}
\begin{align*}
\lccc^{N_C}(\vk;\kG)_{ij} 
&= \sum_{p,q=0}^{N_C}c_{pq} \sum_{\alpha=1}^{\Nb} \Bigg[ \big(U_{\tauorth}\Sigma_{\tauorth}^{1/2} \otimes I_{\Nb}\big) T_p\!\left(\Hbar(\vk)\right) (\Jtilde)_i(\vk) \\
&\qquad\qquad \times T_q\!\left(\Hbar(\vk)\right) (\Jtilde)_j(\vk)\, \big(U_{\tauorth}\Sigma_{\tauorth}^{1/2} \otimes I_{\Nb}\big)^\dagger \Bigg]_{\vzero\alpha,\vzero\alpha}.
\end{align*}
The matrix polynomials $\Tn{\Hbar(\vk)}$ are then evaluated through their action on vectors rather than being formed explicitly.
For any vector $\vv$, their action is generated by the recurrence
\begin{align*}
T_0(\Hbar(\vk))\vv &= \vv, \\[1ex]
T_1(\Hbar(\vk))\vv &= \Hbar(\vk)\vv, \\[1ex]
T_{n+1}(\Hbar(\vk))\vv &= 2\Hbar(\vk)\Tn{\Hbar(\vk)}\vv - T_{n-1}(\Hbar(\vk))\vv.
\end{align*}
This recurrence relation of the Chebyshev polynomials enables an efficient KPM evaluation of local quantities without explicitly diagonalizing $\Htilde(\vk)$. The resulting momentum-resolved local density of states and local correlation function are subsequently averaged using the diffraction-guided PBZ quadratures defined in Sections \ref{subsec:pbz_quadrature} and \ref{subsec:current_current_local_global}. 

At fixed PBZ quadrature, the numerical accuracy is therefore controlled by the scattering-channel cutoff, the truncated canonical orthogonalization threshold $\tauorth$, and the KPM expansion order $N_C$. In our numerical simulations for the density of states and Chern numbers, we empirically use $\tauorth=10^{-14}$ and $N_C=400$, for which the numerical results are stable.

\section{Convergence mechanism of PBZ averaging}
\label{app:pbz-mechanism}

This appendix explains why the diffraction-guided PBZs introduced in Section \ref{subsec:pbz_quadrature} provide more efficient averaging domains than general regions of comparable size.
The PBZ is not merely a large polygon:  
its facets are determined by strong diffraction vectors that connect nearly equivalent scattering channels.  
This geometry produces an additional cancellation between contributions from opposite facets.

From the cut-and-project scheme summarized in Appendix \ref{app:real_reciprocal_geometry}, we have
\begin{equation}
\label{eq:projection-duality}
\ppara^{\rm T} \ppara^* + \pperp^{\rm T} \pperp^*
= 2\pi I_D,
\end{equation}
and hence, for $\vn,\vh\in\Z^D$,
\begin{equation}
\nonumber
\label{eq:scalar-duality}
(\ppara\vn)\cdot(\ppara^{*}\vh)
+ (\pperp\vn)\cdot(\pperp^{*}\vh)
= 2\pi\vn\cdot\vh\in2\pi\Z.
\end{equation}
For $\vG = \ppara^{*}\vh\in\Lambda^{*}$, let
$\vG_{\perp}=\pperp^{*}\vh$ denote its internal  component.  
The standard window formula for the normalized structure factor in \eqref{structure_factor} can also be written as \cite{DuneauKatz1985,BaakeGrimm2013}
\begin{equation}
\label{eq:window-formula}
|F(\vG)| = \frac{1}{\mu_\perp(W)}\left|
\int_W \ee^{\im \vG_\perp\cdot\vy} \diff\mu_\perp(\vy) \right|,
\end{equation}
where $\diff\mu_\perp$ is the Haar measure of the internal space. It represents the average of the internal-space phases over the acceptance window. This average approaches one only when the phase variation across the window vanishes, corresponding to $|\vG_\perp|\to0$. Hence, the strong diffraction vectors selected by $|F(\vG)|\geq 1-\eps$ correspond to momentum shifts with increasingly small internal-space mismatch.

We next connect this observation to the momentum-resolved local density of states $\ldos(\vk;\kg)$ defined in \eqref{eq:ldos_def}.
Using the higher-dimensional representation associated with the cut-and-project construction, we write the local density of states as the restriction of a periodic function $A(\bm\theta;\kg)$ on the higher-dimensional torus.
Its Fourier expansion gives
\begin{equation}
\label{eq:local-Fourier-series}
\ldos(\vk;\kg) = A(\ppara^{\rm T}\vk;\kg)
= \sum_{\vn\in\Z^D} a_{\vn}(\kg)
\ee^{\im(\ppara\vn)\cdot\vk},
\end{equation}
Using the duality relation \eqref{eq:projection-duality}, each Fourier mode transforms under the shift $\vG$ as
\begin{equation}
\label{eq:approximate-period}
\ee^{\im(\ppara\vn)\cdot(\vk+\vG)}
= \ee^{-\im(\pperp\vn)\cdot\vG_{\perp}}
\ee^{\im(\ppara\vn)\cdot\vk}.
\end{equation}
Consequently, a strong diffraction vector acts as an approximate period of the local spectral quantities: translating $\vk$ by $\vG$ changes each mode only through the internal phase $\ee^{-\im(\pperp\vn)\cdot\vG_{\perp}}$.

For a finite domain $\Omega\subset\R^d$, define
\begin{equation}
\nonumber
\label{eq:domain-average}
\dos_{\Omega}(\kg):=\frac{1}{|\Omega|}\int_{\Omega}\ldos(\vk;\kg) \diff\vk,
\qquad
\Phi_{\Omega}(\vr):=\frac{1}{|\Omega|}\int_{\Omega}
\ee^{\im\vr\cdot\vk}\diff\vk. 
\end{equation}
By the quasiperiodic averaging property, the full-space mean retains only the zero Fourier mode.
Substituting \eqref{eq:local-Fourier-series} therefore yields
\begin{equation}
\nonumber
\label{eq:error-identity}
\dos(\kg)=a_{\vzero}(\kg), 
\qquad
\dos_{\Omega}(\kg)-\dos(\kg)
= \sum_{\vn\ne\vzero} a_{\vn}(\kg)
\Phi_{\Omega}(\ppara\vn). 
\end{equation}
This identity separates the Fourier content of the local observable from the geometry of the averaging domain.

For a general regular domain $\Omega_R$ of characteristic radius $R$, the
Gauss-Green formula gives, for $\vr\ne\vzero$,
\begin{equation}
\label{eq:generic-boundary-identity}
\Phi_{\Omega_R}(\vr) = \frac{1}{\im|\Omega_R||\vr|^2} \int_{\partial\Omega_R}(\vr\cdot\bm\nu) \ee^{\im\vr\cdot\vk}\diff S,
\end{equation}
and therefore
\begin{equation}
\nonumber
\label{eq:generic-domain-bound}
|\Phi_{\Omega_R}(\vr)|
\leq C(R|\vr|)^{-1}.
\end{equation}
This is the usual boundary reduction.  
Since a generic boundary is unrelated to the approximate translations in
\eqref{eq:approximate-period}, no further cancellation is built into this estimate.

The Wigner-Seitz geometry of the PBZ provides an additional relation between opposite boundary contributions. 
For the symmetry-related PBZs considered in this work, the active diffraction vectors occur in opposite pairs $\{\pm \vG_{\eps,j}\}_{j=1}^J$, with corresponding opposite facets $E^\pm_{\eps,j}$ satisfying $E_{\eps,j}^{+}=E_{\eps,j}^{-}+\vG_{\eps,j}$.
Defining
$\displaystyle \vu_{\eps,j}:= \vG_{\eps,j}/|\vG_{\eps,j}|$
and grouping the boundary contributions in
\eqref{eq:generic-boundary-identity} into opposite-facet pairs,
we translate $E_{\eps,j}^{+}$ back to $E_{\eps,j}^{-}$ to obtain the exact identity
\begin{align}
\label{eq:paired-facet-identity}
\Phi_{\pbz}(\vr)
=\frac{1}{\im|\pbz||\vr|^2}
\sum_{j=1}^{J} (\vr\cdot\vu_{\eps,j}) \bigl(\ee^{\im\vr\cdot\vG_{\eps,j}}-1\bigr) \int_{E_{\eps,j}^{-}}\ee^{\im\vr\cdot\vk}\diff S.
\end{align}
The factor in parentheses measures the mismatch between the two opposite-facet contributions and would vanish if $\vG_{\eps,j}$ were an exact reciprocal period. 
For the quasiperiodic system, the duality relation expresses this mismatch entirely through the internal-space component
\begin{equation}
\label{eq:phase-mismatch}
\left|\ee^{\im(\ppara\vn)\cdot\vG_{\eps,j}}-1\right|
= \left|\ee^{-\im (\pperp\vn)\cdot \vG_{\eps,j,\perp}}-1\right|
\leq |\pperp\vn|\,|\vG_{\eps,j,\perp}|.
\end{equation}

Set
\begin{equation}
\nonumber
\label{eq:pbz-scales}
R_{\eps}=\operatorname{diam}(\pbz),
 \qquad
\rho_{\eps}=\max_j|\vG_{\eps,j,\perp}|.
\end{equation}
Along the strong-diffraction hierarchy, the selected active vectors have increasingly small internal components, so that $\rho_\eps\rightarrow 0$ as $\eps\rightarrow 0$.
For these regular PBZ geometries, the facet measures and cell volume satisfy the scaling bounds
\[
|E^-_{\eps,j}|\le C R_\eps^{d-1},
\qquad
|\pbz|\ge cR_\eps^d.
\]  
Substituting these estimates and
\eqref{eq:phase-mismatch} into
\eqref{eq:paired-facet-identity} gives
\begin{equation}
\label{eq:pbz-domain-bound}
\left|\Phi_{\pbz}(\ppara\vn)\right|
\leq C\frac{\rho_{\eps} |\pperp\vn|}
 {R_{\eps}|\ppara\vn|}.
\end{equation}
The factor $(R_{\eps}|\ppara\vn|)^{-1}$ is the same boundary reduction available for a regular domain, whereas $\rho_{\eps}|\pperp\vn|$ is the additional PBZ reduction.  The latter
measures how closely the two opposite facets sample the same quasiperiodic
pattern.

For the cut-and-project structures considered here, we assume the standard Diophantine-type control on the physical projection \cite{HaynesKoivusaloWalton2018}, under which $|\ppara \vn|^{-1}$ grows at most algebraically with $|\vn|$.
The spectral smoothing introduced by $\kg$ yields sufficiently rapid decay of the Fourier coefficients $a_{\vn}(\kg)$ to dominate this algebraic small-divisor growth. 
Consequently, the weighted Fourier sum
\begin{equation}
\nonumber
\label{eq:weighted-content}
M_g:=\sum_{\vn\ne\vzero}|a_{\vn}(\kg)|
 \frac{|\pperp\vn|}{|\ppara\vn|}
\end{equation}
is finite. 
Then substituting  \eqref{eq:pbz-domain-bound}  into the exact error identity and summing over the Fourier modes yields
\begin{equation}
\nonumber
\label{eq:pbz-average-error}
|\dos_{\eps}(\kg)-\dos(\kg)| \leq C\frac{\rho_{\eps}}{R_{\eps}}M_g,
\qquad \text{with}\quad \dos_{\eps}(\kg):=\dos_{\pbz}(\kg).
\end{equation}

Moving along the PBZ hierarchy toward stronger diffraction vectors improves the reciprocal-space average in two coupled ways: the Wigner-Seitz cell becomes larger, while the phase mismatch between opposite facets becomes smaller. 
A general regular polygon of the same size benefits only from the former effect.  This additional cancellation, rather than polygonal symmetry alone, explains the efficient PBZ convergence observed in Figure \ref{fig:pbz_convergence}.  
Its stepwise behavior reflects the fact that the active diffraction shell, and hence the PBZ itself, changes only when $1-\eps$ crosses a diffraction amplitude. 
The same geometric cancellation mechanism also applies to PBZ averages of the local current-current correlation function.

\bibliographystyle{apsrev4-2}
\bibliography{apscontrol,bib}

\end{document}